\documentclass[10pt]{iopart}
\usepackage{graphicx}
\usepackage{placeins}
\usepackage{cite}
\usepackage{comment}

\usepackage[bottom]{footmisc}
\usepackage[margin=1in]{geometry}

\expandafter\let\csname equation*\endcsname\relax
\expandafter\let\csname endequation*\endcsname\relax
\usepackage{amsmath}
\usepackage[makeroom]{cancel}

\usepackage[dvipsnames]{xcolor}
\usepackage[bookmarksnumbered,breaklinks=true,colorlinks=true,linkcolor=Red,urlcolor=blue,citecolor=Aquamarine,bookmarksopen=true,driverfallback=dvipdfm]{hyperref}
\usepackage[all]{hypcap}
\usepackage{cleveref}
\hypersetup{
   pdfpagelayout=OneColumn,
   pdfstartview=FitH
}

\usepackage[format=hang,font=small,labelfont=bf,textfont=sf,figurename=Fig.]{caption}[2013/02/03]

\usepackage{ltablex}

\usepackage{makecell}

\usepackage{booktabs,tabularx}

\usepackage{enumerate}

\usepackage{float}

\usepackage{commath}  

\makeatletter
\long\def\@makefntext#1{\parindent 1em\noindent
 \makebox[1em][l]{\footnotesize\rm$\m@th{\arabic{footnote}}$}%
 \footnotesize\rm #1}
\def\@makefnmark{\hbox{$^{\arabic{footnote}}\m@th$}}
\def\@thefnmark{\arabic{footnote}}
\makeatother
\newcommand{\isdraft}{false}

\begin{document}

\setlength{\abovedisplayskip}{18pt}
\setlength{\belowdisplayskip}{18pt}
\setlength{\abovedisplayshortskip}{18pt}
\setlength{\belowdisplayshortskip}{18pt}

\title[Multi-party quantum fingerprinting with weak coherent pulses]{Multi-party quantum fingerprinting with weak coherent pulses: circuit design and protocol analysis}

\author{Hip\'olito G\'omez-Sousa}

\address{Department of Signal Theory and Communications, University of Vigo,\\EI de Telecomunicaci\'on, Campus Universitario, E-36310 Vigo, Spain}

\ead{hgomez@com.uvigo.es}
\vspace{10pt}

\begin{abstract}
Quantum communication has been leading the way of many remarkable theoretical results and experimental tests in physics. In this context, quantum communication complexity (QCC) has recently drawn earnest research attention as a tool to optimize the amounts of transmitted qubits and energy that are required to implement distributed computational tasks. On this matter, we introduce a novel multi-user quantum fingerprinting protocol that is ready to be implemented with existing technology. Particularly, we extend to the multi-user framework a well-known two-user coherent-state fingerprinting scheme. This generalization is highly non-trivial for a twofold reason, as it requires not only to extend the set of protocol rules but also to specify a procedure for designing the optical devices intended for the generalized protocol. Much of the importance of our work arises from the fact that the obtained QCC figures of merit allow direct comparison with the best-known \emph{classical} multi-user fingerprinting protocol, of significance in the field of computer technologies and networking. Furthermore, as one of the main contributions of the manuscript, we deduce innovative analytical upper bounds on the amount of transmitted quantum information that are even valid in the two-user protocol as a particular case. These original analytical bounds are of interest for estimating the realistic protocol performance prior to experimental realizations. Ultimately, comparative results are provided to contrast different protocol implementation strategies and, importantly, to show that, under realistic circumstances, the multi-user protocol can achieve tasks that are impossible by using classical communication alone. Our work provides relevant contributions towards understanding the nature and the limitations of quantum fingerprinting and, on a broader scope, also the limitations and possibilities of quantum-communication networks embracing a node that is accessed by multiple users at the same time.
\end{abstract}

\noindent{\it Keywords\/}: quantum information, quantum fingerprinting, coherent states, multi-party protocol.\\
\hspace{1pt}

\noindent \textit{Journal reference:~~}     New J. Phys. 22, 113004 (2020) \\
\noindent \textit{DOI:~~} \url{https://dx.doi.org/10.1088/1367-2630/abc2e5}
\vspace{10pt}


\section{Introduction}
\label{sec1}
Quantum-communication networks \cite{Merali, Patel, Qiu, Wang, Liu} are widely regarded as prospective central platforms for implementing next-generation distributed information-processing protocols. Such networks will play a crucial role delivering to multiple users enhanced capabilities that are attainable by harnessing the quantum-mechanical properties of light. Many of these quantum-improved capabilities are already well understood nowadays, yet our present-day knowledge and experimental results are mostly limited to the context of restricted frameworks, e.g.  point-to-point communication schemes merely involving two distant users. Extant application examples that exploit quantum enhancements include improved metrology systems \cite{Giovannetti}, machine learning with quantum neural networks \cite{Dunjko1, Biamonte}, and quantum cryptographic protocols that deliver a level of security that is classically unattainable \cite{Bennett, Ekert, HK}.

Quantum communication complexity (QCC) has emerged as a discipline to study the communication cost of distributed quantum protocols. Namely, the minimum amount of information, measured in qubits, that must be transmitted through a quantum network to solve distributed computational problems \cite{Buhrman, Brassard, Buhrman2, Arrazola, Kumar, Xu, Guan}. In this general context, the network users are assumed to have certain quantum resources at their direct disposal, such as entangled-photon sources and quantum channels. Equivalently, in theoretical computer science, (classical) communication complexity deals with bits of transmitted information when the parties are restrained to use classical resources alone \cite{Yao, Kushilevitz}. Both the classical and the quantum cases have customarily corresponded to a scenario in which just two parties, Alice and Bob, receive respective $N\textrm{-bit}$ input sequences $x,y\in {{\left\{ 0,1 \right\}}^{N}}$. Subsequently, under this conventional scenario, the pair of distant users seek to cooperatively compute the value of an arbitrary Boolean function $f(x,y)$. They are subject to the important constraint of keeping as low as possible the amount of transmitted information. Remarkably, a positive quantum advantage was demonstrated in this context, viz. theoretical results \cite{Buhrman, Brassard, Arrazola, Kumar} and experimental implementations  \cite{Xu, Guan, Trojek, Zhong} have been published hitherto reporting distributed information-processing protocols that implement computational tasks that are unfeasible by exclusively using classical communication.

The specific model of interest for communication complexity that we adhere to in this work is called \textit{simultaneous message passing} model and it was also subscribed in all the pertinent references above, both classical and quantum. This model was first described in the classical context by Yao \cite{Yao} and it features two basic requirement assumptions suitably applicable in any real-world technological settings:  (i) Alice and Bob are not permitted to retrieve shared randomness; (ii) Alice and Bob each send messages to a third party, the referee, whose ultimate role is to determine the value of the function $f(x,y)$. Specifically, this latter requirement means that Alice and Bob may not communicate to each other in a direct manner and, additionally, for computational purposes, the communication path to the referee node must be one-way \textit{only.}

\begin{figure}[h]
\centerline{\includegraphics[width=0.7\columnwidth,draft=\isdraft]{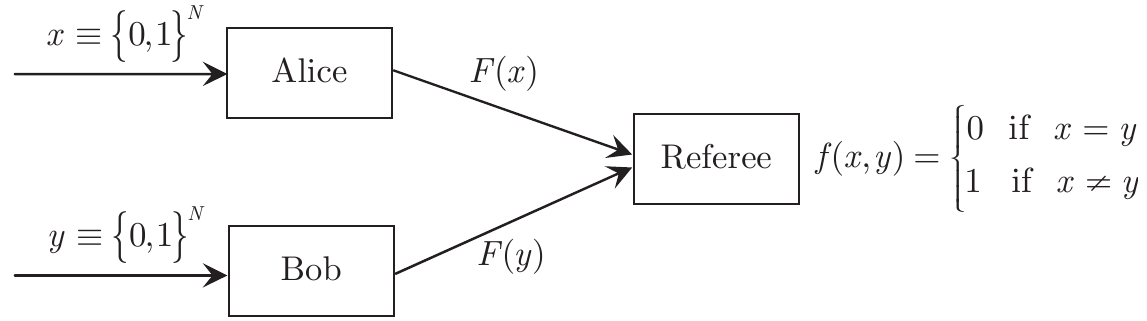}}
\caption{Schematic illustration of a general two-user fingerprinting protocol. The represented scheme is applicable to both classical and quantum protocols. Alice and Bob receive, or already possess in their custody, raw classical binary inputs $x$ and $y$, respectively, comprising messages of $N$ bits. They apply a certain mathematical function $F(\cdot)$ to their respective bit strings $x$ and $y$. The binary outputs of this function represent $M\textrm{-bit}$ fingerprints that both users encode as quantum or classical information. The referee receives the incoming signals from Alice and Bob and, after applying a certain classical or quantum procedure, concludes whether the original $N\textrm{-bit}$ sequences coincide or differ.}
\label{fig1}
\end{figure}

Amidst all the QCC protocols, the class of quantum fingerprinting (QF) protocols is undoubtedly the most actively investigated in recent years \cite{Buhrman, Arrazola, Kumar, Xu, Guan, Zhong, Beaudrap, Horn, Du, Massar, Garcia}. QF was first introduced by Buhrman \emph{et al} in \cite{Buhrman} as a prominent theoretical problem whose classical fingerprinting counterpart had been previously well-established \cite{Yao, Kushilevitz, Babai, Newman, Kremer, Ambainis2}. The seminal inception of the fingerprinting concept naturally appears in the field of (classical) communication complexity as a practical mechanism for solving the \textit{equality} problem in a distributed framework, i.e. for discerning if two separate distant bit strings $x$ and $y$ are effectively the same string. Figure~\ref{fig1} sketches a general two-user fingerprinting protocol. In this two-user framework, the particular Boolean function can be simply defined as
\begin{equation}
f(x,y)=\left\{ \begin{aligned}
  & 0\text{ if }x=y, \\
 & 1\text{ if }x\ne y. \\
\end{aligned} \right.
\label{eq0}
\end{equation}
Obviously, the two users Alice and Bob together with the referee can always trivially achieve the goal of computing (\ref{eq0}) by communicating to the referee the entire $N\textrm{-bit}$ inputs $x$ and $y$. However, if they choose instead to send \textit{fingerprints} $F(x)$ and $F(y)$ of the original inputs $x$ and $y$, they can always succeed with a sought reduced communication cost when an arbitrarily small probability of error is tolerated. We remark that, according to our notation, $F(x)$ (and analogously $F(y)$) may refer to both classical or quantum cases. In a classical protocol, fingerprint $F(x)$ consists of a bit string shorter than $x$ that may be computed as a hash function. Conversely, in the quantum case, $F(x)$ represents a bit string longer than $x$ that is then encoded and transmitted as qubits in the form of quantum states, which are ultimately quantum-processed by the referee. In particular, optimal classical fingerprinting protocols are known to require fingerprints of at least $\Omega\small(\sqrt{N}\small)$  bits  \cite{Babai, Newman}, which is a fundamental lower bound. By sending quantum states, in comparison, Alice and Bob may, under certain conditions, require fingerprints of just $\mathcal{O}\small(\log_2{N}\small)$ qubits to solve the same problem subject to an identical probability of error as in the classical protocol, which represents an exponential reduction \cite{Buhrman, Arrazola}. The direct comparison between bits and qubits is fully justified by virtue of Holevo's theorem \cite{Holevo}, which establishes that $M\textrm{-bit}$ classical messages cannot be encoded into, and then decoded from, quantum messages comprising less than $M$ qubits.

Besides the achievable quantum advantage in the field of communication complexity, research in QF was also sparked by some other relevant attainments that quantum fingerprints can bestow but are beyond the scope of this document. In particular, QF was also applied to construct a theoretic quantum automaton with an exponential improvement in size when compared to a classical randomized automaton \cite{Ambainis1, Ablayev}. Another application consists of utilizing QF as a proposed cryptographic hash function \cite{Gavinsky}. Finally, as an eventual application already noted in \cite{Arrazola}, QF may also play a pivotal role improving certain extant schemes for quantum digital signatures in \cite{Clarke, Dunjko2}.

The first successful experimental demonstrations of QF protocols were reported in \cite{Beaudrap, Horn, Du} and they consisted of distributed implementations of the equality problem (\ref{eq0}). A downside of all these initial experimental efforts lies in the fact that their fingerprint states must be extremely entangled, even when the input size $N$ is small. These experimental demands greatly surpass those that are achievable with current technology, except when restricting the transmitted information to a few qubits per user. Therefore, their current practical interest is very limited. A different approach for implementing quantum fingerprinting was proposed in \cite{Massar, Garcia}. However, these other theoretical approaches demand the preparation of quantum states of fixed photon number, which is still a challenging task from the experimental point of view \cite{Gauthier}. Recently, another innovative theoretical proposal for a QF protocol that is suitable to be implemented with present-day technology without requiring entanglement was published in \cite{Arrazola}. In this avant-garde protocol, Alice and Bob send coherent states of low amplitude that the referee interferes in a balanced beamsplitter. On the basis of this protocol, a pioneering proof-of-principle implementation that needs to send less information than the best-known classical protocol \cite{Babai} was reported in \cite{Xu}. However, this experiment in \cite{Xu} employs an improved referee strategy tacitly accompanied by \textit{numerical} techniques for its analysis, instead of the original \textit{analytical} method in \cite{Arrazola}. Finally, a recent further enhanced version of the experiment in \cite{Xu} was detailed in \cite{Guan}. This enhanced experimental setup makes use of ultralow-noise SNSPDs (superconducting nanowire single-photon detectors) and it beats not only the best-known classical protocol but also the classical theoretical limit discussed in \cite{Babai, Newman}.

\begin{figure}[h]
\centerline{\includegraphics[width=1\columnwidth,draft=\isdraft]{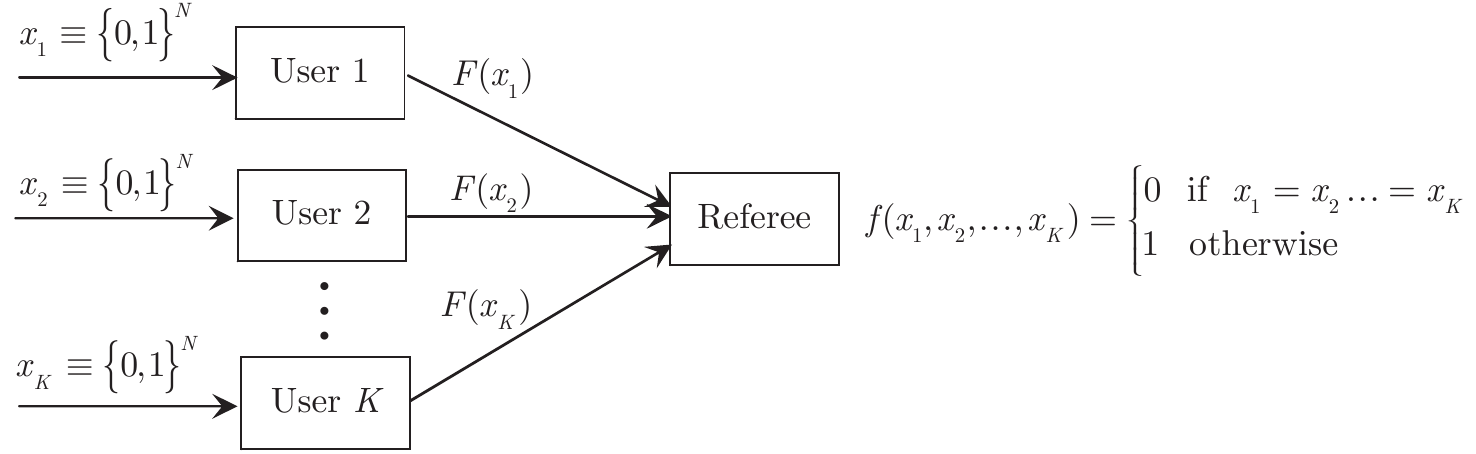}}
\caption{Schematic of a generalized fingerprinting protocol consisting of $K$ users, or parties. Every user receives, or already owns, a raw $N\textrm{-bit}$ sequence $x_k,$ with $1\leq k\leq K$. Afterwards, each party transmits signals that encode $M\textrm{-bit}$ fingerprints $F(x_k)$, as either classical or quantum information. By processing the incoming signals, the referee establishes whether the original binary strings $x_k$ are coincident to each other, or whether at least one of them is different from the rest of strings.}
\label{fig2}
\end{figure}

As our main contribution, the present work is focused on extending the two-user coherent-state QF protocol in \cite{Arrazola} to multiple users ($K\geq2$), including the essence of all the aforementioned enhancements advocated in \cite{Xu, Guan}. In particular, our $K\textrm{-party}$  proposal retains the analytical character of the two-party methods in \cite{Arrazola} while also preserving the benefits of the improved referee's rules in \cite{Xu, Guan}. Just for the sake of clarity, in the general $K\textrm{-party}$ framework for the equality problem, each user receives a binary input sequence $x_k$, with $1\leq k\leq K$. They then send to the referee node their respective fingerprints $F(x_k)$ encoded as either classical or quantum information. The referee's task is to determine if all the original $K$ inputs $x_k$ are the same or not, as sketched in figure \ref{fig2}.
Our proposed extension of coherent-state QF for more than two users has many noteworthy implications of intrinsic importance in the study of quantum networks. First, the communication cost analysis of our extended QF protocol can be directly compared to results that exist for an analogous \emph{classical} $K\textrm{-user}$ protocol \cite{Fischer}, which are of pragmatic interest in the field of distributed computational algorithms involving multiple users. Either in the classical or in the quantum case, the main goal ultimately consists of minimizing the required energy expenditures, which are related to the amount of transmitted information. Furthermore, our work represents a contribution towards the comprehension of multi-user quantum networks, in similar fashion as other unrelated recent proposals such as \cite{Zhu, Grass}, which are committed to introduce multipartite QKD (quantum key distribution) protocols that entail a central network node. Finally, since it is accepted that QCC is connected to some foundational aspects of quantum mechanics \cite{Buhrman2, Arrazola}, our work may also have an impact in helping to expand the knowledge related to some underlying physical phenomena present in the quantum world. This fundamental knowledge includes, as an example, the per-user information-carrying capacity of a quantum channel, and the relationship between multipartite entanglement and nonorthogonality.

Before concluding this introduction, we present the structure of the document's remainder, and, at the same time, we also introduce some other accompanying prominent contributions of this work. The rest of the document starts briefly describing in section 2 the groundwork basics of coherent-state quantum fingerprinting. Most of these preliminaries are essential to become acquainted with notation and concepts used later in the subsequent development of our QF extension. Section 3 provides various linear-optics innovative generalizations of the ordinary 50:50 beamsplitter concept that thus far was used for the two-user protocols in previous works \cite{Arrazola, Xu, Guan}. Section 4 is devoted to the analysis of our multi-user quantum fingerprinting protocol, presenting various suitable referee strategies and making use of the new generalized circuit designs exposed in the preceding section. In this section 4, we also introduce original analytical upper bounds on the amount of transmitted quantum information. Resembling the two-user bounds in \cite{Arrazola}, but unlike the numerical methods employed in \cite{Xu, Guan}, our novel bounding method is entirely analytical in nature. This fact allows us to compute with a low computational cost upper bounds on the amounts, per user, of both transmitted qubits and energy. Remarkably, these new analytical bounds may be applied, as a particular case, to the conventional two-user setup in \cite{Xu, Guan}. Further, another relevant feature is that they can be easily used in an experimental setting, just by taking a few preliminary measurements in the classical optical regime.   Next, section 5 compares our multi-user protocol with the best-known analogous classical protocol described in \cite{Fischer}, and with the classical limit deduced in this manuscript's appendices. Importantly, we assess in this section the protocol resilience against experimental errors, and prove that the quantum protocol can beat, under certain circumstances, the best-known classical protocol and the classical limit. In closing the regular part of the manuscript, the last section is committed to present main conclusions and future work perspectives. In addition, the paper contains three appendices as well. A detailed list of symbols used in the paper is exposed in appendix A. Regarding appendix B, it includes exhaustive mathematical derivations of all the upper bounds in section 4. Finally, the multi-party classical limit used in section 5 is derived in appendix C.

\section{Fundamentals of coherent-state quantum fingerprinting}
\label{sec2}
Besides introducing relevant notation, this section contains an abridged description, including some novel explanatory contributions, of the two-user coherent-state protocol proposed in \cite{Arrazola}, which was later adapted with improvements for the experimental deployments in \cite{Xu, Guan}. Hereof, as a requirement inherited from all the other preceding QF protocols in \cite{Buhrman, Beaudrap, Horn, Du, Massar, Garcia}, this protocol demands the users to apply an error-correcting code (ECC) for the sole purpose of amplifying the Hamming distance between the input bit strings of Alice and Bob\footnote{For example, if the original inputs $x$ and $y$ perfectly match except for one single bit, then the ECC will output binary strings that differ in a much larger number of bits. Unlike traditional applications of ECC in digital communication systems, only the encoding part of the ECC implementation is used in quantum fingerprinting protocols, not the decoding part.}. An ECC can be mathematically modelled as a function $E:{{\left\{ 0,1 \right\}}^{N}}\to {{\left\{ 0,1 \right\}}^{M}}$ such that $E(x)$ is the so-called codeword associated with Alice's input $x$. Ultimately, Alice encodes $E(x)$ as quantum information that she transmits through her channel; the description on Bob's side is analogous. The ratio between the lengths of $E(x)$ and $x$ is called the rate of the ECC and it is defined in this work as $c=\frac{M}{N}>1$. Another important parameter of the ECC is the minimum Hamming distance between any two different codewords. Related to this ECC distance, we define an ECC parameter $\delta$ that designates the maximum fraction of bits in which any two codewords $E(x)$ and $E(y)$, satisfying the requirement $E(x) \neq E(y)$, have the same bit values. As a consequence, the minimum distance of the ECC can be simply computed using $\delta$ as $(1-\delta)M$. Without any loss of generality, we assume, just as in all the previous works on the subject, that an entire $N$-bit input $x$ can always be mapped by the ECC into an $M$-bit codeword $E(x)$. If this were not the case, the same statistical behaviour studied in this paper could be reproduced by slicing the input bit strings into smaller blocks.

The coherent-state QF protocol, as noted above, overcomes all the implementation issues present in earlier QF proposals and it makes quantum fingerprinting practical with current technology. In lieu of requiring either entangled states or a fixed number of photons, Alice and Bob each send a so-named ``coherent state in the fingerprint mode'' \cite{Arrazola}. This coherent fingerprint state can be rigorously defined for Alice as  ${{\left| \alpha  \right\rangle }_{x}}={{D}_{x}}\left( \alpha  \right)\left| 0 \right\rangle $, where $\alpha$ is a complex number and ${{D}_{x}}\left( \alpha  \right) =\exp \left( \alpha a_{x}^{\dagger }-{{\alpha }^{*}}{{a}_{x}} \right)$  is the displacement operator corresponding to the annihilation operator ${{a}_{x}}=\frac{1}{\sqrt{M}}\sum\nolimits_{m=1}^{M}{{{(-1)}^{E{{(x)}_{m}}}}{{b}_{m}}}$. Each term in this summation represents a time-bin mode; furthermore, $b_m$, with $1 \leq m \leq M$, denotes the annihilation operator of the $m\text{th}$ mode and $E(x)_m$ is the $m\text{th}$ bit in codeword $E(x)$. The signal states on Bob's side are analogous, replacing $x$ with $y$. Elementary calculation leads to a simple equivalent expression of the signal states as a train of $M$ weak coherent pulses:
\begin{equation}
{{\left| \alpha  \right\rangle }_{x}}= \bigotimes_{m=1}^M{{\left| {{(-1)}^{E{{(x)}_{m}}}}\frac{\alpha }{\sqrt{M}} \right\rangle }_{m}}.
\label{eq1}
\end{equation}
The total mean photon number corresponding to the entire train of pulses is $\mu=\abs{\alpha}^2$, whereas the mean photon number per each individual pulse in the sequence is $\mu_{\textrm{pulse}}=\frac{\mu}{M}$. It is worth noting that all the coherent states that form together the fingerprint state (\ref{eq1}) have the same amplitude, but their individual phases that encode the information are determined by the specific binary codeword $E(x)$, which itself depends on the particular raw input string $x$.

\begin{figure}[h]
\centerline{\includegraphics[width=0.9\columnwidth,draft=\isdraft]{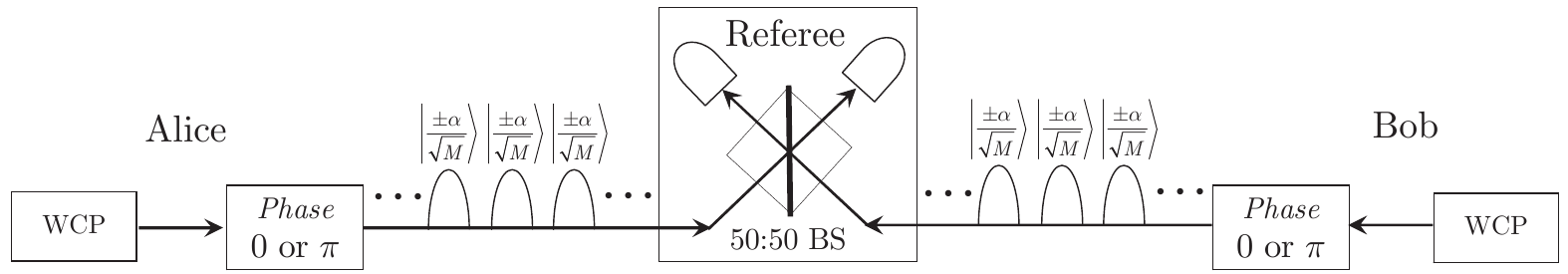}}
\caption{Illustration of a two-user coherent-state quantum fingerprinting protocol. Alice and Bob each send a sequence of $M$ weak coherent pulses whose phases (0 or $\pi$) are modulated according to codewords of an error correcting code (ECC). These codewords are determined depending on the raw binary inputs Alice and Bob want to fingerprint. The incoming individual pulses interfere in a standard 50:50 beamsplitter (BS) located at the referee's circuit. By observing at least one of the two detectors and counting its clicks, the referee infers if the two complete trains of coherent states from Alice and Bob are either the same or different. In an ideal implementation, one of the detectors may only fire if two incoming individual pulses are equal, and the other detector may only fire if two incoming individual pulses are different.}
\label{fig3}
\end{figure}

In the two-user coherent-state QF protocol, the referee must rely on a quantum measurement to verify if the phases of pairs of arriving pulses are equal or different. A simple practical way of implementing such a measurement involves a standard balanced beamsplitter wherein the incoming individual pulses interfere as depicted on figure \ref{fig3}. In the ideal case, whenever a click is recorded on the output detectors, the referee unambiguously knows whether the phases in a pair of incoming pulses are the same or not. It must not escape our notice that, in order to produce a correct interference at the referee's beamsplitter, Alice and Bob need a certain method for establishing a common phase reference. This phase reference may be established before starting the protocol itself or, alternatively, the referee can incorporate phase-locking techniques into her setup. In fact, this latter alternative may be implemented by exploiting different practical methods already developed in \cite{Rubenok,Liu2,Tang} within the mature field of quantum key distribution (QKD).

For the sake of an easy generalization to multi-user instances in this document, we label the two-user protocol detectors as ``1'' and ``2''. By convention, a click in detector ``1'' reveals lack of relative parity in the two phases, whereas a click in detector ``2'' indicates that the two phases are coincident. In this manner, we can now summarize the basic coherent-state QF protocol steps in an ideal implementation:

\begin{itemize}
\item [\emph{(i)}] Alice and Bob agree to use a common ECC and a common value of $\alpha$.
\item [\emph{(ii)}] They prepare coherent fingerprint states $\left|\alpha\right\rangle_x$ and $\left|\alpha\right\rangle_y$ using their respective input sequences $x$ and $y$ according to the quantum state in (\ref{eq1}).
\item [\emph{(iii)}] Both parties send these pulse trains to the referee through their respective quantum channels.
\item [\emph{(iv)}] The referee interferes the individual pulses using a standard 50:50 beamsplitter and she announces that the original inputs $x$ and $y$ are different, i.e. $f(x,y)=1$, if and only if at least one click is observed in detector ``1''. As is apparent from this last statement, it suffices for the referee to observe just one detector when the analysis is constrained to an ideal (defectless) implementation\footnote{Even in nonideal realistic implementations, just one detector is enough as is explained next. Nonetheless, the pioneering referee rules in \cite{Arrazola} demand observing two detectors in the realistic case (and just one detector in the ideal one).}.
\end{itemize}

In the absence of experimental imperfections, such as a flawed beamsplitter or dark counts in the detectors, the referee always announces the correct outcome $f(x,y)=0$ with certainty, whenever the original inputs of Alice and Bob are equal, i.e. $x=y$. This errorless referee behaviour for the case $x=y$ is due to the fact that the only possible detector responses are either clicks in detector ``2'' only or no clicks at all in both detectors. For the other case $x\neq y$, error probability $p_\textrm{error}$ is the same as the probability of obtaining no clicks in detector ``1''. In particular, after the individual pulses interfere in the ideal referee beamsplitter, independently of their relative phases, there will always be a coherent state $\left| \pm \sqrt{2}\frac{\alpha }{\sqrt{M}} \right\rangle $ going into one detector and the vacuum entering the other detector. The click probability is calculated from the Poissonian statistics of the coherent states as ${{p}_{\textrm{click}}}=1-\exp \left( -2\frac{{{\left| \alpha  \right|}^{2}}}{M} \right)$. Accordingly, the worst-case error probability is simply ${{p}_{\textrm{error}}}={{(1-{{p}_{\textrm{click}}})}^{(1-\delta )M}}$, because the minimum amount of pulses that may potentially produce clicks in detector ``1'' is $(1-\delta)M$, as dictated by the distance of the ECC. Introducing into this last equality the above expression of $p_\textrm{click}$ and then solving for ${{\left| \alpha  \right|}^{2}}$, we obtain
\begin{equation}
{{\left| \alpha  \right|}^{2}}=\frac{1}{2(1-\delta )}\ln \left( \frac{1}{{{p}_{\textrm{error}}}} \right).
\label{eq2}
\end{equation}
This equation in (\ref{eq2}) provides the minimum mean photon number of each entire train of pulses that is needed to get a desired error level in the referee outcomes, in the ideal case. Notice that this minimum value of the total mean photon number ${{\left| \alpha  \right|}^{2}}$ depends exclusively on the error probability and on the chosen ECC. Under the ideal premises considered so far, for fixed $p_\textrm{error}$ and $\delta$, mean photon number ${{\left| \alpha  \right|}^{2}}$ remains constant regardless of the raw message length $N$ of Alice's and Bob's binary inputs. Also, we bring into attention that no mathematical approximations were invoked in this work in order to obtain (\ref{eq2}).

Up to this point, we have considered an ideal scenario only; however, any practical QF implementation will inevitably be affected by experimental imperfections. These imperfections render unusable, or at least highly impractical, the decision rule presented above. This is so because detector ``1'' may fire even if the individual input pulses at the beamsplitter are equal. Nonetheless, in case of small imperfections, we may expect the total number of clicks registered in ``1'' when $x\neq y$ to be much larger than when $x=y$. Similarly, we presume the opposite behaviour for detector ``2'', regarding the number of clicks for the respective cases $x\neq y$ and $x=y$. Exploiting these statistical behaviours in the detectors, different decision rules can be contrived to make the QF protocol robust to experimental errors. Particularly, the rule proposed in \cite{Arrazola} is based on calculating a fraction of clicks ${{f}_{2}}=\frac{{{D}_{2}}}{{{D}_{1}}+{{D}_{2}}}$, where $D_k$ is the total number of clicks in detector $k=1,\,2$ for the entire quantum pulses from Alice and Bob. Applying Hoeffding's inequality \cite{Hoeffding} under this rule, an analytical expression analogous to (\ref{eq2}), but including the effects of imperfections, was deduced in \cite{Arrazola}. Though valid, this referee strategy in \cite{Arrazola} was found in \cite{Xu} to be extremely sensitive to small variations in the parameters that quantify the errors caused by imperfections, which hinders its experimental applicability.

The improved referee strategy proposed in \cite{Xu} for coping with experimental errors takes into consideration the amount $D_1$ only; no fraction of clicks is needed, unlike the other rule mentioned above. This referee strategy consists on using a threshold value $r$ such that outcome $f(x,y)=1$ is announced if and only if $D_1>r$ is observed. The value of $r$ is defined as ensuring the equality ${{p}_{\text{error }\!\!|\!\!\text{ }x=y}}={{p}_{\text{error }\!\!|\!\!\text{ }x\ne y}}$, with ${{p}_{\text{error }|\text{ }x=y}}=\Pr \left( D_{1}^{\text{E}}>r \right)$ and ${{p}_{\text{error }|\text{ }x\ne y}}=\Pr \left( D_{1}^{\text{D}}\le r \right)$. In these two probabilities, $D_{1}^{\text{E}}$ ($D_{1}^{\text{D}}$) represents the random variable that models the number of clicks at output ``1'' when the entire input sequences are equal (different). The value of $r$ may be numerically computed by approximating the numbers of clicks $D_{1}^{\text{E}}$ and $D_{1}^{\text{D}}$ by binomial distributions, and then computationally looking up in inverse distribution tables. In particular, the binomial distributions that include imperfection effects may be defined as $D_{1}^{\text{E}}\sim \textsf{Bin}\left( M,p_{\text{click,1}}^{\text{E}} \right)$ and $D_{1}^{\text{D}}\sim \textsf{Bin}\left( M,\ (1-\delta )p_{\text{click,1}}^{\text{D}}+\delta \ p_{\text{click,1}}^{\text{E}} \right)$, where $p_{\text{click,1}}^{\text{E}}$ and $p_{\text{click,1}}^{\textsf{D}}$ are the probabilities of detector ``1'' firing for the cases of equal and different individual input pulses, respectively. These probabilities are given by
\begin{subequations}
\begin{gather}
p_{\text{click,1}}^{\text{E}}=1-{\exp\left({-\frac{2(1-v){{\left| \alpha  \right|}^{2}}}{M}}\right)}+{{p}_{\text{dark}}},
\\[12pt]
p_{\text{click,1}}^{\text{D}}=1-{\exp\left({-\frac{2v{{\left| \alpha  \right|}^{2}}}{M}}\right)}+{{p}_{\text{dark}}}.
\end{gather}
\label{eq3}
\end{subequations}\\[0pt]
In these two equations above, $p_\textrm{dark}$ is the dark count probability and $v$ is the visibility, which quantifies the actual contrast of the interferometer. We emphasize that the traditional definition of interferometric visibility (also known as \emph{fringe contrast}; see, for example, page 12 of \cite{Ellis}) commonly used in optics and quantum photonics is \emph{not} the same that applies in the present case for $v$.  Also, even though visibility $v$ is extensively utilized in \cite{Arrazola,Xu,Guan}, a concise definition is lacking in these references. Here, we provide such a definition for the two-user case, which later in the manuscript is extended to $K$ users with an arbitrary $ K \geq 2 $:
\begin{equation}
v=\frac{1}{2}\left( 1+\frac{g_{1}^{\text{D}}-g_{1}^{\text{E}}}{2} \right).
\label{eq3b}
\end{equation}
This last equation contains the equal-input gain $g_{1}^{\text{E}}=\frac{M\mu _{1}^{\text{E}}}{{{\left| \alpha  \right|}^{2}}}$, which can be expressed as the ratio of the mean photon number $\mu _{1}^{\text{E}}$ at output ``1'' and the nominal photon number $\frac{{{\left| \alpha  \right|}^{2}}}{M}$ at each input. The other gain $g_{1}^{\text{D}}=\frac{M\mu _{1}^{\text{D}}}{{{\left| \alpha  \right|}^{2}}}$ is analogously defined, but for the case of different inputs, i.e. when the phases of two incoming pulses are dissimilar. The fundamental reason why $v$ is defined here in terms of photonic gains is to promote an easy experimental estimation through measurements in the classical optical regime, actually even before starting the protocol itself. In general, for the two-user case only, the relationships between $v$ and the gains are clearly $g_1^\text{D}=2v$ and $g_1^\text{E}=2(1-v)$.

Using all the above notation in this section, we sketch next a computational iterative algorithm for numerically calculating both $r$ and ${\left| \alpha  \right|}^{2}$, assuming the same statistical model and protocol rules introduced in \cite{Xu}. This numerical algorithm, or another equivalent algorithm that produces the same results in \cite{Xu,Guan}, is not explicitly detailed in these references, but we include it in this document in order to facilitate the comparison with our analytical method that requires neither numerical iterations nor solving nonlinear equations. In the algorithm stated below, $F_{\textsf{Bin}}^{-1}$ denotes the binomial inverse cumulative distribution function, which can be calculated, for instance, using the \textsf{BinoInv} function available in \textsc{Matlab}\nobreak\hspace{.08em}\textregistered.

~\\
\noindent \textbf{Algorithm 2.1.} (Computational method for calculating $r$ and ${{\left| \alpha  \right|}^{2}}$ in a realistic two-user coherent-state QF protocol)
\begin{itemize}
\item [\textsf{(1)}] \textsf{Fix ECC parameter} $\delta$\textsf{, ECC rate} $c=M/N$\textsf{, dark count probability} $p_\textrm{dark}$\textsf{, visibility }$v$\textsf{, input size} $N$ \textsf{and target value of} $p_\textrm{error}$ \textsf{. Also, fix, as initial value, }${\left| \alpha  \right|}^{2}=0$.
\item [\textsf{(2)}] \textsf{Compute} ${{r}_{\text{E}}}=F_{\textsf{Bin}}^{-1}\left( 1-{{p}_{\text{error}}},\text{ }M,\text{ }p_{\text{click,1}}^{\text{E}} \right)$.
\item [\textsf{(3)}] \textsf{Compute} ${{r}_{\text{D}}}=F_{\textsf{Bin}}^{-1}\left( {{p}_{\text{error}}},\ M,\ \left[ 1-\delta  \right]p_{\text{click,1}}^{\text{D}}+\delta p_{\text{click,1}}^{\text{E}} \right)-1$.
\item [\textsf{(4)}] \textsf{If} ${{r}_{\text{D}}}={{r}_{\text{E}}},$ \textsf{STOP, take threshold $r={{r}_{\text{D}}}$ and keep as final result the current updated value of} ${{\left| \alpha  \right|}^{2}}$. \textsf{~~If} ${{r}_{\text{D}}}<{{r}_{\text{E}}},$ \textsf{increase} ${{\left| \alpha  \right|}^{2}}$ \textsf{by a small amount $\Delta_\alpha$ and repeat (2) to (4). For input sizes $N\geq10^{-5}$, such as those considered in previous literature and in this work, an increment $\Delta_\alpha=1$, or even greater, is adequate as typically ${{\left| \alpha  \right|}^{2}}>100$.  }
\end{itemize}
~\\
Applying this algorithm, it can be shown that, in general, contrary to what happens for the ideal-case solution in (\ref{eq2}), the minimum required ${\left| \alpha  \right|}^{2}$ is now no longer independent of $M$ for a fixed $p_\textrm{error}$. This observation is apparent from the results in \cite{Xu,Guan}. However, for input sizes below a certain value of $M$ that strongly depends on $p_\text{dark}$, it is also observed that the protocol is still able to countenance a constant ${\left| \alpha  \right|}^{2}$ and concomitantly maintain the desired target error probability, as happens in the ideal scenario.

Prior to ending this explanation of the realistic two-user QF protocol, we note that, in any real implementation, we must also take into consideration the effect of losses. To do so, we define a parameter $\eta$ that combines the effect of the overall losses present in the whole experiment, such as detector efficiencies and losses in the quantum channel. The effect of $\eta$ is equivalent to transforming state $\left|\alpha\right\rangle_x$ in (\ref{eq1}) into another state $\left|\sqrt{\eta}\alpha\right\rangle_x$, which can always be compensated just by increasing the total transmitted mean photon number as ${{\left| \alpha  \right|}^{2}}\to \frac{{{\left| \alpha  \right|}^{2}}}{\eta }$. Thus, the protocol exhibits robustness to losses, in the sense that the scaling properties of ${{\left| \alpha  \right|}^{2}}$ with respect to the rest of protocol parameters remain unchanged if losses rise.

As a concluding outline, we next formally summarize how to quantify the amount of transmitted quantum information $Q$, measured in qubits. In order to do so, once ${{\left| \alpha  \right|}^{2}}$ has been computed,  the information quantification draws upon the following upper bound. This bound is valid for any coherent-state QF protocol, either ideal or realistic, and it can be plainly inferred from the demonstration of Theorem 1 contained in \cite{Arrazola}:
\begin{equation}
\begin{array}{l}
Q= \left( {{\left| \alpha  \right|}^{2}}+\Delta  \right){{\log }_{2}}\left( M+{{\left| \alpha  \right|}^{2}}+\Delta -1 \right)+{{\log }_{2}}\left( 2\Delta  \right) \text{~~[qubits/user]}.\\
\end{array}
\label{eq4}
\end{equation}
In this upper bound, parameter $\Delta$ must be minimized from the following non-linear inequality using a numerical method, such as \textsf{fzero} in \textsc{Matlab}\nobreak\hspace{.08em}\textregistered:
\begin{equation}
2{{\text{e}}^{-{{\left| \alpha  \right|}^{2}}}}{{\left( \frac{\text{e}{{\left| \alpha  \right|}^{2}}}{{{\left| \alpha  \right|}^{2}}+\Delta } \right)}^{{{\left| \alpha  \right|}^{2}}+\Delta }}\le {{\left( \frac{\epsilon }{2} \right)}^{2}}.
\label{eq4b}
\end{equation}
The value of $\epsilon$ in (\ref{eq4b}) is fixed and it may be understood as an indicator of the accuracy of (\ref{eq4}). In our simulations, following the same convention as in previous works, we take $\epsilon=10^{-6}$. This parameter $\epsilon$ should not be confused by any means with the error probability $p_\text{error}$ of the protocol, inasmuch as $\epsilon$ refers solely to the probability of getting a certain inaccurate result in a $Q$ prediction given by (\ref{eq4}) for any complete realization of the QF protocol. We note that a good and simple approximation to the upper bound in (\ref{eq4}), confirmed in many of our simulation results plotted at log-log scale for $\epsilon=10^{-6}$, is simply $Q \approx |\alpha|^2\;\log_2{M}$.

In concluding this section, we provide a brief discussion on the communication cost of the family of QF protocols introduced here. On this subject, because $M$ and $N$ are linearly related by the ECC rate, the result in (\ref{eq4}) stipulates that the scaling properties of $Q$ may be described as $Q\sim \mathcal{O}\small({{\log }_{2}}N\small)$ or, more precisely, as  $Q\sim \mathcal{O}\small({{{\left| \alpha  \right|}^2\log }_{2}}N\small)$, except for an arbitrarily small $\epsilon$. For the discussion in progress, we must also consider, as explained in this section, that as long as $p_\textrm{error}$ stays set to a specific value, ${\left| \alpha  \right|}^2$ remains constant and independent of $N$ in the ideal QF protocol. This statement is also true in the nonideal case for values of $N$ below a certain threshold that depends on imperfections. Consequently, taking for granted a constant mean photon number, it is often found in the literature the recurring assertion that a coherent-state QF protocol provides exponential savings in the transmitted information when compared to classical protocols, which in turn require fingerprints of no less than $\Omega\small(\sqrt{N}\small)$  bits  for a fixed error probability \cite{Babai, Newman}. We remark that, even though this affirmation is factually true for a constant ${\left| \alpha  \right|}^2$, it is still possible to achieve huge savings even if, in order to maintain the desired target $p_\textrm{error}$, imperfections force a variation of ${\left| \alpha  \right|}^2$ as a function of $N$. With regard to this assertion, we present in section \ref{sec5} results that show improvements of several orders of magnitude when compared to the best-known classical protocol, even under circumstances that make the strict exponential savings unattainable.

\section{Multi-party referee circuit designs}
\label{sec3}
This section addresses the non-trivial task of extending the beamsplitter device concept to multiple users, in such ways that our extended device circuits can then be used in the implementation of a QF protocol analogous to the pioneering $K\textrm{-user}$ \emph{classical} protocol in \cite{Fischer}. The generalized optical circuits, for $K \geq 2$ users, must drive the output photonic detectors in a manner that the clicks registered in these detectors provide enough information for the referee. The referee's task is then to conclude if at least one of the binary sequences $x_k$, with $1\leq k\leq K$, differs in at least a single bit when compared to the rest of the sequences.

Before introducing the beamsplitter generalization itself, we describe in essence the signal states that interfere at the referee's circuit. In this regard, we adopt the same phase-encoding scheme prescribed in earlier two-user QF protocols that was previously detailed in section 2, which is based on transmitting coherent states whose phase modulation is furnished by an error correcting code (ECC) \cite{Arrazola, Xu, Guan}. Therefore, each of the $K$ users sends through the respective quantum channel a train of $M$ weak coherent pulses characterized by an adapted version of (\ref{eq1}):
\begin{equation}
{{\left| \alpha  \right\rangle }_{x_k}}= \bigotimes_{m=1}^M{{\left| {{(-1)}^{E{{(x_k)}_{m}}}}\frac{\alpha }{\sqrt{M}} \right\rangle }_{m}},
\label{eq5}
\end{equation}
where, in this case, label $k$, satisfying $1\leq k\leq K$, is assigned to identify each user and also each user's sequence $x_k$. Again, $E(x_k)_m$ tags the $m\textrm{th}$ bit of an ECC codeword corresponding, on this occasion, to a binary string $x_k$.

Given that the generalized protocol involves $K$ users transmitting via $K$ separate channels, it seems mandatory for the referee to employ an optical multiport device in which the number of inputs is $K$ as well. Moreover, we take into consideration the fact that circuits built with linear-optics elements, such as phase shifters and beamsplitters, can always be described by means of a unitary matrix, and vice-versa, if the number of outputs in the multiport is also $K$ \cite{Reck1}. Specifically, restricting ourselves for now to an ideal and lossless scenario, the family of multiport circuits in our proposal can be effectively represented by a general unitary matrix in which one row contains the same element value repeated $K$ times. Further, the rest of this matrix's rows have the trait of adding up to zero, as shown in (\ref{eq5b}). For descriptive purposes, we have chosen in (\ref{eq5b}) an arbitrary row $k$ as the only one that sums up to $\sqrt{K}$ instead of zero:
\begin{equation}
{{U}_{K}}=\left( \begin{matrix}
   {{u}_{1,1}} & {{u}_{1,2}} & \cdots  & \cdots  & \cdots  & \cdots  & {{u}_{1,K}}  \\
   \vdots  & \vdots  & \vdots  & \vdots  & \vdots  & \vdots  & \vdots   \\
   {{u}_{k-1,1}} & {{u}_{k-1,2}} & \cdots  & \cdots  & \cdots  & \cdots  & {{u}_{k-1,K}}  \\
   \frac{1}{\sqrt{K}} & \frac{1}{\sqrt{K}} & \frac{1}{\sqrt{K}} & \frac{1}{\sqrt{K}} & \frac{1}{\sqrt{K}} & \frac{1}{\sqrt{K}} & \frac{1}{\sqrt{K}}  \\
   {{u}_{k+1,1}} & {{u}_{k+1,2}} & \cdots  & \cdots  & \cdots  & \cdots  & {{u}_{k+1,K}}  \\
   \vdots  & \vdots  & \vdots  & \vdots  & \vdots  & \vdots  & \vdots   \\
   {{u}_{K,1}} & {{u}_{K,2}} & \cdots  & \cdots  & \cdots  & \cdots  & {{u}_{K,K}}  \\
\end{matrix} \right)
\begingroup
\small
 \begin{matrix}
   \xrightarrow{\hspace*{20pt}} \sum\nolimits_{\ell =1}^{K}{{{u}_{1,\ell }}=0}  \\
   \vdots   \\
   \;\;\;\xrightarrow{\hspace*{20pt}} \sum\nolimits_{\ell =1}^{K}{{{u}_{k-1,\ell }}=0}  \\
   \vdots   \\
   \;\;\;\xrightarrow{\hspace*{20pt}} \sum\nolimits_{\ell =1}^{K}{{{u}_{k+1,\ell }}=0}  \\
   \vdots   \\
   \xrightarrow{\hspace*{20pt}} \sum\nolimits_{\ell =1}^{K}{{{u}_{K,\ell }}=0}  \\
\end{matrix}
\endgroup
\label{eq5b}
\end{equation}
If we assume monochromatic light with the same polarization in every input beam, the generic unitary scattering matrix $U_K$ describes a classical-optics transformation that is performed on electric fields as $E_{\textrm{out}}=U_K E_{\textrm{in}}$. Analogously, in quantum optics, matrix $U_K$ performs a transformation that linearly relates the creation operators of the input modes to the corresponding operators of the output modes.

\begin{figure}[h]
\centerline{\includegraphics[width=0.6\columnwidth,draft=\isdraft]{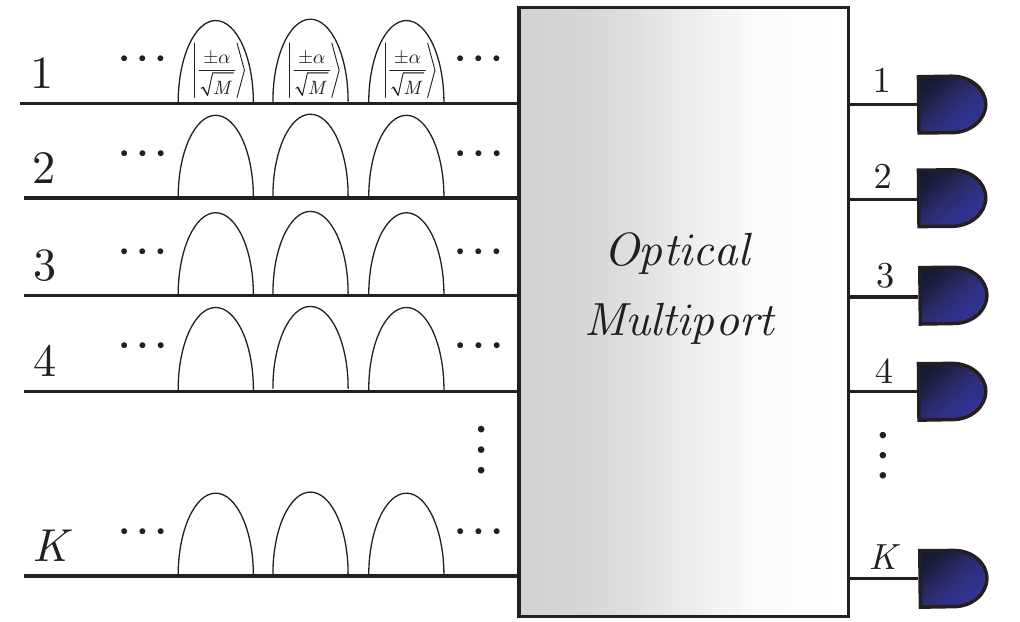}}
\caption{Operation of the referee's optical multiport circuit, portrayed here as a ``black box,'' in a proposed general $K\textrm{-party}$ quantum fingerprinting protocol implemented with weak coherent pulses. Each party transmits a sequence of coherent states  $\bigotimes_{m=1}^M{{\left| \pm \frac{\alpha }{\sqrt{M}} \right\rangle }_{m}}$  whose phases are modulated following an identical procedure as in the two-user coherent-state protocol. Under ideal assumptions, if all the $K$ input pulses  $\bigotimes_{k=1}^K{{\left| \pm \frac{\alpha }{\sqrt{M}} \right\rangle }_{k}}$ that arrive at the referee from the users at a given moment have the same phase, just one of the $K$ detectors may fire. On the contrary, any detector may fire if at least one of these $K$ input pulses is different.}
\label{fig4}
\end{figure}

 Let us assume, for now, that all the $K$ detectors connected to the circuit output ports are, as the device circuit itself, ideal. The only row adding up to $\sqrt{K}$ in the unitary matrix above instinctively corresponds to a multiport output that, in general, is entitled to yield clicks in any case whatsoever, i.e. we do not impose any particular conditions on the individual $K$ input phases. Conversely, the remaining $K-1$ rows that sum up to zero correspond to circuit outputs that cannot lead to clicks when the $K$ incoming phases are the same. Accordingly, our general multiport proposal consists of $K-1$ output detectors that may produce clicks \emph{only} for the case of different inputs $x_k$, and $1$ output detector without restrictions. This latter detector is the single one that may click if all the $K$ individual input pulses that arrive at the referee multiport have the same phase. It may also fire, however, when these phases differ. Thus, in order to summarize the decision rule, the referee ideally announces that the bit sequences are different, i.e. $f(x_1,x_2,\ldots x_K)=1$ (at least one $x_k$ is dissimilar), if and only if she observes at least one click in the $K-1$ detectors associated with the zero-sum rows. The transition matrix in (\ref{eq5b}) may be understood as corresponding to a generalization of a standard 50:50 beamsplitter, for a system described in a $K\text{-dimensional}$ Hilbert space. Notice that a standard 50:50 beamsplitter matrix is just a particular case in two dimensions. The schematic black-box representation of a generic referee's multiport is depicted in figure \ref{fig4}.

Before concluding this section introduction, let us be clear about two details: (i) for a fixed $K$ and keeping into consideration not breaking the unitary condition, the matrix elements in the zero-sum rows of~(\ref{eq5b}) may be chosen in many distinct ways that lead to different multiport circuit designs; (ii) even if all the elements in the unitary matrix are already fixed, diverse design rules can be applied that also produce different multiport circuit implementations, all of them represented by the same matrix \cite{Reck1,Reck2}. Relatedly, the rest of the chapter describes various circuit designs aimed at being used by the referee in the multi-party QF protocol. Suboptimal designs (from the point of view of dealing with experimental imperfections) are also presented in the chapter's remainder, for a twofold reason: (i) for comparison purposes, and (ii) because some of these designs, though not optimal, have some other interesting experimental benefits that we shall discuss in brief.

\subsection{Generalized beamsplitter designs}
\label{sec31}
Generalized beamsplitters, also called multiport beamsplitters or multiport interferometers, were first formally addressed by Zeilinger \emph{et al} in \cite{Zeilinger}. After this description, Reck \emph{et al} released in \cite{Reck1, Reck2} the first known systematic procedure for designing the corresponding device. This method takes a unitary matrix characterization as starting point and then provides an optical network of two-input beamsplitters and phase shifters that implements the unitary transformation. Interestingly, this design proposal was later used in \cite{Zukowski} in order to construct real experiments for testing diverse EPR correlations. In general, a generalized beamsplitter with $K$ input ports (and an equal number of output ports) is characterized by a $K \times K$ matrix $U_K$ built exclusively by taking powers of the $K\textrm{th}$ root of unity $\gamma_K=\exp{(\textrm{i}\frac{2\pi}{K})}$. Explicitly, the matrix elements of $U_K$ are given by
\begin{equation}
{{u}_{ij}}=\frac{1}{\sqrt{K}}\exp \left[ \mathrm{i}\frac{2\pi }{K}(i-1)(j-1) \right],\mathrm{~}i=1,\ldots ,K,\mathrm{~}j=1,\ldots ,K.
\label{eq6}
\end{equation}
It is immediate to check that $U_K$ is in fact a unitary matrix that, furthermore, satisfies all the requirements in (\ref{eq5b}) concerning the family of multiport circuits in our proposal. This demonstration can be done by taking into account the following manifest property of the roots of unity: $\sum\limits_{\ell =1}^{K}{\gamma _{K}^{(i-1)(\ell -1)}}\gamma _{K}^{-(j-1)(\ell -1)}=K{{\delta }_{ij}}$, where $\gamma_K=\exp{(\textrm{i}\frac{2\pi}{K})}$ and $\delta_{ij}$ is a Kronecker delta. Therefore, we can use the optical realizations of this kind of matrices as a possible multi-party referee circuit.

Up to this date, just two systematic methods are known for the design of devices implementing the transformation given by the characteristic matrix in (\ref{eq6}) for any value of $K$. The method by Reck \emph{et al} in \cite{Reck1, Reck2} was devised in 1994, whereas the second known method was made available by Clements \emph{et al} in \cite{Clements}, nearly two decades later. We emphasize the fact that these design methods are universal in the sense that they can provide optical circuit realizations not only for the specific generalized beamsplitter matrices considered here but also for any unitary matrix. We refer the interested reader to the references above for more information about the circuit topologies.

Both universal designs chiefly require an identical number of $\frac{K(K-1)}{2}$ beamsplitters in order to construct the multiport device. However, Clements layout achieves a smaller optical depth, which, as per the exhaustive comparative analysis in \cite{Flamini1}, is a key parameter highly correlated with errors caused by fabricative imperfections and by optical losses inside the multiport. We define the optical depth parameter as the maximum number of beamsplitters, i.e. counted by traversing the longest path across the multiport, considering all paths from any input port to any output port. In particular, Clements design dispenses an optical depth of $K$ beamsplitters, whereas the depth intrinsic to Reck design is $2K-3$. As $K$ grows, the latter requires roughly twice the depth of the former design.

Even though Clements layout presents, in general, a superior error tolerance in any realistic operational conditions, which is indeed extremely important for experimental implementations, Reck design can be advantageous when using programmable circuits for fulfilling  the unitary transformation.  In particular, configuring methods exist for the Reck design that can be applied to program integrated photonic chips, without requiring a full characterization of the internal circuit components \cite{Miller1, Miller2}.

\subsection{Extendable design}
\label{sec32}
As opposed to the generalized beamsplitter designs, which were already-known and we merely found and demonstrated a new application for them, we start here presenting original multiport layouts specifically devised for our QF protocol proposal. For each of these novel designs, we first present our optical circuit realization and then introduce the associated unitary matrix. We remark that the previously discussed Reck and Clements procedures can also be applied to obtain valid designs corresponding to our newly introduced unitary matrices. These designs, however, turn out to be far from optimal when compared with our own specific circuits in terms of both number of beamsplitters and optical depths. Thus, we shall discard these other designs in the upcoming discussions regarding our proposals.

\begin{figure}[h]
\centerline{\includegraphics[width=0.8\columnwidth,draft=\isdraft]{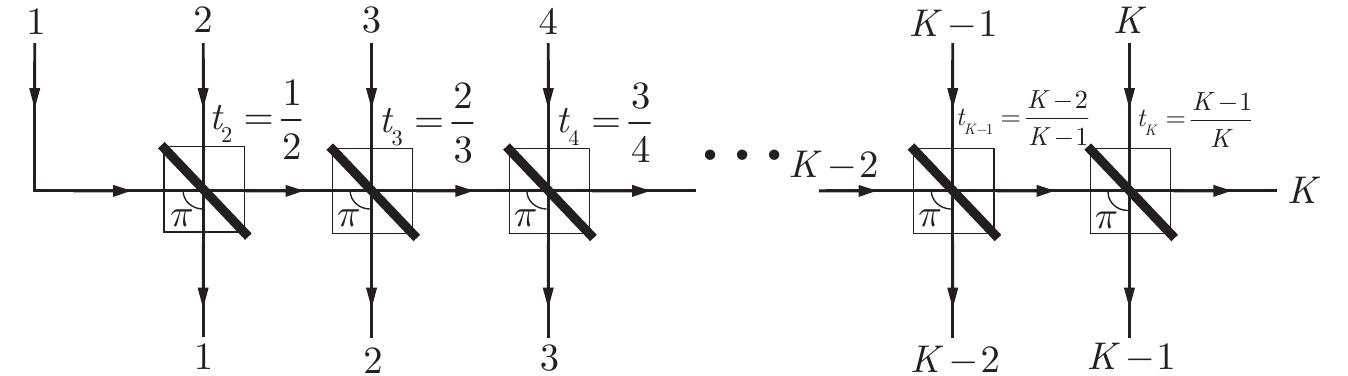}}
\caption{General example, valid for any number of parties $K$, of our proposed extendable design. Power transmittances are represented by $t$.}
\label{fig7}
\end{figure}

As a first innovative design approach for producing novel referee circuit architectures, we present what we call the \emph{extendable design}. This proposed layout consists of a chain of $K-1$ concatenated unbalanced beamsplitters, epitomized by the sequence in figure~\ref{fig7}. As per convention, we assign a label $k$, with $2\leq k\leq K$, to each of them. A beamsplitter $k$ contains the input of the quantum channel from user $k$, except for beamsplitter $k=2$, which also contains the input for the user with label $k=1$. Moreover, each unbalanced beamsplitter $k$ is characterized by an individual power transmittance $t_k=\frac{k-1}{k}$.

An obvious benefit of the extendable design lies in the fact that fingerprinting users can be added and removed without requiring much effort to change the physical layout. For example, new users can be simply added without affecting the extant part of the circuit already in use for the previous users. In return, a serious drawback of this design arises from the asymptotic behaviour of the power transmittances. As the number of users $K$ increases, the required power transmittances tend to be close to 1 and very close to each other. Consequently, in a realistic scenario, an important contribution of errors originating from mismatched transmittances is to be expected. This effect can be also apparent from observing the generic matrix coefficients in (\ref{eq7}), which represent amplitude transmittances. The column and row indices for these coefficients correspond respectively to the input and output channel labels on the schematic view in figure~\ref{fig7}:
\begingroup
\small
\begin{equation}
U_K=\left(
\begin{matrix}
 \frac{-1}{\sqrt{2}} & \frac{1}{\sqrt{2}} & 0 & 0 & \cdots  & 0 & 0  \\
   \frac{-1}{\sqrt{6}} & \frac{-1}{\sqrt{6}} & \sqrt{\frac{2}{3}} & 0 & \cdots  & 0 & 0  \\
   \vdots  & \vdots  & \ddots  & \ddots  & \ddots  & 0 & 0  \\
   \frac{-1}{\sqrt{(K-2)(K-3)}} & \frac{-1}{\sqrt{(K-2)(K-3)}} & \cdots  & \frac{-1}{\sqrt{(K-2)(K-3)}} & \sqrt{\frac{K-3}{K-2}} & 0 & 0  \\
   \frac{-1}{\sqrt{(K-1)(K-2)}} & \frac{-1}{\sqrt{(K-1)(K-2)}} & \frac{-1}{\sqrt{(K-1)(K-2)}} & \cdots  & \frac{-1}{\sqrt{(K-1)(K-2)}} & \sqrt{\frac{K-2}{K-1}} & 0  \\
   \frac{-1}{\sqrt{K(K-1)}} & \frac{-1}{\sqrt{K(K-1)}} & \frac{-1}{\sqrt{K(K-1)}} & \frac{-1}{\sqrt{K(K-1)}} & \cdots  & \frac{-1}{\sqrt{K(K-1)}} & \sqrt{\frac{K-1}{K}}  \\
   \frac{1}{\sqrt{K}} & \frac{1}{\sqrt{K}} & \frac{1}{\sqrt{K}} & \frac{1}{\sqrt{K}} & \frac{1}{\sqrt{K}} & \cdots  & \frac{1}{\sqrt{K}}  \\
   \end{matrix}
\right).
\label{eq7}
\end{equation}
\endgroup

\subsection{Optimal design}
\label{sec33}
In the following, we present an optical circuit topology that minimizes both the number of required beamsplitters and the optical depth. We call this novel topology \emph{optimal design,} albeit this designation is, strictly speaking, a surmise based on the evidence provided by comparing with other known topologies. This optimal design certainly has the minimum depth of all the circuit architectures devised in this work for the multi-user coherent-state QF problem. We may conjecture that no other topology exists that achieves the shortest possible optical depth and the smallest number of standard (two-input) unbalanced (generic) beamsplitters for the problem under consideration.

The objective of finding the optimal layout is motivated in order to reduce fabrication resources. In other words, working with compact circuit designs that need just a few beamsplitters is a key factor for the manufacture of planar waveguide photonic circuits. Additionally, internal propagation losses are reduced when the optical depth is small, and the cumulative effect of errors caused by fabrication imperfections, namely errors when setting the values of transmittances and phase shifts, is also expected to be lower for a shorter optical depth \cite{Flamini1}.

\begin{figure}[h]
\centerline{\includegraphics[width=0.88\columnwidth,draft=\isdraft]{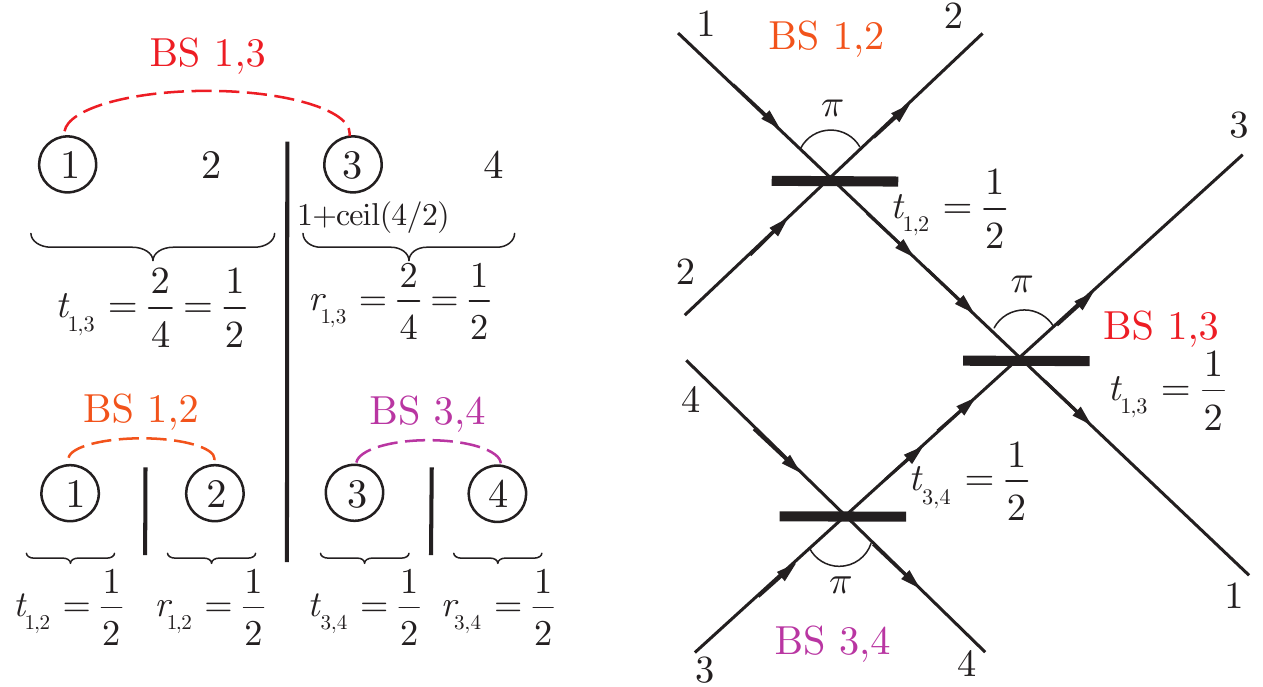}}
\caption{Example of our optimal fingerprinting circuit design for $K=4$ parties. In general, the circuit design follows a tree structure obtained by recursively dividing into two groups the integer labels assigned to identify the players, as exemplified on the accompanying diagram. $r$ denotes power reflectance and $t$ denotes power transmittance.}
\label{fig8}

\end{figure}

\begin{figure}[h]
\centerline{\includegraphics[width=0.88\columnwidth,draft=\isdraft]{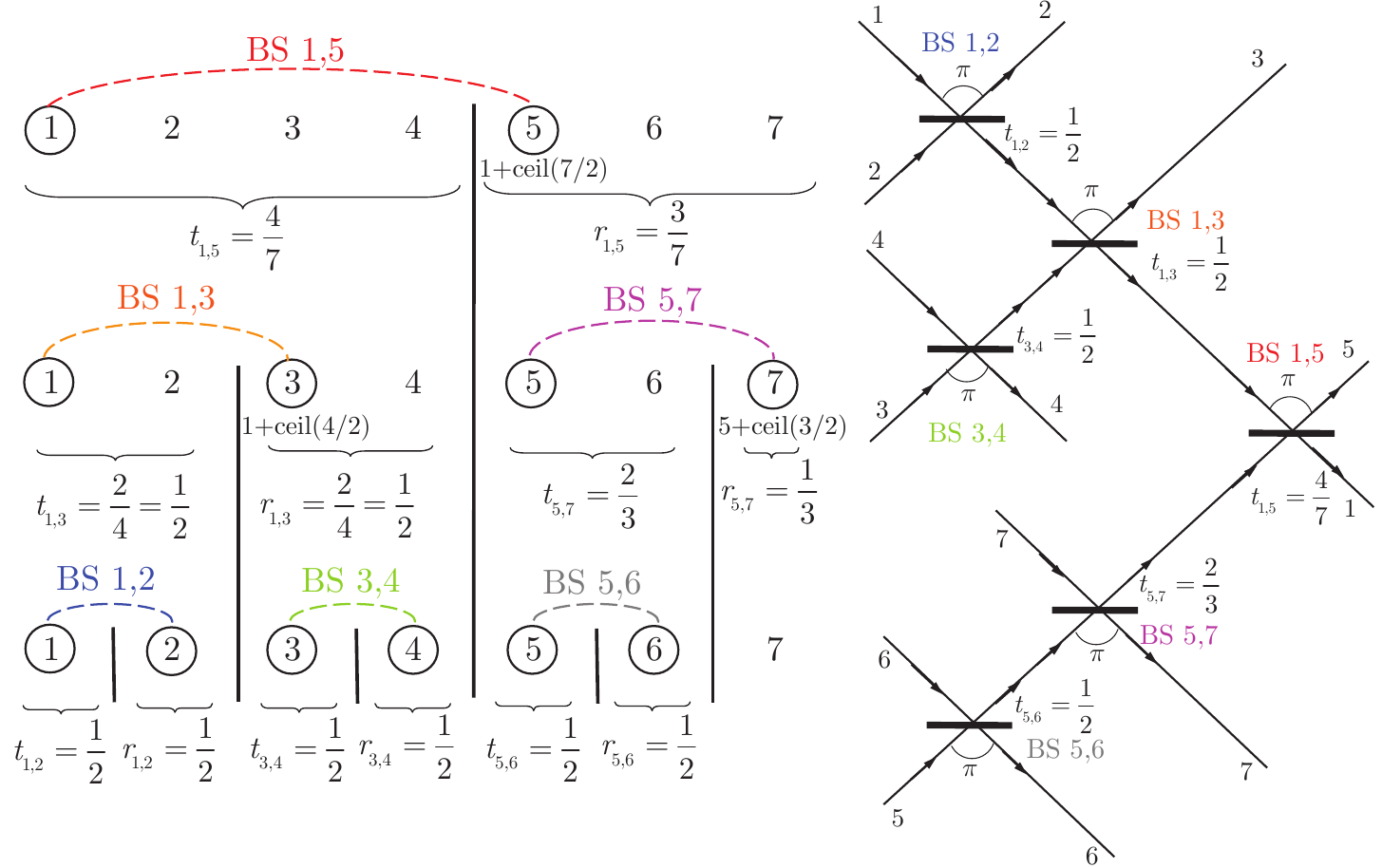}}
\caption{Example of our optimal fingerprinting circuit design for $K=7$ parties. In general, the design consists of a tree structure merely constructed by recursively dividing into two groups the integer labels allocated to the players. $r$ denotes power reflectance and $t$ denotes power transmittance. The tree structure is not symmetrical if $K$ is odd, as it corresponds to the herein depicted explanatory example.   Ceiling division (used in the represented example) or integer division is required to divide the users into blocks. Beamsplitter power transmittances and reflectances are easily calculated as fractions of total users that are included in each block, as recursively typified on the accompanying diagram.}
\label{fig9}
\end{figure}

For a fixed number of users $K$, the construction of the optimized topology starts by assigning an integer label $k=1\ldots\,K$ to each user and then recursively allocating these labels into two groups. At each recursive-division level, a beamsplitter corresponding to these two groups is then placed over the layout, beginning with the highest level of the group division diagram, and forming a characteristic tree arrangement. A simple self-explanatory example for $K=4$ parties that only requires a certain type of 50:50 beamsplitters is illustrated in figure~\ref{fig8}. A more intricate but self-evident example, for the case $K=7$, is contained in figure~\ref{fig9}. When the amount of labels at some division level is an odd number, as in this example for $K=7$, we employ the ceiling division in order to split the labels into two groups. Replacing the ceiling division with integer division in the diagram is also acceptable; it would produce a different equivalent circuit with another similar tree topology. Finally, for each level contained in the division diagram presented on the figures' left, the power transmittances corresponding to the unbalanced beamsplitters must be calculated in the form of fractions obtained as the amount of labels in the first group over the total number of labels in the two groups. The distinctive tree design generated using this method can be described by a unitary matrix satisfying all the requirements in (\ref{eq5b}). In particular, the single matrix row in (\ref{eq5b}) that does not sum up to zero corresponds here to the output with label 1 in the layouts on figures~\ref{fig8} and \ref{fig9}.

Taking into consideration the previously-explained optimal design procedure, we present comparative results in table~\ref{tab1}, where $\lceil \,\log_2K\,\rceil$ is the ceiling function of $\log_2K$. This logarithmic function mathematically arises from the circuit's tree structure. The analytical comparisons with the rest of topologies clearly show that our optimal design inherently provides enormous \emph{exponential} savings in terms of optical depth. Moreover, the reduction in the amount of required beamsplitters exhibits noteworthy quadratic savings when compared to the generalized beamsplitter layouts generated according to the Reck and Clements procedures.

\begin{table}[H]
\vspace{20pt} 
\caption{Number of beamsplitters (BS's) and optical depth in 4 referee optical circuit designs.}
\rm
\begin{tabular*}{\textwidth}{@{}l*{15}{@{\extracolsep{0pt plus12pt}}l}}
\br
Design& Number of beamsplitters & Optical depth\\
\mr
Our optimal design&$K-1$&$\left\lceil\mathrm{~}{{\log }_{2}}K\mathrm{~}\right\rceil$\\
\\Our extendable design&$K-1$&$K-1$\\
\\Generalized BS with Clements design&$\frac{K(K-1)}{2}$&$K$\\
\\Generalized BS with Reck design&$\frac{K(K-1)}{2}$&$2K-3$\\
\br
\end{tabular*}
\label{tab1}
\end{table}

\vspace{15pt} 

In the absence of experimental data, we need a realistic theoretical imperfection model in order to study how the referee multiport circuit behaves, depending on experimental imperfections, i.e. the losses and the fabricative imperfections (fabrication noise) present in the circuit components. In the following, we develop such a model for the optimal design that concerns us here; nonetheless, we note that this model can straightforwardly be generalized to any multiport optical circuit. We first present a matrix decomposition, in which the action of every generic beamsplitter upon the quantum states in the ideal circuit is described by a matrix. Finally, based on this ideal matrix decomposition, we hand over the definitive model that contains the parameters that allow for imperfections.

The effect of each beamsplitter in the optimal design can be described by a unitary matrix expressed as follows:
\vspace{20pt} 
\begingroup
\small
\setcounter{MaxMatrixCols}{20}
\begin{equation}
{{U}_{\text{BS}\text{,IDEAL}}}=\left( \begin{matrix}
   1 & 0 & {} & {} & {} & {} & {} & {} & {} & {} & {}  \\
   0 & 1 & {} & {} & {} & {} & {} & {} & {} & {} & {}  \\
   {} & {} & \ddots  & {} & {} & {} & {} & {} & {} & {} & {}  \\
   {} & {} & {} & \sqrt{t} & 0 & {} & 0 & \sqrt{1-t} & {} & {} & {}  \\
   {} & {} & {} & 0 & 1 & {} & 0 & 0 & {} & {} & {}  \\
   {} & {} & {} & {} & {} & \ddots  & {} & {} & {} & {} & {}  \\
   {} & {} & {} & 0 & 0 & {} & 1 & 0 & {} & {} & {}  \\
   {} & {} & {} & -\sqrt{1-t} & 0 & {} & 0 & \sqrt{t} & {} & {} & {}  \\
   {} & {} & {} & {} & {} & {} & {} & {} & \ddots  & {} & {}  \\
   {} & {} & {} & {} & {} & {} & {} & {} & {} & 1 & 0  \\
   {} & {} & {} & {} & {} & {} & {} & {} & {} & 0 & 1  \\
\end{matrix} \right),
\label{eq8}
\vspace{20pt} 
\end{equation}
\endgroup
where matrix size is $K \times K$ and the indices of the off-diagonal elements correspond to the labels of the input and output channels of each beamsplitter in the design. Also, $t$ tags the power transmittance of the considered beamsplitter. For example, for the layout in figure~\ref{fig8}, the matrix decomposition can be expressed as
\begin{equation*}
{{U}_{4}}=\underbrace{\left( \begin{matrix}
   \sqrt{{{t}_{1,3}}} & 0 & \sqrt{1-{{t}_{1,3}}} & 0  \\
   0 & 1 & 0 & 0  \\
   -\sqrt{1-{{t}_{1,3}}} & 0 & \sqrt{{{t}_{1,3}}} & 0  \\
   0 & 0 & 0 & 1  \\
\end{matrix} \right)}_{\text{BS 1,3}}.\underbrace{\left( \begin{matrix}
   \sqrt{{{t}_{1,2}}} & \sqrt{1-{{t}_{1,2}}} & 0 & 0  \\
   -\sqrt{1-{{t}_{1,2}}} & \sqrt{{{t}_{1,2}}} & 0 & 0  \\
   0 & 0 & 1 & 0  \\
   0 & 0 & 0 & 1  \\
\end{matrix} \right)}_{\text{BS 1,2}}.\underbrace{\left( \begin{matrix}
   1 & 0 & 0 & 0  \\
   0 & 1 & 0 & 0  \\
   0 & 0 & \sqrt{{{t}_{3,4}}} & \sqrt{1-{{t}_{3,4}}}  \\
   0 & 0 & -\sqrt{1-{{t}_{3,4}}} & \sqrt{{{t}_{3,4}}}  \\
\end{matrix} \right)}_{\text{BS 3,4}},
\end{equation*}
and the corresponding unitary matrix that describes the entire optimal circuit in figure~\ref{fig8}, for $K=4$, is
\begin{equation*}
{{U}_{4}}=\left( \begin{matrix}
   \frac{1}{2} & \frac{1}{2} & \frac{1}{2} & \frac{1}{2}  \\[6pt]
   -\frac{1}{\sqrt{2}} & \frac{1}{\sqrt{2}} & 0 & 0  \\[6pt]
   -\frac{1}{2} & -\frac{1}{2} & \frac{1}{2} & \frac{1}{2}  \\[6pt]
   0 & 0 & -\frac{1}{\sqrt{2}} & \frac{1}{\sqrt{2}}  \\
\end{matrix} \right).
\end{equation*}
Similarly, for the case $K=7$ depicted in figure~\ref{fig9}, we can write the decomposition as
\begingroup
\tiny
\begin{equation*}
\begin{matrix}
  {{U}_{7}}=\underbrace{\left( \begin{matrix}
   \sqrt{{{t}_{1,5}}} & 0 & 0 & 0 & \sqrt{1-{{t}_{1,5}}} & 0 & 0  \\
   0 & 1 & 0 & 0 & 0 & 0 & 0  \\
   0 & 0 & 1 & 0 & 0 & 0 & 0  \\
   0 & 0 & 0 & 1 & 0 & 0 & 0  \\
   -\sqrt{1-{{t}_{1,5}}} & 0 & 0 & 0 & \sqrt{{{t}_{1,5}}} & 0 & 0  \\
   0 & 0 & 0 & 0 & 0 & 1 & 0  \\
   0 & 0 & 0 & 0 & 0 & 0 & 1  \\
\end{matrix} \right)}_{\text{BS 1,5}}.\underbrace{\left( \begin{matrix}
   \sqrt{{{t}_{1,3}}} & 0 & \sqrt{1-{{t}_{1,3}}} & 0 & 0 & 0 & 0  \\
   0 & 1 & 0 & 0 & 0 & 0 & 0  \\
   -\sqrt{1-{{t}_{1,3}}} & 0 & \sqrt{{{t}_{1,3}}} & 0 & 0 & 0 & 0  \\
   0 & 0 & 0 & 1 & 0 & 0 & 0  \\
   0 & 0 & 0 & 0 & 1 & 0 & 0  \\
   0 & 0 & 0 & 0 & 0 & 1 & 0  \\
   0 & 0 & 0 & 0 & 0 & 0 & 1  \\
\end{matrix} \right)}_{\text{BS 1,3}}.\underbrace{\left( \begin{matrix}
   1 & 0 & 0 & 0 & 0 & 0 & 0  \\
   0 & 1 & 0 & 0 & 0 & 0 & 0  \\
   0 & 0 & 1 & 0 & 0 & 0 & 0  \\
   0 & 0 & 0 & 1 & 0 & 0 & 0  \\
   0 & 0 & 0 & 0 & \sqrt{{{t}_{5,7}}} & 0 & \sqrt{1-{{t}_{5,7}}}  \\
   0 & 0 & 0 & 0 & 0 & 1 & 0  \\
   0 & 0 & 0 & 0 & -\sqrt{1-{{t}_{5,7}}} & 0 & \sqrt{{{t}_{5,7}}}  \\
\end{matrix} \right)}_{\text{BS 5,7}} \\
  \,\ \quad \ .\underbrace{\left( \begin{matrix}
   \sqrt{{{t}_{1,2}}} & \sqrt{1-{{t}_{1,2}}} & 0 & 0 & 0 & 0 & 0  \\
   -\sqrt{1-{{t}_{1,2}}} & \sqrt{{{t}_{1,2}}} & 0 & 0 & 0 & 0 & 0  \\
   0 & 0 & 1 & 0 & 0 & 0 & 0  \\
   0 & 0 & 0 & 1 & 0 & 0 & 0  \\
   0 & 0 & 0 & 0 & 1 & 0 & 0  \\
   0 & 0 & 0 & 0 & 0 & 1 & 0  \\
   0 & 0 & 0 & 0 & 0 & 0 & 1  \\
\end{matrix} \right)}_{\text{BS 1,2}}.\underbrace{\left( \begin{matrix}
   1 & 0 & 0 & 0 & 0 & 0 & 0  \\
   0 & 1 & 0 & 0 & 0 & 0 & 0  \\
   0 & 0 & \sqrt{{{t}_{3,4}}} & \sqrt{1-{{t}_{3,4}}} & 0 & 0 & 0  \\
   0 & 0 & -\sqrt{1-{{t}_{3,4}}} & \sqrt{{{t}_{3,4}}} & 0 & 0 & 0  \\
   0 & 0 & 0 & 0 & 1 & 0 & 0  \\
   0 & 0 & 0 & 0 & 0 & 1 & 0  \\
   0 & 0 & 0 & 0 & 0 & 0 & 1  \\
\end{matrix} \right)}_{\text{BS 3,4}}.\underbrace{\left( \begin{matrix}
   1 & 0 & 0 & 0 & 0 & 0 & 0  \\
   0 & 1 & 0 & 0 & 0 & 0 & 0  \\
   0 & 0 & 1 & 0 & 0 & 0 & 0  \\
   0 & 0 & 0 & 1 & 0 & 0 & 0  \\
   0 & 0 & 0 & 0 & \sqrt{{{t}_{5,6}}} & \sqrt{1-{{t}_{5,6}}} & 0  \\
   0 & 0 & 0 & 0 & -\sqrt{1-{{t}_{5,6}}} & \sqrt{{{t}_{5,6}}} & 0  \\
   0 & 0 & 0 & 0 & 0 & 0 & 1  \\
\end{matrix} \right)}_{\text{BS 5,6}}, \\
\end{matrix}
\end{equation*}
\endgroup
and the resulting matrix for the complete circuit is
\begin{equation*}
{{U}_{7}}=\left(
\begin{matrix}
   \frac{1}{\sqrt{7}} & \frac{1}{\sqrt{7}} & \frac{1}{\sqrt{7}} & \frac{1}{\sqrt{7}} & \frac{1}{\sqrt{7}} & \frac{1}{\sqrt{7}} & \frac{1}{\sqrt{7}}  \\[6pt]
   -\frac{1}{\sqrt{2}} & \frac{1}{\sqrt{2}} & 0 & 0 & 0 & 0 & 0  \\[6pt]
   -\frac{1}{2} & -\frac{1}{2} & \frac{1}{2} & \frac{1}{2} & 0 & 0 & 0  \\[6pt]
   0 & 0 & -\frac{1}{\sqrt{2}} & \frac{1}{\sqrt{2}} & 0 & 0 & 0  \\[6pt]
   -\frac{3}{2\sqrt{21}} & -\frac{3}{2\sqrt{21}} & -\frac{3}{2\sqrt{21}} & -\frac{3}{2\sqrt{21}} & \frac{2}{\sqrt{21}} & \frac{2}{\sqrt{21}} & \frac{2}{\sqrt{21}}  \\[6pt]
   0 & 0 & 0 & 0 & -\frac{1}{\sqrt{2}} & \frac{1}{\sqrt{2}} & 0  \\[6pt]
   0 & 0 & 0 & 0 & -\frac{1}{\sqrt{6}} & -\frac{1}{\sqrt{6}} & \sqrt{\frac{2}{3}}  \\
\end{matrix}
\right).
\end{equation*}

\begin{figure}[h]
\centerline{\includegraphics[width=0.9\columnwidth,draft=\isdraft]{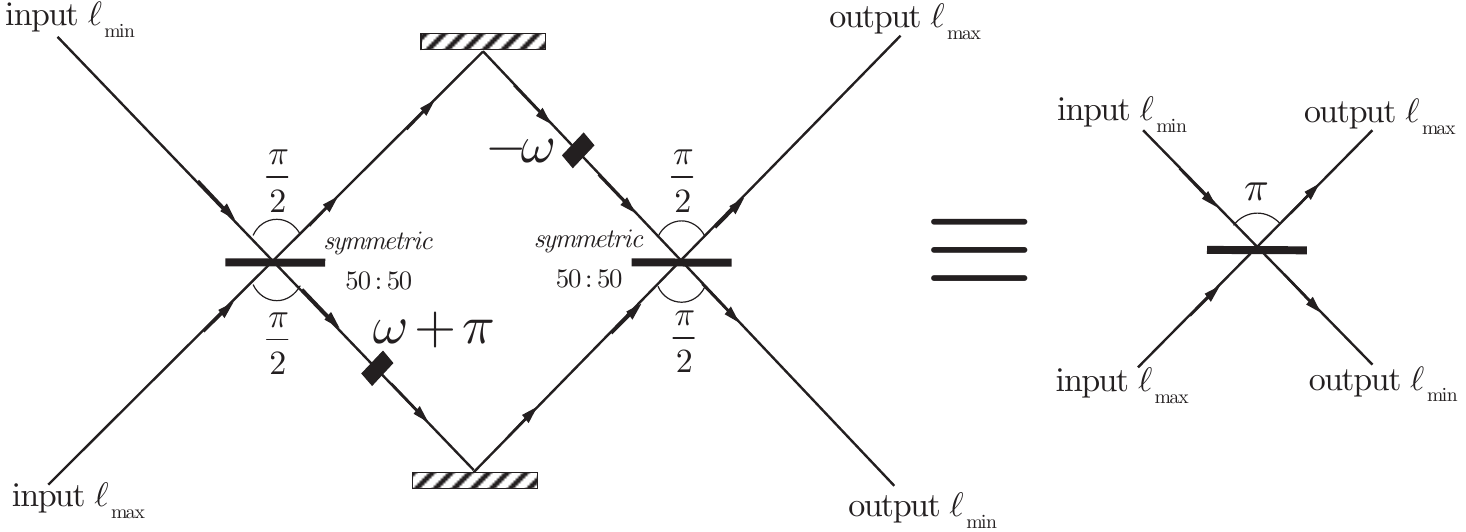}}
\caption{Implementation of each unbalanced beamsplitter (BS) in our optimal circuit design (see, for instance, figures \ref{fig8} and \ref{fig9}), for any pair of port labels in the design ${{\ell }_{\min }},{{\ell }_{\max }}\in \left[ 1,K \right],\text{ }{{\ell }_{\min }}<{{\ell }_{\max }}$. As depicted within this figure, two symmetric 50:50 BS's and two phase shifters are used to mimic the behaviour of any unbalanced BS. In other different circuit designs, the unbalanced BS's may also be implemented in a similar manner.   Phase $\omega$ is related to the unbalanced BS power transmittance as $\omega=\arcsin{(\sqrt{t})}$. According to our design notation convention, phase $\pi$, depicted on the equivalent unbalanced BS (on the right part of this figure), always goes, upon reflection, from the smallest integer label ${\ell }_{\min }$ to the greatest integer label ${\ell }_{\max }$ in any general circuit tree structure, as it is clearly exemplified in figures \ref{fig8} and \ref{fig9}.}
\label{fig10}
\end{figure}

In order to finally include the effects of imperfections in the matrix decomposition, we first need to closely analyze how the generic beamsplitters are implemented in a real photonic circuit. The conventional way for achieving a general optical realization, totally equivalent to a generic beamsplitter, consists of a basic building block, comprising a Mach-Zehnder interferometer built with two cascaded symmetric 50:50 beamsplitters and two phase shifters \cite{Reck1,Reck2,Clements,Flamini1}. For the particular case of our optimal design, this basic building block can be implemented as illustrated in figure~\ref{fig10}. The value of phase $\omega$ in the block is related to the unbalanced power transmittance as $t=\sin^2(\omega)$, with $\omega \in [0,\pi/2]$.

Exploiting the model in figure~\ref{fig10}, the ideal matrix associated to each generic beamsplitter in our design can be further decomposed as
\begin{equation}
\begin{aligned}
   U_{\text{IDEAL}}^{{}}&=\left( \begin{matrix}
   0 & 1  \\
   1 & 0  \\
\end{matrix} \right)\cdot \left( \begin{matrix}
   \frac{1}{\sqrt{2}} & \frac{\text{i}}{\sqrt{2}}  \\
   \frac{\text{i}}{\sqrt{2}} & \frac{1}{\sqrt{2}}  \\
\end{matrix} \right)\cdot \left( \begin{matrix}
   {{\text{e}}^{\text{i}(\omega +\pi )}} & 0  \\
   0 & {{\text{e}}^{-\text{i}\omega }}  \\
\end{matrix} \right)\cdot \left( \begin{matrix}
   \frac{1}{\sqrt{2}} & \frac{\text{i}}{\sqrt{2}}  \\
   \frac{\text{i}}{\sqrt{2}} & \frac{1}{\sqrt{2}}  \\
\end{matrix} \right)= \\
 & =\left( \begin{matrix}
   \sin (\omega ) & \cos (\omega )  \\
   -\cos (\omega ) & \sin (\omega )  \\
\end{matrix} \right)=\left( \begin{matrix}
   \sqrt{t} & \sqrt{1-t}  \\
   -\sqrt{1-t} & \sqrt{t}  \\
\end{matrix} \right),
\end{aligned}
\label{eq9}
\end{equation}
where, for simplicity, we only represent the matrix elements whose indices belong to the non-zero off-diagonal elements in (\ref{eq8}). Flip matrix $\left( \begin{smallmatrix} 0 & 1  \\ 1 & 0  \\ \end{smallmatrix} \right)$ in (\ref{eq9}) is simply for accommodating the output channel labels to those in our design, as typified in figures~\ref{fig8} and \ref{fig9}. Both matrices $\left( \begin{smallmatrix} \frac{1}{\sqrt{2}} & \frac{\text{i}}{\sqrt{2}}  \\ \frac{\text{i}}{\sqrt{2}} & \frac{1}{\sqrt{2}}  \\ \end{smallmatrix}\right)$ correspond to standard symmetric beamsplitters, commonly used in quantum optics. Finally, matrix $\left( \begin{smallmatrix} {{\text{e}}^{\text{i}(\omega +\pi )}} & 0  \\ 0 & {{\text{e}}^{-\text{i}\omega }}  \\ \end{smallmatrix} \right)$ in (\ref{eq9}) models the shifters, characterized by a phase value $\omega$ in figure~\ref{fig10}.

In order to accurately simulate fabricative imperfections and losses, we follow a Monte Carlo method similar to that in \cite{Flamini1}, in which the ideal matrix decomposition in (\ref{eq9}) is replaced with an analogous expression that includes various parameters that quantify imperfection levels:
\begin{equation}
\begin{gathered}
  U_{\text{REALISTIC}}^{{}}=\left( \begin{matrix}
   0 & \eta _{\text{BS}}^{2}  \\
   \eta _{\text{BS}}^{2} & 0  \\
\end{matrix} \right)\cdot \left( \begin{matrix}
   {{\tau }_{1}} & \text{i}\sqrt{1-\tau _{1}^{2}}  \\
   \text{i}\sqrt{1-\tau _{1}^{2}} & {{\tau }_{1}}  \\
\end{matrix} \right)\cdot  \\
  \cdot \left( \begin{matrix}
   {{\text{e}}^{\text{i}(\omega +\pi +{{\sigma }_{\text{P}}}\cdot \textsf{randn})}} & 0  \\
   0 & {{\text{e}}^{\text{i}(-\omega +{{\sigma }_{\text{P}}}\cdot \textsf{randn})}}  \\
\end{matrix} \right)\cdot \left( \begin{matrix}
   {{\tau }_{2}} & \text{i}\sqrt{1-\tau _{2}^{2}}  \\
   \text{i}\sqrt{1-\tau _{2}^{2}} & {{\tau }_{2}}  \\
\end{matrix} \right), \\[6pt]
  \text{with }{{\tau }_{1}},{{\tau }_{2}}={{2}^{-1/2}}(1+{{\sigma }_{\text{T}}}\cdot \textsf{randn}). \\
\end{gathered}
\label{eq10}
\end{equation}
The losses associated to each symmetric beamsplitter are denoted by $\eta _\text{BS}$; we assume in (\ref{eq10}) that both beamsplitters in figure~\ref{fig10} undergo identical losses, even though this restriction may be straightforwardly worked around. The ideal amplitude transmittance $\frac{1}{\sqrt{2}}$ is now substituted in (\ref{eq10}) with its realistic equivalent ${{\tau }_{1}},{{\tau }_{2}}={{2}^{-1/2}}(1+{{\sigma }_{\text{T}}}\cdot \textsf{randn})$, where each of the subscripts $1,2$ corresponds to a different 50:50 beamsplitter. Parameter $\sigma_\text{T}$ accounts for the fabrication noise level affecting the transmittances $\tau_{1,2}$, modelled as the standard deviation of a zero-mean normal random variable \cite{Flamini1}. Similarly, $\sigma_\text{P}$ is the fabrication noise level affecting the phase shifters present in the basic building block, also modelled as Gaussian noise \cite{Flamini1}. We remark that the four different random variables in (\ref{eq10}), dubbed as \textsf{randn}, actually represent four different realizations in a Monte Carlo simulation. To approach our analysis towards realistic values for the above discussed imperfection parameters, we consider femtosecond laser writing as the reference technology used for constructing the optical circuits. Additionally, we assume thermo-optic active control for the phase shifters. Under this fabrication assumptions, it is currently possible to achieve tolerances of $0.01$ for the amplitude transmittances and $0.01~\text{rad}$ for the phase shifters \cite{Crespi,Flamini2}, i.e. present-day fabricative technology permits to describe all the imperfections by a single value as low as ${{\sigma }_{\text{T}}}={{\sigma }_{\text{P}}}=\sigma =0.01$. In consequence, for simplicity, we shall use the same parameter $\sigma$ to denote both phase and amplitude noise levels.

Finally, before concluding this section, we emphasize that the statistical imperfection model developed here is fully independent from, but it can be combined with, the analytical method for upper-bounding the transmitted information, which is introduced in section \ref{sec4} for a realistic multi-party QF scheme. For instance, the upper-bounding method explained in the next section may be applied in a real experiment wherein, once the circuit topology is chosen and implemented, the imperfection model is irrelevant because the performance results rely on measurements only.

\section{Quantum fingerprinting protocol analysis}
\label{sec4}
In this section, we provide a description of various decision rules that govern the referee announcements in the multi-party QF protocol. Based on these rules, we develop the mathematical formalism for upper-bounding both the required amount of abstract information and, importantly towards estimating resource expenditures, also the energy consumption of the protocol. First, we analyze the multi-party ideal case, which allows us to gain an initial insight into the protocol capabilities, in a similar fashion as the ideal analysis carried out in section 2 for the well-known two-user case.

The second part of the section addresses the more interesting realistic scenario, in which we present a method for upper-bounding the transmitted information that takes into account any kinds of circuit imperfections, as well as the detector dark count rates. The application of this method relies on determining certain gains that ultimately include the effects of all imperfections and losses present in the circuit. We present a comprehensive analysis of the method, emphasizing the effect of different parameters on the protocol performance, but leaving for appendix B the mathematical details of the model, which is based on applying a certain form of the Chernoff bounds to the click probabilities at the detectors.

Following the already-detailed general matrix characterization in (\ref{eq5b}) and the generic portrayal in figure~\ref{fig4}, which is applicable to every circuit design in section 3, we consider through this entire section that an integer label $k$, satisfying $1\leq k \leq K$, is assigned so as to identify each output detector. However, in order to make our analysis valid for any circuit without loss of generality, we introduce, from now on, the restriction that label $k=K$ always corresponds to the only detector that loses photons, in comparison with all-equal inputs, when at least one of the input pulses differs from the rest. In other words, label $k=K$ is allocated to the only detector that may click in the ideal case when all the input states are the same. This specific photonic detector corresponds to the row that does not add up to zero in (\ref{eq5b}). The restriction introduced here does not pose any limitations on the circuit design, as it implies a relabeling of the exit ports, which, if applicable, is equivalent to a mere permutation of the rows of the unitary matrix associated to the circuit.

\subsection{Ideal scenario}
\label{sec41}
Under ideal premises, the referee declares that at least one bit sequence $x_k$ is different, i.e. $f(x_1,x_2,\ldots x_K)=1$, if and only if at least one click happens in the $K-1$ detectors related to the zero-sum rows in matrix (\ref{eq5b}). Therefore, an error never occurs if the $K$ original binary inputs $x_k$ are the same. Otherwise, the worst possible situation clearly always corresponds to just one user sending a coherent-state sequence $\bigotimes_{m=1}^M{{\left| \pm \frac{\alpha }{\sqrt{M}} \right\rangle }_{m}}$ that differs from the rest $K-1$ users' sequences. This situation is the most similar to $K$ all-equal sequences of coherent states and, hence, the most difficult to distinguish by the referee. Besides this, for the purpose of carrying out the analytical calculations, the worst case takes place when the number of different coherent states in the sequence that differs is at its minimum. This theoretical minimum is exactly $(1-\delta)M$ as imposed by the minimum distance of the used error correcting code, as detailed in section 2. Thus, we obtain that error probability $p_\text{error}$ can be calculated as
\begin{equation}
\begin{gathered}
{{p}_{\text{error}}}=\prod\limits_{k=1}^{K-1}{{{(1-p_{\text{click},k}^{\text{D}})}^{(1-\delta )M}}}=\prod\limits_{k=1}^{K-1}{{{\text{e}}^{-(1-\delta )M\cdot \mu _{k}^{\text{D}}}}}\\
=\exp \left( -(1-\delta )M\cdot \sum\limits_{k=1}^{K-1}{\mu _{k}^{\text{D}}} \right)=\exp \left( -\frac{4(1-\delta )(K-1){{\left| \alpha  \right|}^{2}}}{K} \right),
\end{gathered}
\label{eq11}
\end{equation}
with $p_{\text{click},k}^\text{D}$ being the click probability at output detector $k$ when just one individual input state differs from the rest, and $\mu_k^\text{D}$ being the corresponding mean photon number of the coherent state impinging this aforesaid detector $k$. For the last equality, we have taken into account
\begin{equation}
\sum\limits_{k=1}^{K-1}{\mu _{k}^{\text{D}}}=\frac{K\cdot {{\left| \alpha  \right|}^{2}}}{M}-\mu _{K}^{\text{D}}=\frac{K\cdot {{\left| \alpha  \right|}^{2}}}{M}-\frac{{{(K-2)}^{2}}\cdot {{\left| \alpha  \right|}^{2}}}{K\cdot M}=\frac{4(K-1)\cdot {{\left| \alpha  \right|}^{2}}}{K\cdot M},
\label{eq12}
\end{equation}
which can be proven multiplying matrix (\ref{eq5b}) by a vector filled with the same repeated value, except for a single entry with an opposite phase. Finally, solving for $|\alpha|^2$ in (\ref{eq11}),
\begin{equation}
{{\left| {{\alpha }_{\text{ideal}}} \right|}^{2}}=\frac{K}{4(1-\delta )(K-1)}\ln \left( \frac{1}{{{p}_{\text{error}}}} \right).
\label{eq13}
\end{equation}
Fixing a specific desired error probability, this latest calculation gives an upper-bound on the mean photon number, and hence also on the energy required per user. The related amount of quantum information per user, suitable to confront a classical protocol, can be directly calculated introducing (\ref{eq13}) into (\ref{eq4}). Again, as in the analysis for deriving (\ref{eq2}) in section 2, here we did not draw upon mathematical approximations concerning the intensity levels of the individual pulses. Also, we note that the known result given in (\ref{eq2}) can be seen as an exact particular case for $K=2$, predicted by the novel generalization in (\ref{eq13}).

A central observation can be made on (\ref{eq13}) by noticing that, as the number of users $K$ increases, the effect of $K$ tends to be less influential on the predicted per-user statistics. Further, we also note that the particular circuit design implemented at the referee node is irrelevant under the ideal assumptions. Finally, we bring attention to the fact that, as in the particular case for $K=2$ anticipated in section 2, the raw message length $N$ (and, consequently, also the number of transmitted pulses $M$) has zero impact on $|\alpha|^2$.

\subsection{Realistic scenario}
\label{sec42}
In this subsection's analysis, we account for any kinds of experimental errors, by means of our analytical method for upper-bounding $|\alpha|^2$. Before entering the analysis, we present the referee decision rules upon which the bounding method depends. The ideal-protocol decision rule in 4.1 is not applicable anymore if subject to realistic constraints, owing to the reasons extrapolated from section 2. As alternatives, we propose here two different referee strategies based on observing different ensembles of detectors attached to the circuit's exit ports. We shall show that either of these two separate rules leads, in general, to different figures of merit when analyzing the QF protocol.

In one of the proposed strategies, the referee counts the number of clicks in the first $K-1$ detectors, labelled $1 \leq k \leq K-1$, which are the detectors that gain impinging photons when some of the input states differ from the rest. Denoting as $D_k$ the total number of clicks observed in every detector $k$ during the entire protocol execution, the referee infers equal sequences if and only if $\sum\nolimits_{k=1}^{K-1}{{{D}_{k}}}\le r$. Parameter $r$ is a certain threshold, below which the outcome ``equal inputs" is announced; we shall provide the details required to calculate $r$ using a closed-form expression. Above the value of $r$, i.e. $\sum\nolimits_{k=1}^{K-1}{{{D}_{k}}}>r$, the referee concludes that the input sequences are different.

The other proposed strategy consists of observing just one detector, labelled $k=K$. This is the single detector that loses photons in the different-sequence situation when compared to the equal-sequence case, under the normal circumstances that we shall mark off. Subject to this decision rule, the referee infers that the input sequences are different if and only if $D_K \leq r$. Complementarily, she announces equal input sequences if and only if $D_K > r$.

In order to deduce the expressions for threshold $r$ and the bounding limit for $|\alpha|^2$, we apply certain types of Chernoff bounds. All the detailed calculations are included in appendix B, but we sketch next the underlying statistical model. In particular, let $X_k^\text{E}$, with $1 \leq k \leq K$, be a random variable with Bernoulli distribution that accounts for the number of individual clicks (0 or 1 click) at detector $k$ when $K$ coherent states arrive at the referee at the same time containing the same phase. In a similar fashion, $X_k^\text{D}$ is an analogous random variable for the case when some of the $K$ input states are different (they contain phase differences). An additional group of random variables $\tilde{X}_{k,m}^{\text{D}}$ is introduced in order to model the effect of the differences present in the $K$ \emph{complete} sequences of $M$ pulses $\bigotimes_{m=1}^M{ \left| \pm \sqrt{\mu_{\rm in}} \right\rangle }_{m}$, as follows:
\begin{equation}
\tilde{X}_{k,m}^{\text{D}}=\left\{ \begin{aligned}
  & X_{k}^{\text{D}}\text{ for any }(1-\delta )\cdot M\text{ indices }m, \\
 & X_{k}^{\text{E}}\text{ for any }\delta \cdot M\text{ indices }m. \\
\end{aligned} \right.
\label{eq14pre}
\end{equation}

The baseline statistical model and the strategies introduced above imply that the referee always provides an erroneous announcement in the following situations. If the referee uses the strategy that consists of observing $K-1$ detectors, an announcement error happens either whenever the input sequences are actually different and $\sum\nolimits_{k=1}^{K-1}{D_{k}^{\text{D}}}=\sum\nolimits_{k=1}^{K-1}{\sum\nolimits_{m=1}^{M}{\tilde{X}_{k,m}^{\text{D}}}}\le r$, or whenever the input sequences are actually equal to each other and $\sum\nolimits_{k=1}^{K-1}{D_{k}^{\text{E}}}=\sum\nolimits_{k=1}^{K-1}{\sum\nolimits_{m=1}^{M}{X_{k}^{\text{E}}}}>r$. In the same way, now under the referee's rule of taking into account just the clicks in detector $k=K$, an error occurs in the following two situations: whenever the input sequences are really equal to each other and $D_{K}^{\text{E}}=\sum\nolimits_{m=1}^{M}{X_{K}^{\text{E}}}\le r$, or whenever they are different and $D_{K}^{\text{D}}=\sum\nolimits_{m=1}^{M}{\tilde{X}_{K,m}^{\text{D}}}>r$. We denote by $p_\text{error}^\text{E}$ the probability of an error happening when the sequences sent by the users are actually equal to each other. Similarly, $p_\text{error}^\text{D}$ is the analogous error probability for different sequences. We remark that, in this work, probabilities $p_\text{error}^\text{E}$ and $p_\text{error}^\text{D}$ are \emph{not} the same as the desired target error probability, which we call $p_\text{error}$. In particular, our specific manner of applying the Chernoff bounds guarantees that $p_\text{error}^\text{E}, p_\text{error}^\text{D} \leq  p_\text{error}$. This is a key difference when comparing with all the methods for calculating $|\alpha|^2$ in previous works \cite{Arrazola, Xu, Guan}, which always secure the equality $p_\text{error}^\text{E}=p_\text{error}^\text{D}=p_\text{error}$ (see, for example, Algorithm 2.1 in this work). As we shall observe in detail, our more relaxed constraint may produce upper-bounds that are not as tight as in the previously published methods. This strict lack of tightness is the price that one has to pay in exchange for a closed-form expression for both $r$ and $|\alpha|^2$ that is instructive and easy to implement for computational purposes. Nonetheless, we shall observe in next section that, for certain ranges of $M$, both approaches essentially yield the same predictions at the logarithmic scale.

Focusing now on the strategy in which the referee observes detectors from 1 to $K-1$, and based on the above statistical description, we get closed-form equations in appendix B.1 that depend on the following two gains:
\begin{subequations}
\begin{gather}
g_{[1,K-1]}^{\text{E}}=\frac{\sum\limits_{k=1}^{K-1}{\mu _{k}^{\text{E}}}}{{{\mu }_{\text{in}}}},\text{~~}0\le g_{[1,K-1]}^{\text{E}}\le K, \label{eq14a}\\[12pt]
g_{[1,K-1],\bar{P}}^{\text{D}}=\frac{\sum\limits_{k=1}^{K-1}{\mu _{k,\bar{P}}^{\text{D}}}}{{{\mu }_{\text{in}}}},\text{~~}0\le g_{[1,K-1],\bar{P}}^{\text{D}}\le K,\label{eq14b}
\end{gather}
\label{eq14}
\end{subequations}\\[0pt]
where $\mu_k^\text{E}$ represents the mean photon number at an output detector $k$ when the $K$ individual input pulses have the same phase. Similarly, $\mu_{k,\bar{P}}^\text{D}$ accounts for the photon number when at least one of the $K$ phases of the individual input pulses is different from the rest. Gain $g_{[1,K-1],\bar{P}}^{\text{D}}$ depends on $\bar{P}$, which is a vector whose elements are phase labels. As an explanatory instance, let us suppose that the referee receives pulses from $K=4$ users and that the individual input states are  ${{\left| \frac{\alpha }{\sqrt{M}} \right\rangle ,\left| \frac{-\alpha }{\sqrt{M}} \right\rangle ,\left| \frac{-\alpha }{\sqrt{M}} \right\rangle ,\left| \frac{\alpha }{\sqrt{M}} \right\rangle}}$. Then, in this particular example, we have $\bar{P}=(1,-1,-1,1).$ In general, as $\bar{P}$ corresponds to different input states, at least 1 component in $\bar{P}$ must have a different phase than the rest of components, i.e. $\bar{P}\ne -\vec{1},\vec{1}$. We denote by $L$ the integer number that indicates the minimum amount of phase labels in $\bar{P}$ that differ from the rest $K-L$ labels. In the previous example, we have $L=2$. Restriction $L \leq \frac{K}{2}$ is imposed, because it is not difficult to realize that values $L > \frac{K}{2}$ introduce zero additional different cases, from the point of view of photon statistics.

Assuming $K{{\mu }_{\text{in}}}=\frac{K{{\left| \alpha  \right|}^{2}}}{M}\ll 1$, which we strictly verify later in the manuscript for the cases of interest, the mathematical development in appendix B.1 gives the following analytical upper bound:
\begin{subequations}
\begin{gather}
 {{\left| \alpha _{[1,K-1]}^{\text{bound}} \right|}^{2}}=\frac{4q+2{{\left[ 4{{q}^{2}}+2{{(1-\delta )}^{2}}{{\left( \min (g_{[1,K-1],\bar{P}}^{\text{D}})-g_{[1,K-1]}^{\text{E}} \right)}^{2}}(K-1)M\mu _{\text{dark}}^{{}}\cdot \ln (1/p_{\text{error}}^{{}}) \right]}^{1/2}}}{\eta \ {{(1-\delta )}^{2}}{{\left( \min (g_{[1,K-1],\bar{P}}^{\text{D}})-g_{[1,K-1]}^{\text{E}} \right)}^{2}}},\label{eq15a}\\[12pt]
  q=\left[ \delta \cdot g_{[1,K-1]}^{\text{E}}+(1-\delta )\cdot \min (g_{[1,K-1],\bar{P}}^{\text{D}}) \right]\cdot \ln (1/p_{\text{error}}^{{}}), \label{eq15b}\\[12pt]
  \text{Fingerprinting possible }\Leftrightarrow \text{ }\min (g_{[1,K-1],\bar{P}}^{\text{D}})>g_{[1,K-1]}^{\text{E}}. \label{eq15c}
\end{gather}
\label{eq15}
\end{subequations}\\[0pt]
Parameter $\eta$ in (\ref{eq15a}) is the combined efficiency that includes the losses of the quantum channel and the detector efficiencies. It does not include, however, the effects of the insertion losses for the beamsplitters, because these are unbalanced losses corresponding to different paths across the multiport circuit. The effect of beamsplitter losses is fully incorporated within the gains in (\ref{eq14}). On another note, we shall clearly show later in this section that the minimization required for (\ref{eq15a}) and (\ref{eq15b}) can be accomplished just by simulating (\ref{eq14b}) for the $K$ vectors $\bar{P}$ that correspond to $L=1$ and then taking the smallest of these $K$ simulated values of $g_{[1,K-1],\bar{P}}^{\text{D}}$. This process is identical when the gains are measured in a real experiment.

We observe in (\ref{eq15a}) that, unlike the ideal case, this bound depends on the number of input pulses $M$. Further, we can also notice that the dark count rate $\mu_\text{dark}$ foists a strong influence that, moreover, is aggravated when both $M$ and the number of users $K$ grow. This worsening consists of an increase in the predicted mean photon number provided by (\ref{eq15a}) and, hence, also in an increment of the energy consumption per user. On a separate note, the condition in (\ref{eq15c}) emerges from the core of the Chernoff bounds themselves (see appendix B.1). This restriction is completely congruent with the desired behaviour of the detectors attached to the exit ports, provided that the experimental error level in the circuit is low enough. We may refer, therefore, to the restriction in (\ref{eq15c}) as the ``normal circumstances of operation''. To conclude our commentaries about (\ref{eq15}), we notice that, once $|\alpha|^2$ has been calculated according to (\ref{eq15a}), it is straightforward to compute the transmitted information, measured in qubits per user, just by applying the result in (\ref{eq4}). Most of the comments provided in this paragraph for the strategy involving $K-1$ detectors are also relevant for the other referee strategy considered in this work, with a few exceptions that we shall note soon.

The referee threshold that corresponds to the rule analyzed so far is
\begin{equation}
r=\frac{1}{2}{{\left| \alpha  \right|}^{2}}\left[ (1+\delta )\cdot g_{[1,K-1]}^{\text{E}}+(1-\delta )\cdot \min(g_{[1,K-1],\bar{P}}^{\text{D}}) \right]+(K-1)M\mu _{\text{dark}}^{{}}.
\label{eq16}
\end{equation}

In the following, we move on to presenting the final results deduced in appendix B.2 for the other rule, in which the referee takes into account detector $k=K$ only. The governing gains for just one detector are
\begin{subequations}
\begin{gather}
g_{K}^{\text{E}}=\frac{\mu _{K}^{\text{E}}}{{{\mu }_{\text{in}}}},\text{~~}0\le g_{K}^{\text{E}}\le K,
\label{eq17a}\\[12pt]
g_{K,\bar{P}}^{\text{D}}=\frac{\mu _{K,\bar{P}}^{\text{D}}}{{{\mu }_{\text{in}}}},\text{~~}0\le g_{K,\bar{P}}^{\text{D}}\le K.\label{eq17b}
\end{gather}
\label{eq17}
\end{subequations}\\[0pt]
Under the same assumptions as in the other decision rule, the bound is now
\begin{subequations}
\begin{gather}
  {{\left| \alpha _{K}^{\text{bound}} \right|}^{2}}=\frac{4q+2{{\left[ 4{{q}^{2}}+2{{(1-\delta )}^{2}}{{\left( g_{K}^{\text{E}}-\max (g_{K,\bar{P}}^{\text{D}}) \right)}^{2}}M\mu _{\text{dark}}^{{}}\cdot \ln (1/p_{\text{error}}^{{}}) \right]}^{1/2}}}{\eta \ {{(1-\delta )}^{2}}{{\left( g_{K}^{\text{E}}-\max (g_{K,\bar{P}}^{\text{D}}) \right)}^{2}}}, \label{eq18a} \\[12pt]
  q=g_{K}^{\text{E}}\cdot \ln (1/p_{\text{error}}^{{}}), \label{eq18b} \\[12pt]
 \text{Fingerprinting possible }\Leftrightarrow \text{ }g_{K}^{\text{E}}>\max (g_{K,\bar{P}}^{\text{D}}). \label{eq18c}
\end{gather}
\label{eq18}
\end{subequations}\\[0pt]
An evident statement can be made by observing (\ref{eq18a}) and comparing it with (\ref{eq15a}). It is clear that, unlike the rule with $K-1$ detectors, the effect of dark counts provided by $\mu_\text{dark}$ in (\ref{eq18a}) is \emph{not} directly worsened as the number of users rises. However, as in the other rule, this effect of $\mu_\text{dark}$ is made worse by the action of $M$ albeit now not aggravated by $K$.  Again, the condition imposed by the Chernoff bounds that enables fingerprinting feasibility, summarized here in (\ref{eq18c}), is in full agreement with our expected behaviour of detectors, as long as a reasonable experimental error level is kept in the referee circuit. As occurs in the minimization for (\ref{eq15}), the maximization required for (\ref{eq18}) is practicably achievable with near zero computational cost. Finally, the corresponding threshold for the referee strategy is given now by
\begin{equation}
r=\frac{1}{2}{{\left| \alpha  \right|}^{2}}\left[ (1+\delta )\cdot g_{K}^{\text{E}}+(1-\delta )\cdot \max(g_{K,\bar{P}}^{\text{D}}) \right]+M\mu _{\text{dark}}^{{}}.
\label{eq19}
\end{equation}

We have hitherto presented a method that allows us to compute upper bounds on $\left| \alpha  \right|^2$, subject to a maximum desired error level $p_\text{error}$ in any realistic case. Of course, this computation also depends on the protocol parameters, such as $K$, $M$ and $\delta$, and on the physical characteristics and imperfections of the photonic components (beamsplitters, detectors, etc). As an interesting remark, table~\ref{tab2} shows how previous approaches compare to our innovative method, regarding various aspects: numbers of users considered in the protocol, referee strategies, and the procedures for calculating the mean photon number.
\begin{table}
\caption{Feature comparison of 3 methods for estimating  $\left| \alpha  \right|^2$ in a realistic scenario.}
\rm
\begin{tabular*}{\textwidth}{@{}l*{15}{@{\extracolsep{0pt plus12pt}}l}}
\br
Method& No. of users & No. of detectors observed  &Type of procedure for calculating $\left| \alpha  \right|^2$  \\
\mr
Proposed in this work&$K\geq2$&1 or $K-1$&Analytical (upper bound)\\
\\Xu's \cite{Xu}&$K=2$&1&Numerical\\
\\Arrazola's\cite{Arrazola}&$K=2$&2&Analytical (upper bound)\\
\br
\end{tabular*}
\label{tab2}
\end{table}

In the rest of the manuscript, we pay a particular attention, amongst other aspects, to the figures of merit (transmitted quantum information, amount of energy, etc.) computed when the raw input size $N$, and consequently also $M$, is arbitrarily large. In particular, this regime corresponds to the situation in which the term with $M$ and $\mu_\text{dark}$ inside the square roots of (\ref{eq15a}) and (\ref{eq18a}) is the leading addend in the sum. This assumption is specially relevant because it is very well known, from all the two-user coherent-state protocols in  \cite{Arrazola, Xu, Guan}, that the dark count rate is a dominant limiting factor as $M$ grows, and identifying and mitigating its effects is still a pressing issue. In our particular model, these premises  make (\ref{eq15a}) and (\ref{eq18a}) more dependent on the subtractions of gains $\left[ \min (g_{[1,K-1],\bar{P}}^{\text{D}})-g_{[1,K-1]}^{\text{E}} \right]$ and $\left[ g_{K}^{\text{E}}-\max (g_{K,\bar{P}}^{\text{D}}) \right]$, respectively. These gain differences thus become more relevant rather than the absolute levels of the gains. We point out that this special devotion for the case of arbitrarily large $M$, however, does not imply, by any means, that we are restricting our study to the limit of an infinite input size where the protocol operates in the asymptotic regime.

For the referee strategy involving $K-1$ detectors, a dominant term in $M$ and $\mu_\text{dark}$ happens when
\begin{equation}
M\gg \frac{2{{\left[ \delta \cdot g_{[1,K-1]}^{\text{E}}+(1-\delta )\cdot \min (g_{[1,K-1],\bar{P}}^{\text{D}}) \right]}^{2}}\cdot \ln (1/{{p}_{\text{error}}})}{{{\mu }_{\text{dark}}}(K-1){{(1-\delta )}^{2}}{{\left( \min (g_{[1,K-1],\bar{P}}^{\text{D}})-g_{[1,K-1]}^{\text{E}} \right)}^{2}}},
\label{eq20pre1}
\end{equation}
and, for the strategy involving a single detector $k=K$, a dominant term happens when
\begin{equation}
M\gg \frac{2{{(g_{K}^{\text{E}})}^{2}}\cdot \ln (1/{{p}_{\text{error}}})}{{{\mu }_{\text{dark}}}{{(1-\delta )}^{2}}{{\left( g_{K}^{\text{E}}-\max (g_{K,\bar{P}}^{\text{D}}) \right)}^{2}}}.
\label{eq20pre2}
\end{equation}
Inasmuch as the quantities $\left[ \min (g_{[1,K-1],\bar{P}}^{\text{D}})-g_{[1,K-1]}^{\text{E}} \right]$ and $\left[ g_{K}^{\text{E}}-\max (g_{K,\bar{P}}^{\text{D}}) \right]$ become relevant, we may easily provide $K\text{-party}$ generalizations of the two-user visibility $v$ in (\ref{eq3b}) that depend on these above-stated subtractions. In this way, these visibility generalizations may be seen as additional figures of merit not only for the quantum protocol by itself, but also with regards to choosing the best suitable design for the referee circuit. A different visibility generalization must be provided for each of the referee strategies analyzed in this work:
\begin{equation}
{{v}_{[1,K-1]}}=\frac{1}{2}\left( 1+\frac{\min (g_{[1,K-1],\bar{P}}^{\text{D}})-g_{[1,K-1]}^{\text{E}}}{\min (g_{[1,K-1],\bar{P}}^{\text{D,ideal}})-g_{[1,K-1]}^{\text{E,ideal}}} \right),\text{~~~~}{{v}_{K}}=\frac{1}{2}\left( 1+\frac{g_{K}^{\text{E}}-\max (g_{K,\bar{P}}^{\text{D}})}{g_{K}^{\text{E,ideal}}-\max (g_{K,\bar{P}}^{\text{D,ideal}})} \right).
\label{eq20}
\end{equation}
These visibilities ${v}_{[1,K-1]}$ and $v_K$ extend (\ref{eq3b}) in such a way that they are calculated by taking the ratios between the realistic gain differences and the ideal ones. Thus, using the fact that the ideal gain values are
\begin{equation}
\min(g_{[1,K-1],\bar{P}}^{\text{D}\text{,ideal}})=\frac{4(K-1)}{K},\text{~~~~}g_{[1,K-1]}^{\text{E}\text{,ideal}}=0,\text{~~~~}
g_{K}^{\text{E}\text{,ideal}}=K,\text{~~~}\max(g_{K,\bar{P}}^{\text{D}\text{,ideal}})=\frac{{{(K-2)}^{2}}}{K},
\label{eq21}
\end{equation}
we may eventually write
\begin{equation}
{{v}_{[1,K-1]}}=\frac{1}{2}\left( 1+\frac{K\left[ \min (g_{[1,K-1],\bar{P}}^{\text{D}})-g_{[1,K-1]}^{\text{E}} \right]}{4(K-1)} \right),\text{~~~~}{{v}_{K}}=\frac{1}{2}\left( 1+\frac{K\left[ g_{K}^{\text{E}}-\max (g_{K,\bar{P}}^{\text{D}}) \right]}{4(K-1)} \right).
\label{eq22}
\end{equation}

\begin{figure}[h]
\centerline{\includegraphics[width=1.0\columnwidth,draft=\isdraft]{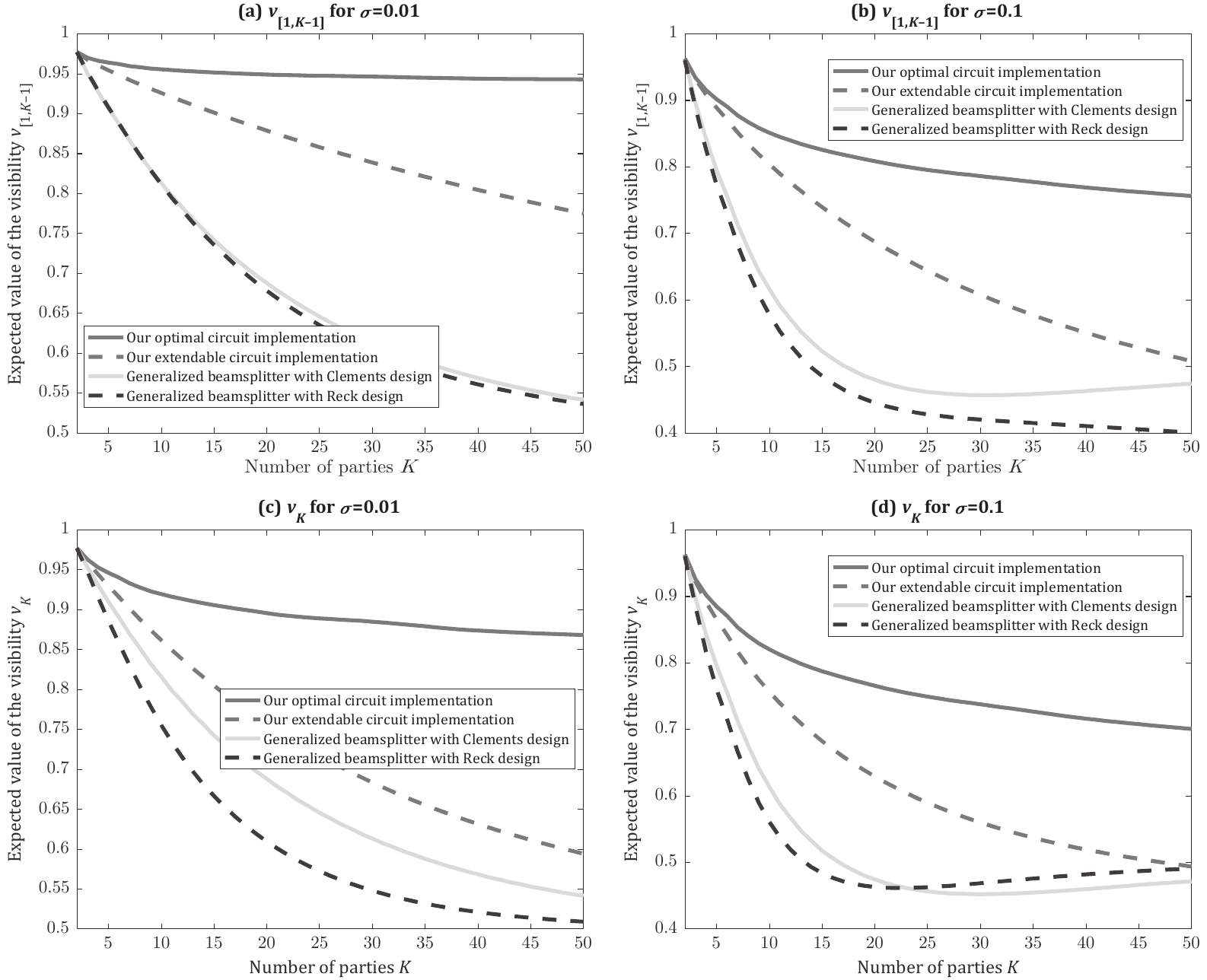}}
\caption{Variation of the mean values of the visibility as a function of the number of parties $K$ in four different referee circuit implementations. Just one input state was taken with a phase different from the rest, i.e. $L=1$. Losses per beamsplitter (BS) were taken to be $\eta_{\mathrm{BS}}=-0.2~\mathrm{dB/BS}$. Different types of visibilities and fabrication noise levels were considered: (a, b) visibility for detectors $[1,K-1]$ and (a) $\sigma=0.01$, (b) $\sigma=0.1$; (c, d) visibility for detector $K$ and (c) $\sigma=0.01$, (d) $\sigma=0.1$. The results were computed as the average of 500 realizations using the model in subsection 3.3. The computed standard deviation (SD, not represented) is $\sim {{10}^{-3}}$ for all the visibility values corresponding to $\sigma=0.01$, and $\sim {{10}^{-2}}$ for $\sigma=0.1$.}
\label{fig11}
\end{figure}

We provide in figure~\ref{fig11} graphical representations that show how the generalized visibilities ${v}_{[1,K-1]}$ and $v_K$ vary with respect to the number of parties $K$, for the two groups of exit ports ($k \in [1,K-1]$ and $k=K$) considered in the referee's decision rules. Two representative values $\sigma=0.01$ and $\sigma=0.1$ were selected for the fabricative error-level parameter $\sigma$, introduced in (\ref{eq10}) and in the subsequent explanation there. Either of these two values can be accomplished with present-day optical circuit fabrication technology, as discussed at the end of section~3. Losses per beamsplitter were chosen to be $\eta_\text{BS}=-0.2\;\text{dB/BS}$, which is a standard value reported in contemporary experiments and practical implementations; see for example \cite{Clements, Carolan}. In the context of this work, this value of $\eta_\text{BS}$ is applicable to the symmetric 50:50 beamsplitters in figure~\ref{fig10}, which are used as the building blocks necessary to implement  the generic unbalanced beamsplitters in the circuit designs. The results plotted in figure~\ref{fig11} highlight the clear superiority of our optimal design, introduced in subsection 3.3, when compared to the rest of circuit designs in section 3. Actually, this is not surprising at all, judging by the exponential savings in optical depth displayed in table~\ref{tab1} for our optimal layout. A small depth is extremely advantageous because the smaller the number of beamsplitters crossed by different internal paths through the circuit, the less error level is carried into the photonic gains of (\ref{eq14}) and (\ref{eq17}).

\begin{figure}[h]
\centerline{\includegraphics[width=0.65\columnwidth,draft=\isdraft]{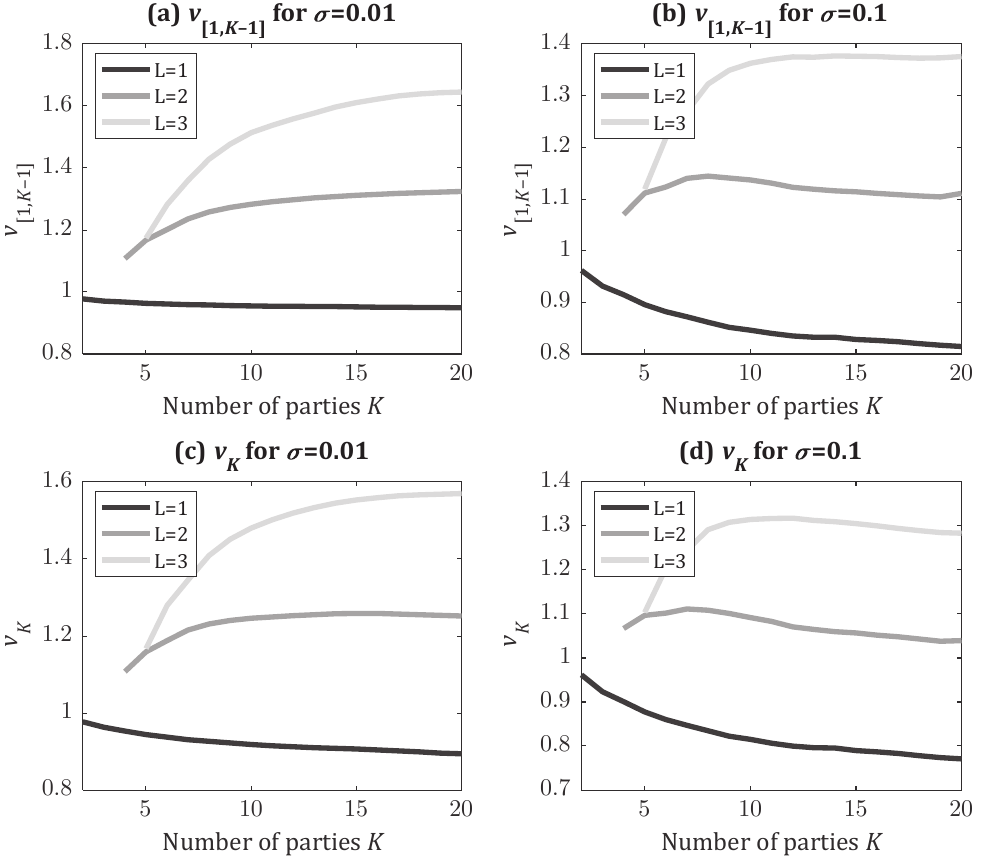}}
\caption{Variation of the mean values of the visibility as a function of the number of parties $K$ in the optimal circuit implementation, for three different values of $L$. The value of $L \leq \frac{K}{2}$ specifies the number of input states whose phases are different from the rest $K-L$ states.  Losses per beamsplitter (BS) were taken to be $\eta_{\mathrm{BS}}=-0.2~\mathrm{dB/BS}$. Different types of visibilities and fabrication noise levels were considered: (a, b) visibility for detectors $[1,K-1]$ and (a) $\sigma=0.01$, (b) $\sigma=0.1$; (c, d) visibility for detector $K$ and (c) $\sigma=0.01$, (d) $\sigma=0.1$. These results clearly show that, under the considered fabrication parameters, the worst-case scenario (smallest visibility) always corresponds to $L=1$.}
\label{fig12}
\end{figure}

In concluding this section, figure~\ref{fig12} contains various visibility plots calculated with different values of the gains $g_{[1,K-1],\bar{P}}^{\text{D}}$ and $g_{K,\bar{P}}^{\text{D}}$ obtained changing the number of input phases that differ. That is, calculated with different sets of labels in vector $\bar{P}$. This variation in vector $\bar{P}$ impacts the maximization and minimization procedures required for (\ref{eq15}) and (\ref{eq18}), and also for the visibilities in (\ref{eq22}). In particular, different values of the relevant parameter $L$ were chosen for the plots in figure~\ref{fig12}. We recall that $L \leq \frac{K}{2}$ is the minimum amount of phase labels in $\bar{P}$ that differ from the rest $K-L$ labels. It is easy to perceive in the plots that, for realistic fabrication noise levels characterized by parameter $\sigma$, the worst case unequivocally corresponds to $L=1$. This observation is to be expected, as all these situations, where a single input phase ($L=1$) of the $K$ individual pulses is different, are the most similar to the case of $K$ identical phases. Accordingly, in all these situations corresponding to $L=1$, gain $g_{[1,K-1],\bar{P}}^{\text{D}}$ reaches the closest value to $g_{[1,K-1]}^{\text{E}}$, and gain $g_{K,\bar{P}}^{\text{D}}$ reaches the most similar value to $g_{K}^{\text{E}}$. As a common sense conclusion, for computing quantities $\min (g_{[1,K-1],\bar{P}}^{\text{D}})$ and $\max (g_{K,\bar{P}}^{\text{D}})$, it suffices to calculate the values of the gains using the $K$ different vectors $\bar{P}$ that have $L=1$. Then, the minimum or maximum of these $K$ calculated values should be taken, as corresponds to each gain. This procedure entails an insignificant computational time. A final observation can be made on figure~\ref{fig12} concerning the fact that some of the plotted visibilities are greater than 1. This is so because, in order to provide a fair level comparison amongst the distinct values of $L$ in the realistic cases, we employed the ideal-case gains for $L=1$ in (\ref{eq20}) to (\ref{eq22}), even for the nonideal cases where $L>1$.

\clearpage 

\vspace{-10 pt} 

\section{Comparative results}
\label{sec5}
A plethora of plot results is presented and discussed in brief here. These results can be reproduced by applying the methodology in the prior section endowed with the statistical outcomes, in the form of simulated photonic gains, from the imperfection model in subsection 3.3. When computing the plot results here, we include as sources of experimental errors: fabricative imperfections (phase shifter and transmittance mismatches), beamsplitter insertion losses, channel losses, detector efficiencies, and the dark count rates present in the detectors. The first two error sources mentioned above are modelled through the statistical imperfection model. The rest of the sources are directly handled by the equations of the analytical methods for upper-bounding the mean photon number. In a realistic scenario, however, additional sources of errors, such as polarization and phase mismatches, can also be directly considered within the statistical model.

The Monte Carlo method underlying the imperfection model was applied averaging the results simulated with 500 unitary matrices corresponding to our \emph{optimal design}.  This is the same number of stochastic realizations used for the main results in \cite{Clements, Flamini1}. In our simulations, each unitary matrix is randomly modified according to the statistical imperfection model. Given the relatively low order of magnitude of the standard deviations compared to the absolute mean magnitudes at the logarithmic scale (see caption in figure~\ref{fig11}), they are not represented on the plots.

In the following, some common values used for the simulations presented in this section are discussed. Except otherwise stated, the combined efficiency, which excludes beamsplitter losses, was set to $\eta=0.5$. This value might seem quite unrealistic; nevertheless, we note that all the results can be straightforwardly scaled because of the fact that $|\alpha|_\eta^2=\frac{1}{2\eta}|\alpha|_{\eta=\frac{1}{2}}^2$, where $|\alpha|_\eta^2$ is the result with an arbitrary $\eta$ and $|\alpha|_{\eta=\frac{1}{2}}^2$ is the specific result used for our calculations. With regard to the foregoing, the main goal of plotting the results is \emph{not} to provide precise quantitative estimations, but rather to provide a qualitative overview of how the different involved variables affect the protocol performance and to prove that a positive quantum advantage is, in principle, already experimentally achievable.

With regard to the dark count probabilities, we use the two discreet values $p_{\mathrm{dark}}=10^{-9}$ and $p_{\mathrm{dark}}=10^{-11}$, except when analyzing the effects of a continuous distribution of $p_{\mathrm{dark}}$, in which case values as high as  $p_{\mathrm{dark}}=10^{-7}$ were used in the computations. Again, these values, specially $10^{-11}$, may seem difficult to achieve in practice with today's technologies. However, we shall observe that these stringent (low) values for $p_{\mathrm{dark}}$ are chosen mainly with the purpose of beating the classical limit. The requirement on the dark counts is much less stringent when beating the best-known classical protocol. Just to put all these dark count rates in perspective, the dark count probability of the QF experiment in \cite{Guan} is about $4.4 \times 10^{-9}$.

For the ECC, we chose the same optimized random linear code (RLC) in \cite{Xu}, whose generator matrix is a Toeplitz matrix. In particular, for this type of ECCs, the relationship between the rate $c > 1$ and the minimum-distance parameter $\delta$ is determined as
\begin{equation}
c={{\left[ 1+\delta \cdot{{\log }_{2}}\delta +(1-\delta )\cdot{{\log }_{2}}(1-\delta ) \right]}^{-1}}.
\label{eq23prev}
\end{equation}
The particular ECC values selected for this work are $\delta=0.78$ and $c=\frac{M}{N}=4.17$.

In our plots, for the purpose of adequately confronting the represented QF communication cost, we need on hand the expressions for the analogous cost of a classical protocol. The best classical fingerprinting protocol known to date, valid for $K=2$ only, is detailed in \cite{Babai} and its communication cost can be expressed in closed-form as
\vspace{-10pt} 
\begin{equation}
{{C}_{\text{best}}}=\left\lceil \frac{\log ({{p}_{\text{error}}})}{\log \left(\tfrac{3}{4}\right)} \right\rceil \times 2\sqrt{N}\text{~~~~[bits/user]}.
\label{eq23}
\end{equation}
We remark that there are other works \cite{Kremer, Ambainis2}, independent from \cite{Babai}, that lead to the result in (\ref{eq23}) as well. The protocols explained in \cite{Kremer, Ambainis2}, though, are not the same as the simple protocol in \cite{Babai}.

The best classical fingerprinting protocol known to date that is valid for any $K \geq 2$ was recently reported in \cite{Fischer}. It is based on a generalization of a 2-user protocol in \cite{Babai}. This 2-user protocol, however, is not the same as the protocol that gives the result in (\ref{eq23}). In the generalization, each user sends 4 randomly chosen blocks of size $\left\lceil \frac{3N}{K} \right\rceil $ bits. For each pair of blocks from the users, the node applies a 4-time repeated version of the so-called ``2-user symmetric protocol'' described in \cite{Babai}. Additionally, in order to identify the blocks, every user also sends labels comprising $\left\lceil~{{\log }_{2}}(3N/\left\lceil 3N/K \right\rceil )~\right\rceil$ bits. The resulting ${K}\text{-user}$ communication cost can be stated as
\vspace{-8pt} 
\begin{equation}
{{C}_{\text{best}}}=\left\lceil \frac{\log ({{p}_{\text{error}}})}{\log \left[1-\tfrac{1}{9}(1-{{\text{e}}^{-\frac{1}{2}}})\right]} \right\rceil \times \left[ 8\sqrt{2\left\lceil \frac{3N}{K} \right\rceil }+4\left\lceil {{\log }_{2}}\left( 3N/\left\lceil \frac{3N}{K} \right\rceil  \right) \right\rceil  \right]\text{~~~~[bits/user]}.
\label{eq24}
\end{equation}

As a classical ``no-go'' result, there is a classical limit on the communication cost, below which it is known that no classical protocol may operate, even protocols that could be unknown to date. For $K=2$, this limit was found in incomplete form (some multiplicative factors are missing) in \cite{Babai, Newman}. The complete closed-form version was first provided in the supplementary material of \cite{Guan}. A possible generalization to $K \geq 2$ users is derived in appendix C of the present work and it is given by
\begin{equation}
{{C}_{\text{limit}}}=\frac{\left( 1-2\sqrt{{{p}_{\text{error}}}} \right)\sqrt{N}}{2\sqrt{K\ln 2}}-\frac{1}{K}\text{~~~~[bits/user]}.
\label{eq25}
\end{equation}
Together with the cost of the best-known classical protocol, we shall also plot the limit in (\ref{eq25}) for the sake of completeness when comparing quantum communication costs. For our quantum upper bounds and for the classical quantities in (\ref{eq23}) to (\ref{eq25}), we set the target error probability to a common value $p_\text{error}=10^{-5}$.

\subsection{Transmitted information comparison with previously published two-user referee strategy}
\label{sec52}

As our first set of protocol simulation results, and serving the purpose of strengthening the correctness verification of our analytical method in section 4, we particularize the referee strategy to the case $K=2$ and then compare the results to those dispensed by other methods in previous publications. In this regard, our referee decision strategy for the case with $K-1$ detectors, when particularized to $K=2$, is mostly the same as in the paper by Xu \emph{et al} \cite{Xu}. The only difference lies in the referee threshold. In particular, for a target error probability $p_\text{error}$, Xu's referee strategy uses a threshold $r$ whose value satisfies the equalities ${{p}_{\text{error}}}=\Pr (D_{1}^{\text{E}}>r)=\Pr (D_{1}^{\text{D}}\le r)$. In contrast, we recall that, in our strategy, we can only guarantee ${{p}_{\text{error}}} \geq \Pr (D_{1}^{\text{E}}>r),\Pr (D_{1}^{\text{D}}\le r)$. This less strict condition, in turn, allows for a compact mathematical analysis extended to the multi-user case.

The specific method for ultimately calculating $r$ is not explicitly detailed in \cite{Xu}. To this regard, we employed in our simulations the algorithm in section 2 of this paper. This algorithm assumes the same conditions to calculate $r$ as in \cite{Xu}, so it should always produce the same results as the method actually used in this aforesaid reference, even if different. For comparison purposes, we simulated our two multi-user strategies for the separate detector ensembles $k \in [1,K-1]$ and $k=K$. When particularized to $K=2$, the case $k \in [1,K-1]$ implies observing detector ``1'', just as in Xu's strategy according to our notation, and the case $k=K$ implies observing detector ``2''.

\begin{figure}[h]
\centerline{\includegraphics[width=0.85\columnwidth,draft=\isdraft]{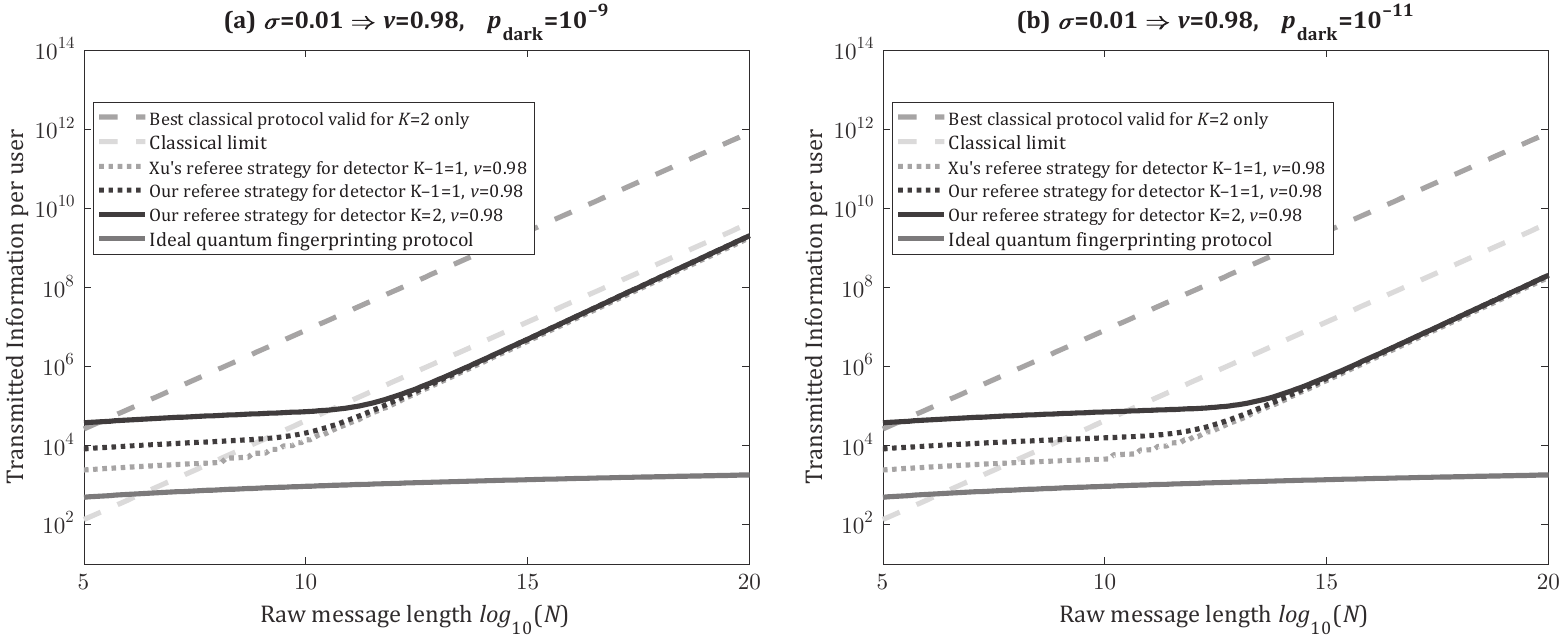}}
\caption{Transmitted information comparison between a previously-published referee strategy that is valid for $K=2$ only, and our two multi-party referee strategies particularized to $K=2$. The parameters for the two-user realistic quantum fingerprinting protocols for all three strategies were $p_{\rm error}=10^{-5}$, combined efficiency excluding beamsplitter (BS) losses $\eta=0.5$, $\eta_{\rm BS}=-0.2~{\rm dB/BS}$, $\sigma=0.01$ ($\Rightarrow v=0.98$ for all the strategies, in this particular scenario), (a)~$p_{\mathrm{dark}}=10^{-9}$, (b)~$p_{\mathrm{dark}}=10^{-11}$.}
\label{fig14}
\end{figure}

Figure~\ref{fig14} shows the simulation results, where the transmitted information is represented as a function of $N$ at log-log scale. The fabrication noise level $\sigma=0.01$ and the beamsplitter losses $\eta_\text{BS}=-0.2~\text{dB/BS}$ utilized here provide a visibility $v=0.98$ for all the realistic protocols. The best-protocol classical information was calculated with (\ref{eq23}), and the classical limit corresponds to (\ref{eq25}). The ideal QF protocol bound can be either from (\ref{eq2}) or from (\ref{eq13}) and it assumes zero losses, zero imperfections and no dark counts.

In view of the results in figure~\ref{fig14}, it is clear that Xu's upper bound is tighter than those provided by our two separate strategies. This behaviour occurs when our strategies are really ensuring an actual error probability below the target $p_\text{error}$, and the users need to send more information than required by $p_\text{error}$. This has to do with our more relaxed condition on the calculation of $r$. However, after the ``elbow'' of the curves, where the slopes become more vertical, all the three plotted functions are basically undistinguishable. This situation corresponds in our analytical method to a dominant term of $M=c\,N$ and $p_\text{dark}$, which occurs when conditions (\ref{eq20pre1}) and (\ref{eq20pre2}) are satisfied. Another relevant comment on the results has to do with the strong influence of $p_\text{dark}$ on the required transmitted information. In particular, a more favorable dark count rate pushes the curve's ``elbow'' towards a point where the input size $N$ is larger. Finally, we perceive that, in these particular simulations, all the QF protocol strategies beat the classical limit for most of the range of $N$, while the best-known classical protocol is beaten nearly for all the represented range of $N$.

\subsection{Transmitted information comparison with a na\"ive K-user protocol}
\begin{figure}[h]
\centerline{\includegraphics[width=1.0\columnwidth,draft=\isdraft]{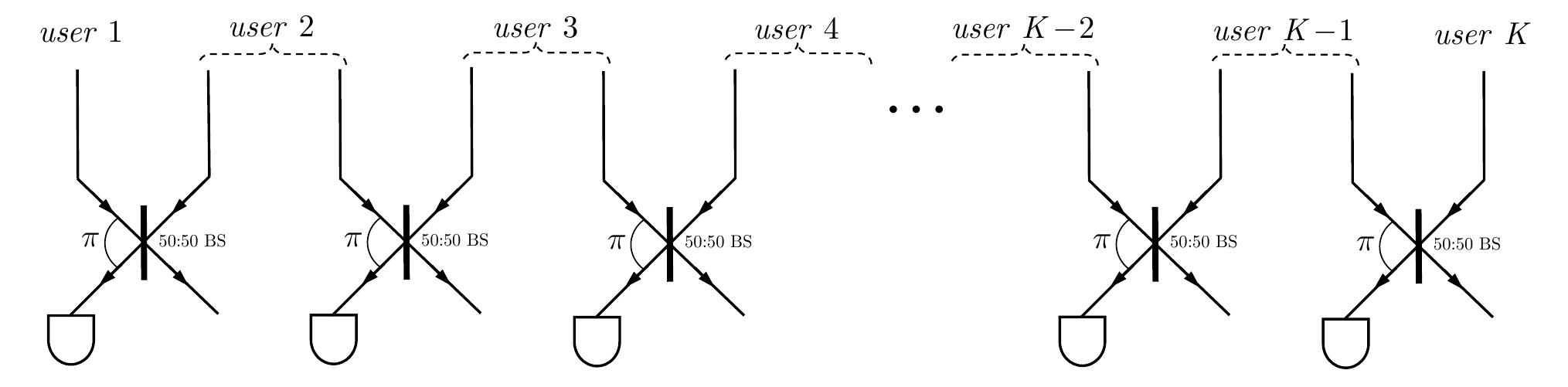}}
\vspace{-10pt} 
\caption{Schematic representation of the referee circuits in the na\"ive $K\text{-user}$ QF protocol. Under the same rules in \cite{Xu, Guan}, used here for each of the $K-1$ individual 2-user protocols, just one detector is required for monitoring a single 50:50 beamsplitter. In general, this na\"ive implementation is simpler than using our optimal multiports introduced in subsection 3.3. However, in the na\"ive approach, most users must send twice the same signal states. The na\"ive protocol performance in terms of information and energy can be well approximated by our (\ref{eq15}). }
\label{fig14b}
\end{figure}

A simple na\"ive $K\text{-user}$ quantum protocol can be implemented by repeating $K-1$ times the 2-user protocol by Xu \emph{et al} that was confronted in the preceding subsection. This na\"ive scheme, made with the prior art protocol, requires that each user, except two of them (say, users $k=1$ and $k=K$), sends twice the train of coherent light pulses that encodes the information. In this scheme, the referee implements the 2-party protocol between users 1 and 2, between users 2 and 3, and so forth as depicted in figure~\ref{fig14b}.
\vspace{-10pt} 

Whichever number of users is involved, the implementation of the  na\"ive approach requires only standard 50:50 beamsplitters, and zero phase shifters, which is indeed a remarkable benefit. Additionally, the transmitted information per user in the \emph{ideal} scenario is roughly only a factor of 2 over the case with only two users. To calculate the transmitted information in a \emph{realistic} scenario, we may take advantage of the fact that Xu's protocol has a worst-case error probability that is the same for equal and different input sequences. Thus, if $p^\prime_\text{error}$ corresponds here to an individual 2-user protocol, then the worst-case error probability of the na\"ive protocol can be calculated as $p_\text{error}=1-(1-p^\prime_\text{error})^{K-1}$. Accordingly, in order to compute an upper bound for $|\alpha|^2$, we may use algorithm 2.1, which includes dark counts and beamsplitter imperfections, by fixing the following error probability in the aforesaid algorithm:
\vspace{-8pt} 
\begin{equation}
p^\prime_\text{error}=1-(1-p_\text{error})^\frac{1}{{K-1}}.
\label{eq25b}
\end{equation}
The value of $|\alpha|^2$ obtained from the algorithm needs to be multiplied by a factor $\frac{2(K-1)}{K}$ to get an accurate information per user in~(\ref{eq4}).

It is important to bring attention to the fact that (\ref{eq25b}) corresponds to a worst-case scenario where all the $K$ users send the same strings. Furthermore, we assumed that the threshold constraints of each individual 2-party protocol in figure~\ref{fig14b} guarantee that the individual error rate is the same for both equal and different 2-user input sequences, as stated in \cite{Xu, Guan} and assured by applying our algorithm 2.1.   Therefore, (\ref{eq25b}) represents a pessimistic upper-bound estimation for the error rate of the overall na\"ive multiparty protocol, considering that the overall error probability would be much smaller if at least one of the $K$ strings were different. Keeping in mind that the target error probability is an error rate limit that cannot be surpassed in any execution of the complete $K\text{-user}$ protocol and that neither the referee nor the protocol designer has previous information about the input sequences, we must always calculate $|\alpha|^2$ assuming a worst-case scenario. This is why in the na\"ive protocol the right choice is to consider all-equal inputs.

\begin{figure}[h]
\centerline{\includegraphics[width=0.85\columnwidth,draft=\isdraft]{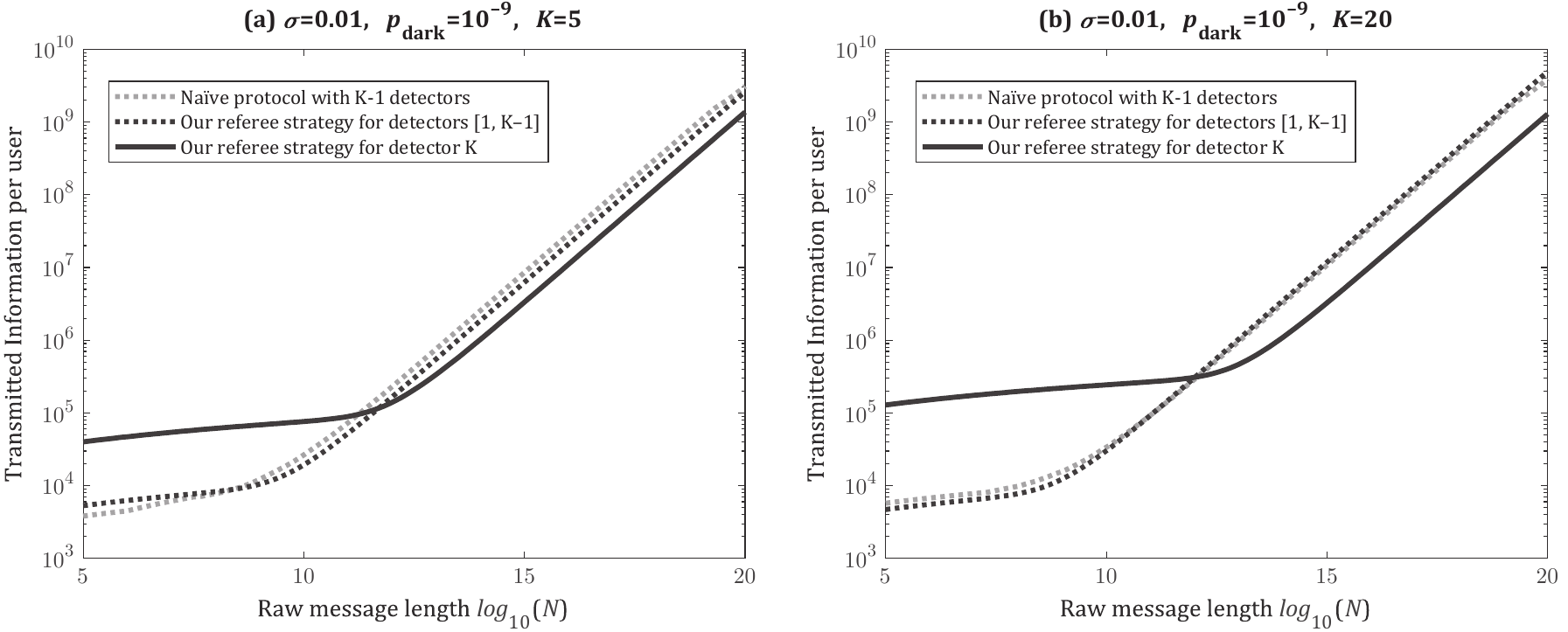}}
\caption{Transmitted information comparison between a na\"ive approach based on Xu's previously-published referee strategy for $K=2$ and our two multi-party referee strategies. The parameters used in the simulations were $p_{\rm error}=10^{-5}$, combined efficiency excluding beamsplitter (BS) losses $\eta=0.5$, $\eta_{\rm BS}=-0.2~{\rm dB/BS}$, $\sigma=0.01$, $p_{\mathrm{dark}}=10^{-9}$, (a)~$K=5$, (b)~$K=20$.}
\label{fig14c}
\end{figure}

We found that the required amount of transmitted information in the na\"ive protocol is roughly the same as in our multi-user protocol when the referee observes $K-1$ detectors in our strategy. This similarity strengthens as the number of users $K$ becomes larger, as shown in figure~\ref{fig14c}. These findings do not override our multi-party protocol analysis, owing to the following important reasons: (i) our multi-user strategy involving just one detector is always superior when $N$ is arbitrarily large, as it is clear from the plots; (ii) we have empirically shown that our analytical formulas for the case with $K-1$ detectors may serve as a decent approximation for predicting the behaviour of a na\"ive protocol, which may be convenient for inferring how the different involved parameters affect the protocol execution in a simple experimental setup.  In addition, a drawback of the na\"ive approach is that it increases the execution time of the protocol, unless two channels per user are available or the information can be encoded in different modes such as polarization modes. The latter approach would require, however, to make the referee circuits more complicated by adding polarizing beamsplitters, whose imperfections would contribute as additional sources of errors.

So far, we have assumed that, for the individual 2-user protocols that form together the na\"ive approach, the equal-input error probability $p^\prime_\text{error,E}$ is the same as the different-input error probability $p^\prime_\text{error,D}$, as inherited from previous works. We may wonder if we can alter the 2-user decision threshold in such a way that $p^\prime_\text{error,E} \neq p^\prime_\text{error,D}$ and the overall $K\text{-party}$ protocol performance is significantly improved. A precise calculation of $p_\text{error,D}$ for the overall $K\text{-user}$ protocol would require scenario assumptions in which some prior knowledge about the probability distributions of the input strings is available. This scenario stays totally ahead of the goals of the present paper; however, we may assume a worst-case scenario for the case of different input strings. Under this worst-case assumption, $p_\text{error,D}$ must be calculated for a situation where just one the $K$ users is sending a different sequence. If we assume that this user is in one of the two ends of the queue in figure~\ref{fig14b}, then
\vspace{-8pt} 
\begin{equation}
p_\text{error,D}=p^\prime_\text{error,D}(1-p^\prime_\text{error,E})^{(K-2)}.
\label{eq25c}
\end{equation}
Also, we can directly obtain $p^\prime_\text{error,E}$ from (\ref{eq25b}) as
\vspace{-8pt} 
\begin{equation}
p^\prime_\text{error,E}=1-(1-p_\text{error,E})^\frac{1}{{K-1}}.
\label{eq25d}
\end{equation}
Keeping in mind that in the worst-case estimation, without prior knowledge, equality $p_\text{error}=p_\text{error,D}=p_\text{error,E}$ must hold, we may introduce (\ref{eq25d}) into (\ref{eq25c}) in order to get
\vspace{-8pt} 
\begin{equation}
p^\prime_\text{error,D}=\frac{p_\text{error}}{(1-p_\text{error})^\frac{K-2}{{K-1}}}.
\label{eq25e}
\end{equation}
In the previous equation, $p_\text{error}$ is the target error probability of the overall protocol. Using this probability, we may calculate both (\ref{eq25d}) (with $p_\text{error}=p_\text{error,E}$) and (\ref{eq25e}), and then take $p^\prime_\text{error,E}$ into step 2 of algorithm 2.1, and $p^\prime_\text{error,D}$ into step 3. We found, however, that the values of $|\alpha|^2$ obtained from this slightly modified algorithm are virtually the same already plotted in figure~\ref{fig14c}, because the effect of $p^\prime_\text{error,E}$ is dominant in algorithm 2.1. As a consequence, this protocol with different 2-user probabilities does not represent a significant improvement over (\ref{eq25b}).

\clearpage 
\subsection{Assessment of the impact of dark counts and visibility on the transmitted information}
\label{sec53}

\begin{figure}[h]
\centerline{\includegraphics[width=1.0\columnwidth,draft=\isdraft]{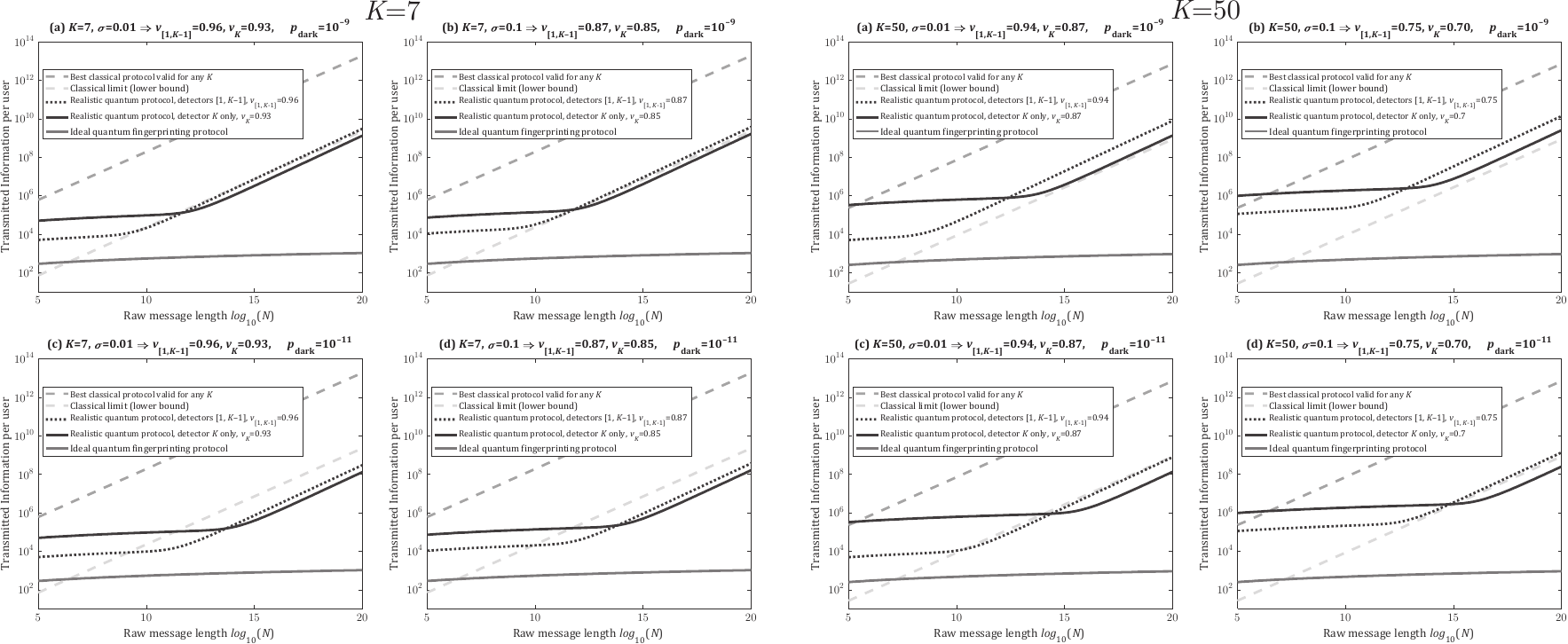}}
\caption{Variation of the transmitted information per user as a function of the raw message length $N$ for $K=7$ (four plots on the left) and $K=50$ users (plots on the right). For comparison purposes, we provide the classical lower bound and also the amount of bits per user required by the best-known classical protocol valid for any $K\geq2$. The realistic multi-party quantum fingerprinting protocols were analyzed for the following common parameters: $p_{\rm error}=10^{-5}$, combined efficiency excluding beamsplitter (BS) losses $\eta=0.5$, $\eta_{\rm BS}=-0.2~{\rm dB/BS}$. In order to appraise the influence of manufacturing imperfections and dark count rates, the following four cases were computed for each value of $K$: (a)~$\sigma=0.01$, $p_{\mathrm{dark}}=10^{-9}$; (b)~$\sigma=0.1$, $p_{\mathrm{dark}}=10^{-9}$; (c)~$\sigma=0.01$, $p_{\mathrm{dark}}=10^{-11}$; (d)~$\sigma=0.1$, $p_{\mathrm{dark}}=10^{-11}$.}
\label{fig15}
\end{figure}

Focusing now only on our protocol realizations with more than two users, we assess in this subsection 5.3 the impact of the fabricative imperfections and the dark count rates on the amounts of transmitted information. To this intent, we provide diverse plots in figure~\ref{fig15} resulting from taking $K=7$ and $K=50$, and then varying $\sigma$ and $p_\text{dark}$ to gather some general conclusions. Each plot shows the evolution of the transmitted quantum information for our two referee strategies, as a function of $N=\frac{M}{c}$ at log-log scale. The amount of information required by the best classical protocol for $K \geq 2$ is also represented using (\ref{eq24}), whereas the classical limit comes from (\ref{eq25}). For the case $K=7$, the fabricative noise $\sigma=0.01$ provides visibilities $v_{[1,K-1]}=0.96$ and $v_K=0.93$, whereas $\sigma=0.1$ provides $v_{[1,K-1]}=0.87$ and $v_K=0.85$. For the other case $K=50$ on the figure's right, the visibilities diminish when compared to $K=7$, as shown on the graphs.

As a first evident assertion, the effect of increasing imperfections (and hence reducing the visibility) has a very small impact on the communication cost for the smallest number of users. It becomes much more noticeable for the largest number of users, specially in the region of the curves before the ``elbow'', where the term of $M$ and $p_\text{dark}$ is not the dominant one in (\ref{eq15a}) and (\ref{eq18a}). Additionally, this effect of increasing $\sigma$ is more prominent when the referee uses $K-1$ detectors than when she uses just one.

As a second observation, in all the simulated cases, the strategy involving $K-1$ detectors is clearly superior than the other one before the ``elbow''. However, as the term of $M$ and $p_\text{dark}$ becomes dominant, the strategy with just detector $k=K$ provides the smallest communication cost. This is to be expected by comparing (\ref{eq15a}) and (\ref{eq18a}). These differences, between the two referee strategies in our work, are much more noticeable as the number of users rises. Further, it can be also observed that the strategy with $K-1$ detectors reaches the elbow point for a smallest raw message length $N$.

Also, we note in figure~\ref{fig15} that overcoming the classical limit is much more difficult than beating the best multi-party classical protocol known to date. When attempting to beat the classical limit, dark counts are a key limiting factor, much more dominant than $\sigma$. In particular, for our simulation with $K=50$, we can only achieve less information than the classical limit if we use $p_\text{dark}=10^{-11}$.

\subsection{Relationship between transmitted information and transmitted energy}
\label{sec54}

\begin{figure}[h]
\centerline{\includegraphics[width=1.0\columnwidth,draft=\isdraft]{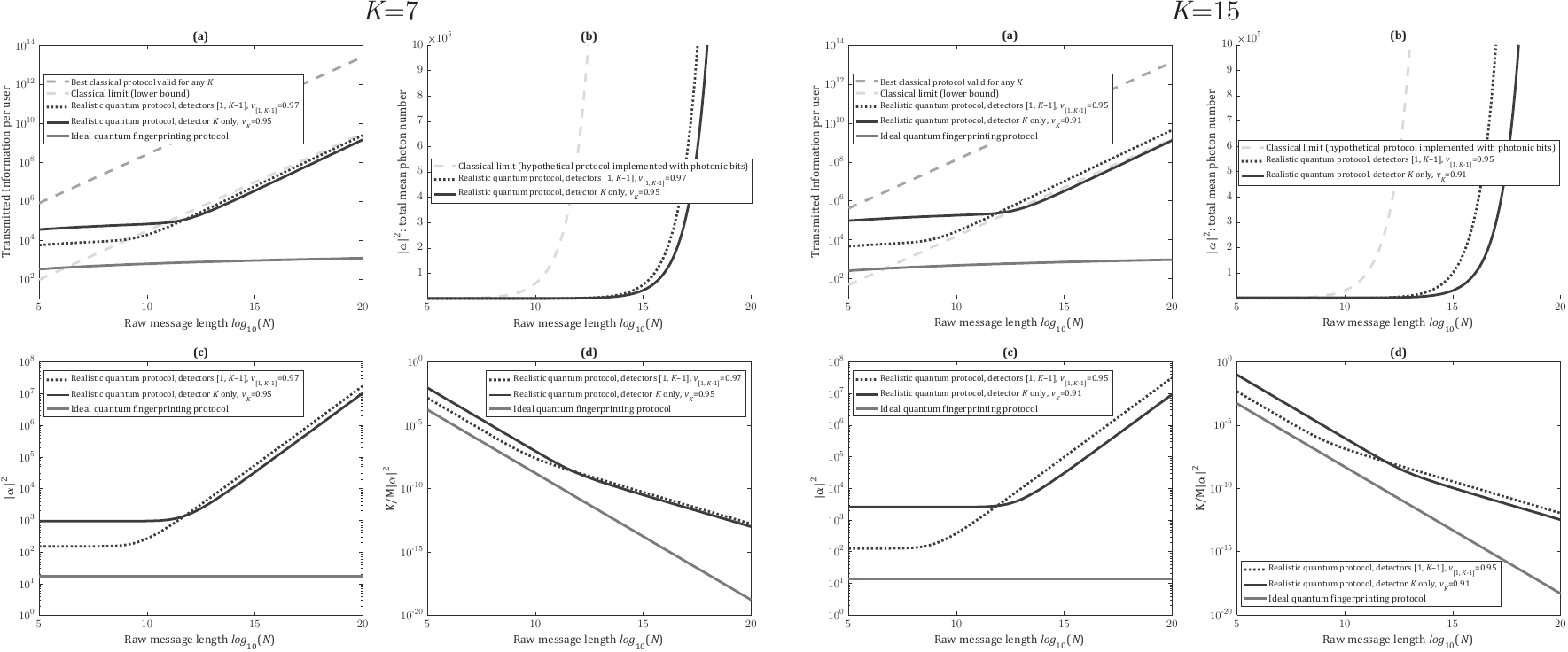}}
\caption{Influence of the raw message length $N$ on the transmitted information and on the photon statistics for $K=7$ (four plots on the left) and $K=15$ users (plots on the right). The following parameters were chosen for the realistic multi-party quantum fingerprinting protocols: $p_{\rm error}=10^{-5}$, combined efficiency excluding beamsplitter (BS) losses $\eta=0.5$, $\eta_{\rm BS}=-0.2~{\rm dB/BS}$, $\sigma=0.01$, $p_{\mathrm{dark}}=10^{-9}$. (a) Comparison in terms of transmitted information, providing the classical limit and also the amount of bits per user required by the best-known classical protocol. (b) Comparison of the total number of photons transmitted per user that are needed in two quantum protocol referee strategies and in a hypothetical classical limit protocol. (c) Comparison of the total number of photons per user required in two quantum protocol referee strategies and in an ideal quantum protocol. (d) Expected total number of photons in $K$ simultaneous individual pulses arriving at the referee input ports at the same time, provided here to assess the validity of the assumption $K{{\mu }_{in}}=\frac{K{{\left| \alpha  \right|}^{2}}}{M}\ll 1$ in the analytical model for upper bounding ${\left| \alpha  \right|}^{2}$.}
\label{fig16}
\end{figure}

We study here the relationship between the transmitted information and the required energy. To this end, figure~\ref{fig16} includes the following plots as functions of $N$: (a)~information per user at log-log scale; (b)~total mean photon number $|\alpha|^2$ at natural scale compared to the amount of photonic bits required at the classical limit; (c)~$|\alpha|^2$ at log-log scale compared to an ideal quantum protocol; (d)~amount $\frac{K \,|\alpha|^2}{M}$ at log-log scale. The plots on the leftmost part of figure~\ref{fig16} correspond to $K=7$, whereas the rightmost part plots were computed using $K=15$. The photonic classical fingerprinting protocol refers to a classical protocol in which a bit is assigned to a photon, hence the term ``photonic bit,'' as introduced in \cite{Arrazola, Xu}. The amount $\frac{K \,|\alpha|^2}{M}$ is represented in order to confirm the strict validity of the assumption $K{{\mu }_{\text{in}}}=\frac{K{{\left| \alpha  \right|}^{2}}}{M}\ll 1$ upon which the mathematical results in (\ref{eq15}) and (\ref{eq18}) rely. All these simulations in figure~\ref{fig16} were carried out taking a reasonable dark count rate $p_\text{dark}=10^{-9}$.

We observe in figure~\ref{fig16}, for the region of non-dominant term of $M$ and $p_\text{dark}$ before the ``elbow'' in (a), that the total mean photon number $|\alpha|^2$ required per user remains constant in (b) and (c). This is the same behaviour exhibited by the ideal protocol in (c), although the $|\alpha|^2$ level of this latter protocol is much lower. After the ``elbow'', the increment in $|\alpha|^2$ becomes exponential in the realistic protocols. This happens because the combined effect of $M$ and $p_\text{dark}$ becomes dominant and the users need to send more energy to keep the error probabilities below the target value $p_\text{error}$. In other words, when the term of $M$ and $p_\text{dark}$ governs the required value of $|\alpha|^2$, the clicks at the detectors become dominated by $p_\text{dark}$ and the gains in (\ref{eq14}) and (\ref{eq17}) become close to each other.

Interestingly enough, there is a region in plots (b), for both values of $K$ under consideration, in which $|\alpha|^2$ remains practically constant for the QF protocol while the amount of energy of the photonic-bit classical protocol grows exponentially, even at the classical limit. Thus, before the ``elbow'', the QF protocol requires an exponential reduction in terms of energy consumption, which is indeed remarkable. Finally, we also observe in plots (d) that the premise $K{{\mu }_{\text{in}}}=\frac{K{{\left| \alpha  \right|}^{2}}}{M}\ll 1$ is comfortably met. The greater the raw message length $N$, the strongest the validity of the assumption on which our analytical model is constructed.

\vspace{-8pt} 

\subsection{Quantum advantages in terms of transmitted information}
\label{sec55}

Starting from this subsection, we focus exclusively on the strategy involving just one detector. This choice is made because such strategy delivers the best performance in terms of energy consumption and transmitted information when the raw input size $N$ is arbitrarily large. To the purpose of intuitively represent on a 2D plane, as a color plot, how the protocol behaves, we define the quantum advantages in terms of information as $C_\text{limit}/Q$ and $C_\text{best}/Q$. Here, $Q$ is the quantum information defined in (\ref{eq4}), $C_\text{best}$ is the number of bits per user in the best-known $K\text{-user}$ classical protocol given in (\ref{eq24}), and $C_\text{limit}$ is the classical limit in (\ref{eq25}).

\begin{figure}[h]
\centerline{\includegraphics[width=1.0\columnwidth,draft=\isdraft]{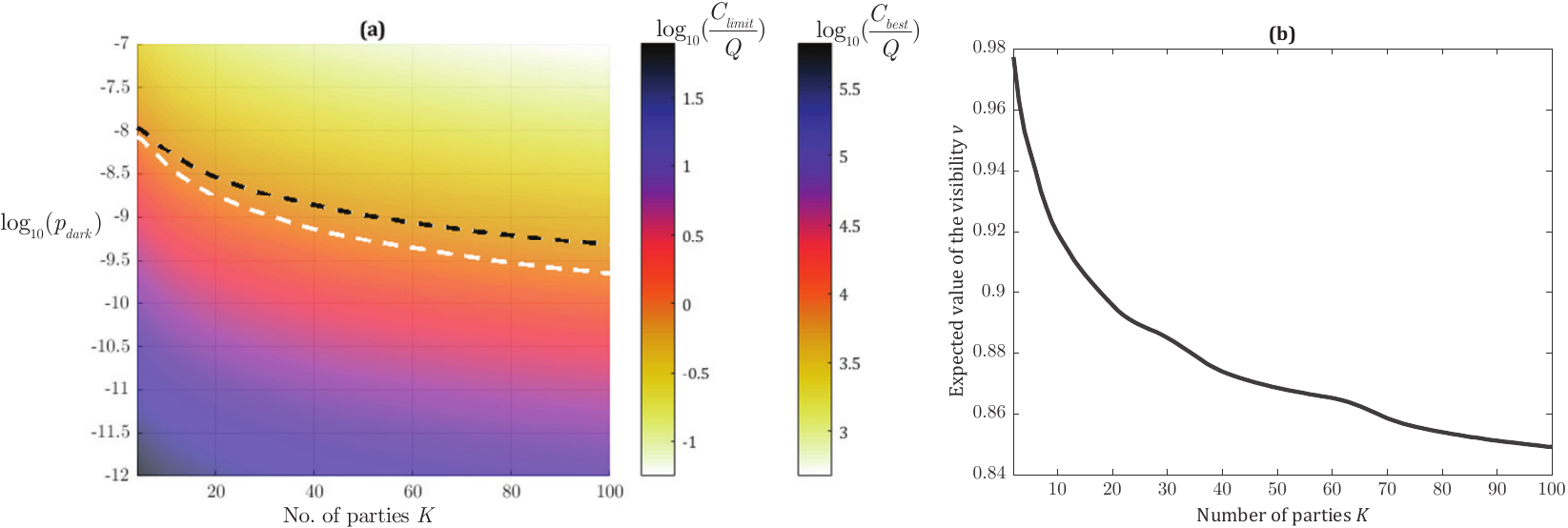}}
\caption{(a) Maximum quantum advantages regarding transmitted information as a function of the number of parties $K$ and dark count rate. Two ratios are presented: between the classical limit information $C_{\rm limit}$ and the quantum information $Q$, and between the best-known classical protocol information $C_{\rm best}$ and $Q$. The following parameters were used together with the optimal referee strategy: $p_{\rm error}=10^{-5}$, combined efficiency $\eta=0.5$ ($Q$ may be easily scaled for any $\eta$ as $Q_{\eta}=\frac{1}{2\eta}Q_{\eta=\frac{1}{2}}$), $\eta_{\rm BS}=-0.2~{\rm dB/BS}$, $\sigma=0.01$. The white dashed curve indicates a lower bound below which a positive quantum advantage $C_\text{limit}/Q$ is attainable. For comparison purposes, the black dashed curve corresponds to an \emph{ideal} referee circuit with $\eta_{\rm BS}=0~{\rm dB}$, $\sigma=0$. (b) Visibility as a function of $K$ corresponding to the realistic optical circuit implementation used in (a).}
\label{fig17}
\end{figure}

Figure~{\ref{fig17}a} exhibits a representation on a 2D plane of the maximum quantum advantages as a function of $K$ and $p_\text{dark}$ when $\sigma=0.01$. The white dashed curve represents a lower bound below which a positive quantum advantage $C_\text{limit}/Q$ may be achievable. The black dashed curve is analogous to the white one, but for an \emph{ideal} circuit with $\eta_\text{BS}=0~\text{dB}$ and $\sigma=0$. Note that this ideal case is \emph{not} the same as in previous figures, because here we solely consider an ideal circuit and the detector dark counts are still on. These two dashed curves call again attention to the fact that, with today's technology, the dark counts are a much more limiting factor for the QF protocol than the fabrication defects in the circuit. Related to this fact, we can check in the graph of figure~{\ref{fig17}b} how the fabrication noise level degrades the visibility as the number of users is increased, from $v_K=0.98$, for $K=2$, down to a still relatively high value $v_K=0.85$ for $K=100$.

\vspace{-8pt} 
\vspace{-8pt} 

We bring a special attention again to the bound represented by the white curve on figure~{\ref{fig17}a}. This curve seems to impose a strong burden on the practicality of the scheme. We recall, however, that the apparent stringent requirement on the dark counts is only for the quantum advantage $C_\text{limit}/Q$, defined in terms of the classical limit. If we consider the quantum advantage with regard to the best-known classical protocol (see the colorbar for $C_\text{best}/Q$ on the same figure), the dark count rate is not a key limiting factor anymore. We emphasize that the classical limit is a theoretical lower bound, below which no classical protocol may operate, as detailed in appendix C. To this date, no general mathematical proof is known that guarantees that a protocol operating at this limit must exist. The best multi-user classical protocol found so far, reported in \cite{Fischer}, requires an information per user that is several orders of magnitude above the limit. As a consequence, from today's perspective, our quantum protocol can be seen as a practical scheme compared to the best classical protocol, even for off-the-shelf common photonic detectors. If a better classical protocol is found sometime, we may also expect to have detectors with better dark count rates in the future. With the purpose of giving perspective on ultra low dark count values that can be achieved at present in experimental demonstrations of quantum protocols, we focus now on \cite{Shibata}. This reference reports a quantum key distribution experiment using SNSPDs with $p_{\mathrm{dark}}=10^{-11}$, for a system detection efficiency of 4.4\% and a system clock rate of 1 GHz.

\subsection{Quantum advantages in terms of transmitted energy}
\label{sec56}

\begin{figure}[h]
\centerline{\includegraphics[width=1.0\columnwidth,draft=\isdraft]{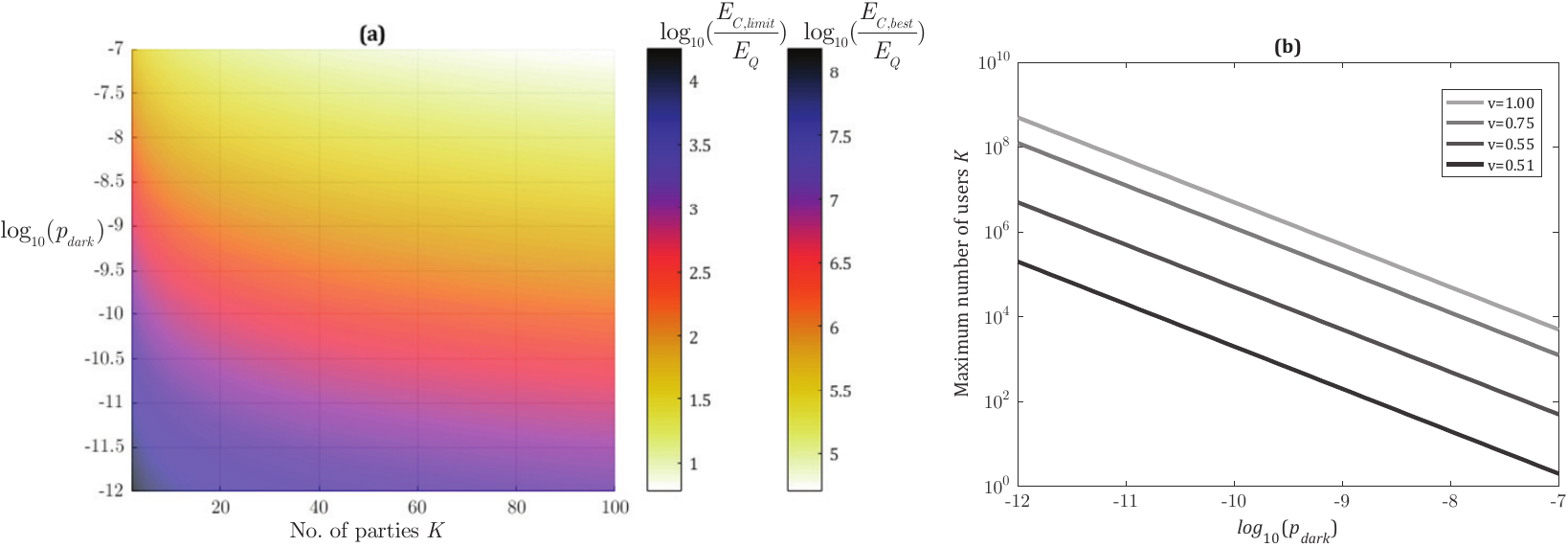}}
\caption{(a) Quantum advantages concerning transmitted power, or energy, as a function of the number of parties $K$ and dark count rate. Two ratios are presented: between the classical limit protocol energy $E_{\rm C,limit}$ and the quantum protocol energy $E_{\rm Q}$, and between the best-known classical protocol energy $E_{\rm C,best}$ and $E_{\rm Q}$. The following parameters were used together with the optimal referee strategy: $p_{\rm error}=10^{-5}$, combined efficiency $\eta=0.5$ (the results here may be easily scaled for any value of $\eta$), $\eta_{\rm BS}=-0.2~{\rm dB/BS}$, $\sigma=0.01$. (b) Lower bounds on the maximum number of users that allows a positive quantum advantage in terms of energy as a function of the dark count rate, for four assumed worst-case visibility values.}
\label{fig18}
\end{figure}

Quantum advantages in terms of transmitted energy, analogous to those defined above for the information, are analyzed here. Figure~{\ref{fig18}a} shows that, indeed, a positive quantum advantage for energies is commonplace even for ordinary photonic detectors when comparing to the classical limit. This reality represents huge energy savings of several orders of magnitude compared to any classical protocol implemented using photonic bits.

In the following, we deduce an approximate expression for calculating the maximum number of users $K$ for which a positive quantum advantage is achievable in terms of classical limit energy, as a function of $\mu_\text{dark}$, $p_\text{error}$, $c$, $\delta$ and visibility $v_K$. We assume that the condition in (\ref{eq20pre2}), for arbitrarily high $M=c\,N$, holds. Then, we rewrite $|\alpha|^2$ in (\ref{eq18a}) as a function of visibility $v_K$ in (\ref{eq22}) instead of as a function of the gains in (\ref{eq17}):
\vspace{-18pt} 
\begin{equation}
{{\left| \alpha _{K}^{\text{bound}} \right|}^{2}}\simeq \frac{{{\left[ 2cN\mu _{\text{dark}}^{{}}\cdot \ln (1/p_{\text{error}}^{{}}) \right]}^{1/2}}}{2\eta \ (1-\delta )(2{{v}_{K}}-1)}.
\label{eq26}
\end{equation}
Now, assuming $K \gg 1$, the version with photonic bits of the classical limit in (\ref{eq25}) can be well approximated, leaving out the term in $1/K$, as
\vspace{-18pt} 
\begin{equation}
{{\left| \alpha _{\text{limit}}^{\text{classical}} \right|}^{2}}\simeq \frac{\left( 1-2\sqrt{p_{\text{error}}^{{}}} \right)\sqrt{N}}{2\eta \sqrt{K\ln 2}}.
\label{eq27}
\end{equation}
Finally, equating these two previous expressions and solving for $K$, we arrive at
\begin{equation}
K\simeq \frac{{{(1-\delta )}^{2}}{{(2{{v}_{K}}-1)}^{2}}{{(1-2\sqrt{p_{\text{error}}^{{}}})}^{2}}}{2\mu _{\text{dark}}^{{}}c\ln (2+1/p_{\text{error}}^{{}})}.
\label{eq28}
\end{equation}
This latest equation takes into account the supposition that $v_K$ is independent from $K$. In practice, this is not the case. However, we may fix an expected worst-case experimental value of the visibility and then obtain a lower bound on the maximum number of users for which the quantum protocol requires less energy than a hypothetical classical protocol matching the classical limit energy. Figure~{\ref{fig18}b} shows an example of the result in (\ref{eq28}) at work. This result may be also of interest for determining the required dark count probability as a function of the desired number of users $K$ in an experimental realization.
\vspace{-8pt} 

\section{Conclusions and future perspectives}
\label{sec6}

In this paper, we have proposed and investigated a $K\text{-user}$ QF protocol based on coherent states. One of the main incentives is on the fact that an analogous classical protocol is known and can be used for comparison purposes to attest quantum advantages. Our work constitutes a step towards a deeper understanding of quantum networks embracing a central processing node.

As one of the main contributions of this work, we have provided innovative optical circuit designs required for the protocol, and have discussed the benefits and issues of each design. Then, we have proposed and detailed two separate referee strategies for the central node. Also, we have introduced a fully-analytical method to perform the estimations of the amounts of qubits and energy required for each user. These analytical expressions are, indeed, very convenient to understand how the different involved quantities affect the protocol execution, even at the experimental level. Further, simulations are presented that certify positive quantum advantages under certain circumstances. This superiority of the quantum protocol over the classical one is especially noticeable when comparing energy consumptions, which paves the way for the deployment of quantum networks implementing data-processing ``green'' protocols. In doing the simulations, we also determined under which conditions one of the proposed strategies is more efficient than the other.

An instinctive approach to continue our research would exploit the fact that the protocol sends pulses with very low amplitude, mostly empty coherent states. This means that the expected time between clicks at each individual detector is large, and detector dead times are usually not a problem. Besides this benefit, the referee node can be adjusted to process many signals in parallel. This idea was first proposed in \cite{Kumar} for the two-user protocol and, very recently, it was experimentally demonstrated with improvements in \cite{Zhong}.

Another natural step for the continuation of this work would investigate different ways of defining the trains of pulses sent by the users. In the present work, we stuck to the same scaling properties and to the same number of pulses per user as in the standard two-user protocol. Perhaps, some improvements in the communication cost can be attained by redefining the coherent states in such a way that exploits more efficiently the peculiarities of the multi-party scenario.

There is possibly another interesting research direction that would consider scenario assumptions where the referee or the protocol designer has some prior information about the probabilities of the different strings sent by the users. An important question here would be to determine, for all input sizes, if the na\"ive protocol in subsection 5.2 is able to beat our multi-party referee strategy involving the observation of just one detector.

Finally, we bring attention to the fact that our optimal multiport design in 3.3, or some modified version of it, might be useful to improve other prominent quantum protocols, such as the multipartite QKD (quantum key distribution) protocol in \cite{Grass}. The current version of this particular QKD protocol uses the generalized beamsplitter designs reviewed in subsection 3.1 of this paper.

\ack
The author is grateful for very helpful discussions with Prof. Marcos Curty and Prof. Javier Fraile from the University of Vigo. He is particularly indebted to Prof. Hoi-Kwong Lo for invaluable discussions and feedback, and for hospitality and support during his stay at the University of Toronto, where research for the present work was carried out and part of this manuscript was finished. The author thankfully acknowledges administrative support from Prof. Oscar Rubi\~nos (University of Vigo), and financial support of Postdoctoral Fellowship Grant ED481B 2017/038 from Xunta de Galicia (Ministry of Education and University Planning of a Regional Government in Spain). Additionally, this work was partially supported by the European Regional Development Fund (ERDF), and by the Galician Regional Government under project GRC2019/025 and under agreement for funding AtlantTIC (Atlantic Research Center for Information and Communication Technologies).

\appendix

\section{List of symbols}
\label{apA}
The table below contains detailed descriptions of the foremost notation symbols used in the present manuscript (table~\ref{tabA}).

\begin{tabularx}{\linewidth}{@{}lX@{}}
\caption{List of symbols used in the manuscript.}\\
\toprule
Symbol & Description\\[2pt]
\midrule
\vspace{-6pt}~&~\\
\endhead

$N$ & Raw message length:~~ number of bits in the raw binary strings that each user receives (or already has in his/her custody) in order to fingerprint, per user.\\[7pt]

$M$ & Transmitted sequence length:~~ number of optical pulses sent by each user in a coherent-state quantum protocol in the sequence $\bigotimes_{m=1}^M{{\left| \pm \frac{\alpha }{\sqrt{M}} \right\rangle }_{m}}$; binary length of the fingerprints in a classical protocol.\\[7pt]

$m$&Integer label $1\le m\le M$ assigned to each coherent state or to each mode in the transmitted sequence of coherent states or modes.\\[7pt]

$c=\frac{M}{N}>1$ & Rate of the ECC (error correcting code) used for amplifying differences between different raw message binary strings in a quantum protocol.\\[7pt]

$\delta$& ECC (error correcting code) parameter: maximum fraction of bits in which two codewords have the same bit values. The minimum distance of the ECC is $(1-\delta)M$.\\[7pt]

$Q$&Amount of transmitted information measured in \textsf{qubits/user} that is required in a quantum fingerprinting protocol.\\[7pt]

$C$&Amount of transmitted information measured in \textsf{bits/user} that is required in a classical fingerprinting protocol. \\[7pt]

$E_Q$&Transmitted energy that is required for each user in a quantum fingerprinting protocol.\\[7pt]

$E_C$&Transmitted energy that is required for each user in a classical fingerprinting protocol implemented with photonic bits.\\[7pt]

$K$&Total number of users.\\[7pt]

$k$&Integer label $1\le k\le K$ assigned to each user, and also label allocated to each referee input port and output port.\\[7pt]

${{\left| \alpha  \right|}^{2}}$&Mean photon number of \emph{all} the pulses in the transmitted sequence $\bigotimes_{m=1}^M{{\left| \pm \frac{\alpha }{\sqrt{M}} \right\rangle }_{m}}$ that each user sends towards the referee, in a quantum fingerprinting protocol.\\[7pt]

${{\mu }_{\text{in}}}=\frac{{{\left| \alpha  \right|}^{2}}}{M}$&Mean photon number of each individual input pulse (coherent state $\left| \frac{\pm \alpha }{\sqrt{M}} \right\rangle$, without including any losses in this notation) at the referee.\\[7pt]

$t$&Power transmittance of each unbalanced beamsplitter (BS) in the referee optical circuit. The corresponding power reflectance of the BS is $r=1-t$.\\[7pt]

$\tau$&Amplitude transmittance of a symmetric 50:50 beamsplitter (BS). Ideally, $\tau=2^{-1/2}$.\\[7pt]

$\sigma_{\rm T}$&Fabrication noise level affecting transmittance $\tau$ of each of the 50:50 beamsplitters (BS). These 50:50 BS are used for implementing all the unbalanced beamsplitters present in the referee circuit. $\sigma_{\rm T}$ is modelled as the standard deviation of a zero-mean normal random variable affecting $\tau$ as $\tau=2^{-1/2}(1+\sigma_{\rm T}\cdot\textsf{randn})$.\\[7pt]

$\sigma_{\rm P}$&Fabrication noise level affecting phase shifters inside the referee circuit. If $\phi_{\rm ideal}$ represents an ideal phase, then $\sigma_{\rm P}$ models a phase deviation as $\phi_{\rm realistic}=\phi_{\rm ideal}+\sigma_{\rm P}\cdot\textsf{randn}$.\\[7pt]

$\sigma$&General fabrication noise level of the referee circuit. Present-day technology allows achieving minimum values $\sigma_{\rm T}=\sigma_{\rm P}=0.01$. Accordingly, for simplicity, we always consider that both tolerances have the same value, and we simply define $\sigma=\sigma_{\rm T}=\sigma_{\rm P}$.\\[7pt]

$\mu _{k}^{\text{E}}$&Mean photon number at referee circuit output port $k$, with $1\le k\le K$, when all the individual input states (one state from each user) have the same phase.\\[7pt]

$\mu _{k,\bar{P}}^{\text{D}}$&Mean photon number at referee circuit output $k$, with $1\le k\le K$, when $L$ ($1\leq L \leq \frac{K}{2}$) input phases are different from the rest $K-L$ phases. It depends on $\bar{P}$, which is a vector with phase labels. For example, in a protocol with $K=4$ users:  $\left| \frac{\alpha }{\sqrt{M}} \right\rangle ,\left| \frac{-\alpha }{\sqrt{M}} \right\rangle ,\left| \frac{-\alpha }{\sqrt{M}} \right\rangle ,\left| \frac{\alpha }{\sqrt{M}} \right\rangle \text{ }\Rightarrow \text{ }\bar{P}=(1,-1,-1,1),~L=2$.
\\[7pt]

$\bar{P}$&See previous definition.\\[7pt]

$L$&Integer number $1\leq L \leq \frac{K}{2}$ that indicates the number of phase labels in $\bar{P}$ that are different from the rest $K-L$ labels. Restriction $L \leq \frac{K}{2}$ is introduced because, as an example, individual input states at the referee  $\{\left| \frac{\alpha }{\sqrt{M}} \right\rangle ,\left| \frac{-\alpha }{\sqrt{M}} \right\rangle ,\left| \frac{-\alpha }{\sqrt{M}} \right\rangle ,\left| \frac{-\alpha }{\sqrt{M}} \right\rangle\}$ and $\{\left| \frac{-\alpha }{\sqrt{M}} \right\rangle ,\left| \frac{\alpha }{\sqrt{M}} \right\rangle ,\left| \frac{\alpha }{\sqrt{M}} \right\rangle ,\left| \frac{\alpha }{\sqrt{M}} \right\rangle \}$ both produce identical photon statistics at the referee output ports.\\[7pt]

\makecell[tl]{$p_{\text{click},k}^{\text{E}}$,\\$p_{\text{click},k,\bar{P}}^{\text{D}}$}&Click probabilities at output detector $k$ ($1\le k\le K$), when all the individual input states (one state from each user) at the referee are equal (E) to each other, or when some of them are different (D). The maximum theoretical value that $p_{\text{click}}^{\text{E,D}}$ can take $\forall k$ is $p_{\text{click}}^{\text{E,D}}=1-\exp \left( -\frac{K{{\left| \alpha  \right|}^{2}}}{M} \right)\simeq \frac{K{{\left| \alpha  \right|}^{2}}}{M}$. This approximation holds if $K{{\mu }_{\text{in}}}=\frac{K{{\left| \alpha  \right|}^{2}}}{M}\ll 1$. As a consequence, if $K{{\left| \alpha  \right|}^{2}}/M\ll 1$, then it is always true that $p_{\text{click},k}^{\text{E}}=1-\exp \left( -\mu _{k}^{\text{E}} \right)\simeq \mu _{k}^{\text{E}}$ and $p_{\text{click},k,\bar{P}}^{\text{D}}=1-\exp \left( -\mu _{k,\bar{P}}^{\text{D}} \right)\simeq \mu _{k,\bar{P}}^{\text{D}}$.\\[7pt]

$D^{\rm E}_k,~D^{\rm D}_k$&Number of clicks at output detector with label $k$ when all the $K$ sequences of users' transmitted states are equal (E) to each other, or at least one of them is different (D) from the rest. These numbers of clicks correspond to complete sequences of coherent states $\bigotimes_{m=1}^M{{\left| \pm \frac{\alpha }{\sqrt{M}} \right\rangle }_{m}}$, not to $K$ individual input states $\left| \frac{\pm \alpha }{\sqrt{M}} \right\rangle$.\\[7pt]

$D_k$&Number of clicks observed by the referee at detector $k$, without any knowledge about whether all the users' state sequences are equal to each other, or whether some of them are different.\\[7pt]

$r$&Referee threshold in a quantum protocol. In one strategy, the referee concludes that all the $K$ sequences of transmitted states are equal to each other if $\sum\nolimits_{k=1}^{K-1}{{{D}_{k}}}\le r$. In another strategy, the referee concludes that they are different if ${{D}_{k=K}}\le r$. The calculation of $r$ depends on the particular strategy. Without loss of generality, we have assumed here that label $k=K$ always corresponds to the only detector that loses photons, compared to the all-equal inputs, when at least one of the input states at the referee circuit multiport differs from the rest. \\[7pt]

$p_{\rm error}$&Target upper bound on the desired error probability in any fingerprinting protocol, either classical or quantum.\\[7pt]

$\mu_{\rm dark}$, $p_{\rm dark}$&Dark count rate of the photon detectors.\\[7pt]

\makecell[tl]{$g_{[1,K-1]}^{\text{E}}$,\\$g_{[1,K-1],\bar{P}}^{\text{D}}$,\\ $g_{K}^{\text{E}},g_{K,\bar{P}}^{\text{D}}$}&Gains at the referee optical circuit, defined as $g_{[1,K-1]}^{\text{E}}=\frac{1}{{{\mu }_{\text{in}}}}\sum\nolimits_{k=1}^{K-1}{\mu _{k}^{\text{E}}}$, $g_{[1,K-1],\bar{P}}^{\text{D}}=\frac{1}{{{\mu }_{\text{in}}}}\sum\nolimits_{k=1}^{K-1}{\mu _{k,\bar{P}}^{\text{D}}}$, $g_{K}^{\text{E}}=\frac{\mu _{K}^{\text{E}}}{{{\mu }_{\text{in}}}}$, $g_{K,\bar{P}}^{\text{D}}=\frac{\mu _{K,\bar{P}}^{\text{D}}}{{{\mu }_{\text{in}}}}$. All these gains $g$ are theoretically bounded as $0\le g\le K$. In a real experimental setup, these gains may be measured in the classical optical regime, before starting the quantum protocol.\\[7pt]

\makecell[tl]{${{v}_{[1,K-1]}}$,\\$v_K$}&\makecell[tl]{Visibilities (figure of merit for referee circuits) defined as\\ ${{v}_{[1,K-1]}}=\frac{1}{2}\left( 1+\frac{K\left[ g_{[1,K-1],{{{\bar{P}}}^{*}}}^{\text{D}}-g_{[1,K-1]}^{\text{E}} \right]}{4(K-1)} \right)$, ${{v}_{K}}=\frac{1}{2}\left( 1+\frac{K\left[ g_{K}^{\text{E}}-g_{K,{{{\bar{P}}}^{*}}}^{\text{D}} \right]}{4(K-1)} \right)$.}\\
$$&${{\bar{P}}^{*}}$ above is a worst-case vector $\bar{P}$ that minimizes visibility values. In practice, the minimization may be carried out by calculating (brute force, low computational cost) or measuring (if real experiment) for all $K$ vectors $\bar{P}$ that have $L=1$, and then taking the smallest visibility.\\[7pt]

$\eta$&Combined efficiency that includes losses of the quantum channel and detector efficiencies. It does not include beamsplitter (BS) losses, as these BS losses affect differently each path from any circuit input to any circuit output. The effect of beamsplitter losses is fully included (either by simulation or by measurement) in gains $g_{[1,K-1]}^{\text{E}}$, $g_{[1,K-1],\bar{P}}^{\text{D}}$, $g_{K}^{\text{E}}$, $g_{K,\bar{P}}^{\text{D}}$ and in visibilities ${{v}_{[1,K-1]}}$, $v_K$.\\[7pt]

$\eta_{\rm BS}$&Losses of each of the 50:50 beamsplitters that are used for implementing all the unbalanced beamsplitters in the referee circuit. The effect of these losses $\eta_{\rm BS}$ is fully included (either by simulation or by measurement) in all the gains and in the visibilities defined above. \\[7pt]

$X_{k}^{\text{E}}$ & Random variables with Bernoulli distribution for each output $k$, with $1 \leq k \leq K$. Variables $X_{k}^{\text{E}}$  model the click or no click at output detector $k$ when $K$ input states equal (E) to each other arrive at the referee.\\[7pt]

$X_{k}^{\text{D}}$ & Analogous to the previous definition, but pertaining to the situation when at least one input state is different (D) from the rest.\\

\bottomrule
\label{tabA}
\end{tabularx}

\section{Upper bounds on the total mean photon number per user}
\label{apB}
This appendix, mostly self-contained in nature, covers the detailed mathematical steps required to obtain closed-form analytical upper bounds on the total mean photon number ${{\left| \alpha  \right|}^{2}}$, per user, required for a successful implementation of a  multi-party quantum fingerprinting protocol. Two different upper bounds are derived in \ref{apB1} and \ref{apB2} that are applicable to separate referee strategies (decision rules).  We assume a realistic optical circuit at the referee involving imperfections of any kinds. In our circuit model, the combined effect of such general imperfections is fully included in certain gains that establish relationships between the mean photon number at any circuit input and the mean photon number of certain sets of outputs.

In order to accomplish the aforesaid goal of upper-bounding ${{\left| \alpha  \right|}^{2}}$, we employ a particular version of the Chernoff bounds \cite{Chernoff} as described in detail in \cite{Mitz,Goemans}. We first describe Chernoff bounds as applied to generic random variables. Afterwards, we define the concise random variables that are required in our physical model.

~\\
\noindent \textbf{Theorem B.1} (Chernoff Bounds) \emph{Let $X=\sum\nolimits_{i=1}^{n}{{{X}_{i}}}$ be a random variable obtained as the sum of $X_i,$~$1\le i\le n,$ independent Bernoulli random variables. Let $\mu ={\rm E}(X)$ be the mean, or expected value, of $X$. Then}
\begin{enumerate}[(i)]
\item Upper tail: $\Pr [X\ge (1+{{\lambda }_{\text{upper}}})\mu ]\le \exp \left( -\frac{\lambda _{\text{upper}}^{2}}{2+\lambda_{\text{upper}} }\mu  \right)\text{  }\forall {{\lambda }_{\text{upper}}}>0;$
\item Lower tail: $\Pr [X\le (1-{{\lambda }_{\text{lower}}})\mu ]\le \exp \left( -\frac{\lambda _{\text{lower}}^{2}}{2}\mu  \right)\text{  }\forall \text{ }0<{{\lambda }_{\text{lower}}}<1.$
\end{enumerate}
~\\

In our particular application, we shall need to use an identical threshold value $r$ for the two tails stated in Theorem B.1 above, in order to calculate both the upper tail as $\Pr [X \ge r]$ and the lower tail as $\Pr [X \le r]$. Furthermore, we shall apply each tail to a different random variable; hence we write below $X_{\text{upper}}$ and $X_{\text{lower}}$, using the subscript labels to emphasize the fact that both still-generic random variables $X$ are different. To sum up, we may rewrite Theorem B.1 in a more convenient and clear way for our specific purposes, as follows:
\begin{equation}
\left. \begin{matrix}
  r=(1+{{\lambda }_{\text{upper}}}){{\mu }_{\text{upper}}} \\
  r=(1-{{\lambda }_{\text{lower}}}){{\mu }_{\text{lower}}} \\
\end{matrix} \right\}\Rightarrow \left\{ \begin{aligned}
  & \Pr [X_{\text{upper}}\ge r]\leq\exp \left( -\frac{{{(r-{{\mu }_{\text{upper}}})}^{2}}}{r+{{\mu }_{\text{upper}}}} \right)\text{  }\forall r>{{\mu }_{\text{upper}}}; \\
 & \Pr [X_{\text{lower}}\le r]\leq\exp \left( -\frac{{{(r-{{\mu }_{\text{lower}}})}^{2}}}{2{{\mu }_{\text{lower}}}} \right)\text{  }\forall \text{ }0<r<{{\mu }_{\text{lower}}}. \\
\end{aligned} \right.
\label{eqB1}
\end{equation}

Hereinafter, we describe in brief the circuit at the referee node and define the physical random variables that are required to judiciously apply the above-explained Chernoff bounds. The referee's circuit comprises $K$ optical input ports and $K$ optical output ports, with $K$ being also the total number of parties involved in the protocol. An integer label $k$ with $1 \leq k \leq K$ is assigned to each input and output port. Without any loss of generality, we assume in this appendix that the last label $k=K$ always corresponds to the only output that loses photons when \emph{not} all $K$ input states $\left| \pm \sqrt{{{\mu }_{\text{in}}}} \right\rangle $ have the same phase (see section \ref{sec3} for a detailed description of the referee's circuit). Now, let $X_{k}^{\text{E}}$, with $1 \leq k \leq K$, be a random variable with Bernoulli distribution for the number of clicks (0 or 1 click) at output detector $k$ when $K$ coherent states with identical phases arrive at the referee at the same time from $K$ users. Similarly, $X_{k}^{\text{D}}$ is an analogous random variable for the case when some of the $K$ input states are different, i.e. have phases that differ from the rest. The $K$ different random variables $X_{k}^{\text{E}}$ are independent from each other, for any fixed coherent states that are inputted to the referee at the same arrival time, because the average photon number at each output is also fixed. The same argument is also applicable to the other set of variables $X_{k}^{\text{D}}$.

Random variables $X_{k}^{\text{D}}$ depend on a vector $\bar{P}$ that contains the phases of $K$ simultaneous input pulses $\left| \pm \sqrt{{{\mu }_{\text{in}}}} \right\rangle$, as detailed in \ref{apA}. This dependency is not explicitly included in the notation of $X_{k}^{\text{D}}$ just for the sake of simplicity. We additionally introduce an integer $L$ to specify the number of phases in $\bar{P}$ that are different from the rest $K-L$ phases. Throughout the mathematical development in this appendix, we do not anticipate an analytical worst-case value for $L$. However, numerical evaluation for determining the worst-case $L$ on the grounds of analytical visibility models is carried out in subsection \ref{sec42}. The results there clearly show that, considering present-day technology parameters in any realistic referee circuit design, the worst-case scenario consistently corresponds to $L=1$. As a consequence, we may assert that it suffices to take into account in our upper-bound analysis the $K$ instances of vector $\bar{P}$ that contain just one phase difference.

Let us remark the fact that variables $X_{k}^{\text{D}}$ are used for modelling the effect of differences in $K$ \emph{individual} pulses arriving at the same time at the referee, that is, $\bigotimes_{k=1}^K{ \left| \pm \sqrt{\mu_{\rm in}} \right\rangle }_{k}$. As this set $X_{k}^{\text{D}}$ is not enough for our purposes, an additional ensemble of random variables $\tilde{X}_{k,m}^{\text{D}}$ needs to be introduced. These latter variables are for modelling the effect of differences in the $K$ \emph{complete} sequences of $M$ pulses $\bigotimes_{m=1}^M{ \left| \pm \sqrt{\mu_{\rm in}} \right\rangle }_{m}$ sent by the users, and not just in $K$ \emph{individual} simultaneous input pulses entering the circuit:
\begin{equation}
\tilde{X}_{k,m}^{\text{D}}=\left\{ \begin{aligned}
  & X_{k}^{\text{D}}\text{ for any }(1-\delta )\cdot M\text{ indices }m, \\
 & X_{k}^{\text{E}}\text{ for any }\delta \cdot M\text{ indices }m. \\
\end{aligned} \right.
\label{eqB4}
\end{equation}
Parameter $\delta$ in the definitions above represents the ``distance parameter'' of the error correcting code (ECC). The ECC is used in the quantum protocol for amplifying differences in the transmitted coherent-pulse sequences. $\delta$ is the maximum fraction of bits in which two ECC codewords have the same bit values. The minimum distance of the ECC can be simply expressed as $(1-\delta)M$. Equation (\ref{eqB4}) corresponds to any worst-case scenarios in which the number of instances of $K$ simultaneous dissimilar states arriving at the referee is the same as the minimum ECC distance. In other words, this worst-case different-input scenario intuitively corresponds to the case where the differing sequences are the most similar to all equal sequences. Thus, this described situation is the most difficult to distinguish by the referee.

Before fully entering into mathematical elaboration, we present in detail the two separate referee strategies that we consider in the analytical developments in \ref{apB1} and  \ref{apB2}. In order to simplify the explanation of such strategies,  it is convenient to first define some final notation.  We denote as $D_{k}^{\text{E}}=\sum\nolimits_{m=1}^{M}{X_{k}^{\text{E}}}$ the total number of clicks at any output detector $k$ when the $K$ complete sequences of $M$ coherent states are equal to each other. In the same way, $D_{k}^{\text{D}}=\sum\nolimits_{m=1}^{M}{\tilde{X}_{k,m}^{\text{D}}}$ represents the amount of clicks at detector $k$ when at least one of the input sequences differs from the rest. We simply denote as $D_k$ the total number of clicks in a real scenario where the referee has no previous knowledge of whether the input sequences are different or are the same. The referee utilizes a certain threshold value $r$ that she compares to certain values of $D_{k}$ in order to conclude if the input sequences are different or not. In the remainder part of the appendix, we provide analytical methods for computing both $r$ and the sought upper bounds for ${{\left| \alpha  \right|}^{2}}$. In particular, according to the general operation of the circuit described in section \ref{sec3}, the referee may implement two different decision rules depending on the detectors that she observes, as we summarize next:

\vspace{10pt}
~\\
\noindent\textsf{Strategy observing $K-1$ detectors with labels $1\le k\le K-1$:}
\begin{itemize}
\item [$\bullet$] \textsf{Referee infers equal input sequences if  $\sum\nolimits_{k=1}^{K-1}{{{D}_{k}}}\le r$}.
\item [$\bullet$] \textsf{Referee infers different input sequences if  $\sum\nolimits_{k=1}^{K-1}{{{D}_{k}}}>r$}.
\begin{itemize}
\item Error happens when the input sequences are different and the referee announces ``equal,'' if $\sum\nolimits_{k=1}^{K-1}{D_{k}^{\text{D}}}=\sum\nolimits_{k=1}^{K-1}{\sum\nolimits_{m=1}^{M}{\tilde{X}_{k,m}^{\text{D}}}}\le r$.
\item Error happens when the input sequences are equal and the referee announces ``different,'' if $\sum\nolimits_{k=1}^{K-1}{D_{k}^{\text{E}}}=\sum\nolimits_{k=1}^{K-1}{\sum\nolimits_{m=1}^{M}{X_{k}^{\text{E}}}}>r$.
\end{itemize}
\end{itemize}
~\\
\noindent\textsf{Strategy observing $1$ detector with label $k=K$:}
\begin{itemize}
\item [$\bullet$] \textsf{Referee infers different input sequences if ${{D}_{K}}\le r$}.
\item [$\bullet$] \textsf{Referee infers equal input sequences if  ${{D}_{K}}>r$}.
\begin{itemize}
\item Error happens when the input sequences are equal and the referee announces ``different,'' if $D_{K}^{\text{E}}=\sum\nolimits_{m=1}^{M}{X_{K}^{\text{E}}}\le r$.
\item Error happens when the input sequences are different and the referee announces ``equal,'' if $D_{K}^{\text{D}}=\sum\nolimits_{m=1}^{M}{\tilde{X}_{K,m}^{\text{D}}}>r$.
\end{itemize}
\end{itemize}
\vspace{10pt}

\subsection{Referee strategy observing the detectors that gain photons in the case of different individual input states}
\label{apB1}

In this subsection, we assume that the referee counts clicks in those output detectors with labels $1\le k\le K-1$ and she does not observe detector $k=K$. We use notation $p_{\text{error}}^{\text{E}}$ for the probability of error that occurs when the input sequences are all the same but the referee wrongly announces they are different. Likewise, $p_{\text{error}}^{\text{D}}$ corresponds to an error that happens when at least one of the input sequences is different but the referee incorrectly concludes they are equal.

Applying the upper tail inequality in (\ref{eqB1}) to the first $K-1$ random variables $X_{k}^{\text{E}}$,  previously described in the introduction of the present appendix, we may upper bound error probability $p_{\text{error}}^{\text{E}}$ as follows, with notation $\mathrm{E}(\cdot)$ designating statistical  mean values:
\begin{equation}
p_{\text{error}}^{\text{E}}=\Pr \left[ \sum\limits_{k=1}^{K-1}{\sum\limits_{m=1}^{M}{X_{k}^{\text{E}}}}> r \right]\leq \exp \left( -\frac{{{\left( r-\sum\limits_{k=1}^{K-1}{\sum\limits_{m=1}^{M}{\text{E(}X_{k}^{\text{E}})}} \right)}^{2}}}{r+\sum\limits_{k=1}^{K-1}{\sum\limits_{m=1}^{M}{\text{E(}X_{k}^{\text{E}})}}} \right)\text{  valid if  }r>\sum\limits_{k=1}^{K-1}{\sum\limits_{m=1}^{M}{\text{E}(X_{k}^{\text{E}})}}.
\label{eqB2}
\end{equation}
Using now the first $K-1$ random variables $\tilde{X}_{k,m}^{\text{D}}$ in index $k$, which are described in (\ref{eqB4}), on the lower tail inequality in (\ref{eqB1}), the upper bound on error probability $p_{\text{error}}^{\text{D}}$ satisfies
\begin{equation}
p_{\text{error}}^{\text{D}}=\Pr \left[ \sum\limits_{k=1}^{K-1}{\sum\limits_{m=1}^{M}{\tilde{X}_{k,m}^{\text{D}}}}\le r \right]\le \exp \left( -\frac{{{\left( r-\sum\limits_{k=1}^{K-1}{\sum\limits_{m=1}^{M}{\text{E(}\tilde{X}_{k,m}^{\text{D}})}} \right)}^{2}}}{2\ \sum\limits_{k=1}^{K-1}{\sum\limits_{m=1}^{M}{\text{E(}\tilde{X}_{k,m}^{\text{D}})}}} \right)\text{  valid if   }0<r<\sum\limits_{k=1}^{K-1}{\sum\limits_{m=1}^{M}{\text{E(}\tilde{X}_{k,m}^{\text{D}})}}.
\label{eqB3}
\end{equation}

We must opt now for defining a particular referee threshold value $r$ as a function of the mean values of $X_{k}^{\text{E}}$ and $\tilde{X}_{k,m}^{\text{D}}$. The following definition was chosen because it provides a very easy comparison between the separate error upper bounds in (\ref{eqB2}) and (\ref{eqB3}). This comparison allows us to pick out the worst-case bound. Additionally, the chosen definition for $r$ provides a closed-form threshold expression as a function of relevant parameters of the quantum protocol, as it will become clear later:
\begin{equation}
r=\frac{1}{2}\sum\limits_{k=1}^{K-1}{\sum\limits_{m=1}^{M}{\left[ \text{E(}X_{k}^{\text{E}})+\text{E(}\tilde{X}_{k,m}^{\text{D}}) \right]}}.
\label{eqB5}
\end{equation}
We rewrite (\ref{eqB2}) and (\ref{eqB3}) using the definition for $r$ in (\ref{eqB5}):
\begin{equation}
p_{\text{error}}^{\text{E}}\le \exp \left( -\frac{{{\left( \sum\limits_{k=1}^{K-1}{\sum\limits_{m=1}^{M}{\text{E(}\tilde{X}_{k,m}^{\text{D}})}}-\sum\limits_{k=1}^{K-1}{\sum\limits_{m=1}^{M}{\text{E(}X_{k}^{\text{E}})}} \right)}^{2}}}{2\ \sum\limits_{k=1}^{K-1}{\sum\limits_{m=1}^{M}{\text{E(}\tilde{X}_{k,m}^{\text{D}})}+6\ \sum\limits_{k=1}^{K-1}{\sum\limits_{m=1}^{M}{\text{E(}X_{k}^{\text{E}})}}}} \right),
\label{eqB6}
\end{equation}
\begin{equation}
p_{\text{error}}^{\text{D}}\le \exp \left( -\frac{{{\left( \sum\limits_{k=1}^{K-1}{\sum\limits_{m=1}^{M}{\text{E(}\tilde{X}_{k,m}^{\text{D}})}}-\sum\limits_{k=1}^{K-1}{\sum\limits_{m=1}^{M}{\text{E(}X_{k}^{\text{E}})}} \right)}^{2}}}{8\ \sum\limits_{k=1}^{K-1}{\sum\limits_{m=1}^{M}{\text{E(}\tilde{X}_{k,m}^{\text{D}})}}} \right).
\label{eqB7}
\end{equation}
By combining validity conditions shown in (\ref{eqB2}) and (\ref{eqB3}), it is clear that our ongoing mathematical elaboration based on Chernoff bounds can only be used if   $\sum\nolimits_{k=1}^{K-1}{\sum\nolimits_{m=1}^{M}{\text{E}(X_{k}^{\text{E}})}}<\sum\nolimits_{k=1}^{K-1}{\sum\nolimits_{m=1}^{M}{\text{E(}\tilde{X}_{k,m}^{\text{D}})}}$. The meaning of this inequality in the physical world establishes that $K-1$ circuit outputs must gain photons when we switch from equal input states to different input states. This is the desired behaviour of the circuit under normal realistic circumstances. Moreover, the aforesaid inequality enables an easy comparison between the denominators inside the exponentials in (\ref{eqB6}) and (\ref{eqB7}). The worst-case bound clearly corresponds always to  $p_{\text{error}}^{\text{D}}$ in (\ref{eqB7}), as it provides the greatest upper bound for the error probability. Consequently, the remainder of this \ref{apB1} is aimed at obtaining an upper bound for ${{\left| \alpha  \right|}^{2}}$ based on (\ref{eqB7}), and we dismiss (\ref{eqB6}).

Henceforth, we assume that condition $K{{\mu }_{\text{in}}}=\frac{K{{\left| \alpha  \right|}^{2}}}{M}\ll 1$ holds, where $\mu_{\rm in}$ is the photon number of each individual input pulse. This approximation is always correct if $K\ll M$, which corresponds to the cases of interest addressed in this manuscript, and was checked to be valid for all the realistic scenarios analyzed in Section \ref{sec5}.  Under the considered assumption, we can approximate click probabilities at output detectors as  $p_{\text{click},k}^{\text{E}}\simeq \mu _{k}^{\text{E}}$, $p_{\text{click},k,\bar{P}}^{\text{D}}\simeq \mu _{k,\bar{P}}^{\text{D}}$. A subscript $\bar{P}$ is used to emphasize the fact that the photon number at each output $k$ depends on vector $\bar{P}$ that contains the information of the $K$ input pulse phases. For simplicity, detector efficiencies and dark count rates are not yet specified in the definitions of the click probabilities; below, we introduce a combined efficiency quantity that includes detector efficiencies as well as channel losses.

By defining now the gains of the first $K-1$ circuit outputs as
\begin{equation}
g_{[1,K-1]}^{\text{E}}=\frac{\sum\limits_{k=1}^{K-1}{\mu _{k}^{\text{E}}}}{{{\mu }_{\text{in}}}},\text{~~~~~}g_{[1,K-1],\bar{P}}^{\text{D}}=\frac{\sum\limits_{k=1}^{K-1}{\mu _{k,\bar{P}}^{\text{D}}}}{{{\mu }_{\text{in}}}},
\label{eqB8}
\end{equation}
we can easily express the expected quantities concerning amounts of clicks in (\ref{eqB7}) as
\begin{equation}
\sum\limits_{k=1}^{K-1}{\sum\limits_{m=1}^{M}{\text{E(}X_{k}^{\text{E}})}}\simeq M\cdot \sum\limits_{k=1}^{K-1}{\mu _{k}^{\text{E}}}+(K-1)M\mu _{\text{dark}}^{{}}=\,g_{[1,K-1]}^{\text{E}}{{\left| \alpha  \right|}^{2}}+(K-1)M\mu _{\text{dark}}^{{}},
\label{eqB9}
\end{equation}
\begin{equation}
\begin{gathered}
  \sum\limits_{k=1}^{K-1}{\sum\limits_{m=1}^{M}{\text{E(}\tilde{X}_{k,m}^{\text{D}})}}\simeq \left[ \delta \cdot \sum\limits_{k=1}^{K-1}{\mu _{k}^{\text{E}}}+(1-\delta )\cdot \sum\limits_{k=1}^{K-1}{\mu _{k,\bar{P}}^{\text{D}}} \right]M+(K-1)M\mu _{\text{dark}}^{{}}= \\
  =\delta \cdot g_{[1,K-1]}^{\text{E}}{{\left| \alpha  \right|}^{2}}+(1-\delta )\cdot g_{[1,K-1],\bar{P}}^{\text{D}}{{\left| \alpha  \right|}^{2}}+(K-1)M\mu _{\text{dark}}^{{}}. \\
\end{gathered}
\label{eqB10}
\end{equation}

By using again the validity conditions in (\ref{eqB2}) and (\ref{eqB3}), this time on the two preceding equations, we can obtain an interesting condition that both gains must satisfy:
\begin{equation}
\sum\limits_{k=1}^{K-1}{\sum\limits_{m=1}^{M}{\text{E}(X_{k}^{\text{E}})}}<\sum\limits_{k=1}^{K-1}{\sum\limits_{m=1}^{M}{\text{E(}\tilde{X}_{k,m}^{\text{D}})}}\text{~~~}\Rightarrow \text{~~~}g_{[1,K-1]}^{\text{E}}<g_{[1,K-1],\bar{P}}^{\text{D}}.
\label{eqB11}
\end{equation}

We express now the probability bound in (\ref{eqB7}) as a function of the gains in (\ref{eqB8}) and  of other protocol parameters, by replacing with the expected values in (\ref{eqB9}) and (\ref{eqB10}):
\begin{equation}
\begin{gathered}
  p_{\text{error}}^{{}}\le \exp \left( -\frac{{{\left( \sum\limits_{k=1}^{K-1}{\sum\limits_{m=1}^{M}{\text{E(}\tilde{X}_{k,m}^{\text{D}})}-\sum\limits_{k=1}^{K-1}{\sum\limits_{m=1}^{M}{\text{E(}X_{k}^{\text{E}})}}} \right)}^{2}}}{8\ \sum\limits_{k=1}^{K-1}{\sum\limits_{m=1}^{M}{\text{E(}\tilde{X}_{k,m}^{\text{D}})}}} \right)= \\
  =\exp \left( -\frac{1}{8}\cdot \frac{{{(1-\delta )}^{2}}{{(g_{[1,K-1],\bar{P}}^{\text{D}}-g_{[1,K-1]}^{\text{E}})}^{2}}{{\left| \alpha  \right|}^{4}}}{\left[ \delta \cdot g_{[1,K-1]}^{\text{E}}+(1-\delta )\cdot g_{[1,K-1],\bar{P}}^{\text{D}} \right]{{\left| \alpha  \right|}^{2}}+(K-1)M\mu _{\text{dark}}^{{}}} \right). \\
\end{gathered}
\label{eqB12}
\end{equation}
Gain $g_{[1,K-1]}^{\text{E}}$ in (\ref{eqB12}) is theoretically a constant magnitude, since we assume that all the input pulses have the same or approximately the same amplitude. Conversely, gain $g_{[1,K-1],\bar{P}}^{\text{D}}$ depends on the particular phases of the $K$ input states. In order to infer a worst-case value for gain $g_{[1,K-1],\bar{P}}^{\text{D}}$, we next perform an optimization assuming that $g_{[1,K-1],\bar{P}}^{\text{D}}$ is a continuous variable denoted as $g^{\rm D}$. This is just a ``mathematical license'' taken to analyze the behaviour of the varying gain. A function $f({{g}^{\text{D}}})$ is introduced in the phase argument of (\ref{eqB12}) as $p_{\text{error}}^{{}}\le \exp \left( -\frac{1}{8}f({{g}^{\text{D}}}) \right)$:
\begin{equation}
f(g_{{}}^{\text{D}})=\frac{{{(1-\delta )}^{2}}{{(g_{{}}^{\text{D}}-g_{{}}^{\text{E}})}^{2}}{{\left| \alpha  \right|}^{4}}}{\left[ \delta \cdot g_{{}}^{\text{E}}+(1-\delta )\cdot g_{{}}^{\text{D}} \right]{{\left| \alpha  \right|}^{2}}+(K-1)M\mu _{\text{dark}}^{{}}}.
\label{eqB13}
\end{equation}
By equating the derivative to zero, $\frac{\partial f(g_{{}}^{\text{D}})}{\partial g_{{}}^{\text{D}}}=0$, it is easy to find two critical points. One of these points is $g_{1}^{\text{D}}=g_{{}}^{\text{E}}$ and the other critical point $g_{2}^{\text{D}}$ verifies
\begin{equation}
g_{2}^{\text{D}}-g_{{}}^{\text{E}}=\frac{-2\left[ g_{{}}^{\text{E}}{{\left| \alpha  \right|}^{2}}+(K-1)M\mu _{\text{dark}}^{{}} \right]}{(1-\delta ){{\left| \alpha  \right|}^{2}}}<0.
\label{eqB14}
\end{equation}
This inequality in (\ref{eqB14}) clearly poses a contradiction on condition (\ref{eqB11}) and, as a result, only critical point $g_{1}^{\text{D}}=g_{{}}^{\text{E}}$ stands in our analysis. Moreover, it is easy to prove that function $f({{g}^{\text{D}}})$ decreases as $g^{\rm D}$ shrinks closer to the other gain $g^{\rm E}$. As a consequence of this analysis, the worst-case value of $g_{[1,K-1],\bar{P}}^{\text{D}}$ that minimizes function (\ref{eqB13}) and maximizes error probability bound (\ref{eqB12}) corresponds to the minimum value of $g_{[1,K-1],\bar{P}}^{\text{D}}$. Finally, solving for ${{\left| \alpha \right|}^{2}}$ in (\ref{eqB12}) and including the combined efficiency $\eta$, we get
\begin{equation}
{{\left| \alpha _{[1,K-1]}^{\text{bound}} \right|}^{2}}=\frac{4q+2{{\left[ 4{{q}^{2}}+2{{(1-\delta )}^{2}}{{\left( \min (g_{[1,K-1],\bar{P}}^{\text{D}})-g_{[1,K-1]}^{\text{E}} \right)}^{2}}(K-1)M\mu _{\text{dark}}^{{}}\cdot \ln (1/p_{\text{error}}^{{}}) \right]}^{1/2}}}{\eta \ {{(1-\delta )}^{2}}{{\left( \min (g_{[1,K-1],\bar{P}}^{\text{D}})-g_{[1,K-1]}^{\text{E}} \right)}^{2}}},
\label{eqB15}
\end{equation}
with
\begin{equation}
q=\left[ \delta \cdot g_{[1,K-1]}^{\text{E}}+(1-\delta )\cdot \min (g_{[1,K-1],\bar{P}}^{\text{D}}) \right]\cdot \ln (1/p_{\text{error}}^{{}}).
\label{eqB16}
\end{equation}
Combined efficiency $\eta$ takes into account detector efficiencies and channel losses. Detector efficiencies are assumed to be the same for all detectors. In practice, this may be a rather good realistic approximation; nevertheless, different quantum efficiencies may also be easily considered just by transferring their effects from $\eta$ to the gains $g_{[1,K-1]}^{\text{E}}$ and $g_{[1,K-1],\bar{P}}^{\text{D}}$.

A closed-form expression for the referee threshold is obtained by taking (\ref{eqB9}) and (\ref{eqB10}) into (\ref{eqB5}) and by including the minimum value of $g_{[1,K-1],\bar{P}}^{\text{D}}$ (the value that maximizes error probability, as proven above):
\begin{equation}
r=\frac{1}{2}{{\left| \alpha  \right|}^{2}}\left[ (1+\delta )\cdot g_{[1,K-1]}^{\text{E}}+(1-\delta )\cdot \min(g_{[1,K-1],\bar{P}}^{\text{D}}) \right]+(K-1)M\mu _{\text{dark}}^{{}}.
\label{eqB17}
\end{equation}

We remark again the fact that quantum fingerprinting with this particular strategy of observing $K-1$ detectors is only possible if $\min (g_{[1,K-1],\bar{P}}^{\text{D}})>g_{[1,K-1]}^{\text{E}}$.

\subsection{Referee strategy observing the detector that loses photons in the case of different individual input states}
\label{apB2}
In this second subsection, we address the referee decision rule consisting of counting clicks in just the last detector. By convention, this last detector has a label $k=K$ assigned. The notation employed throughout the present mathematical elaboration is identical to that in \ref{apB1}.

We apply the upper tail case in (\ref{eqB1}) to random variable $\tilde{X}_{K,m}^{\text{D}}$ (note the subscript $k=K$ corresponding to the last variable in the ensemble $\tilde{X}_{k,m}^{\text{D}}$, $1 \leq k \leq K$, defined in (\ref{eqB4})) in order to upper bound error probability $p_{\text{error}}^{\text{D}}$, corresponding to the case of at least one input sequence differing from the rest:
\begin{equation}
p_{\text{error}}^{\text{D}}=\Pr \left[ \sum\limits_{m=1}^{M}{\tilde{X}_{K,m}^{\text{D}}}> r \right]\le \exp \left( -\frac{{{\left( r-\sum\limits_{m=1}^{M}{\text{E(}\tilde{X}_{K,m}^{\text{D}})} \right)}^{2}}}{r+\sum\limits_{m=1}^{M}{\text{E(}\tilde{X}_{K,m}^{\text{D}})}} \right)\text{  valid if  }r>\sum\limits_{m=1}^{M}{\text{E(}\tilde{X}_{K,m}^{\text{D}})}.
\label{eqB18}
\end{equation}
Using now random variable $X_{K}^{\text{E}}$ on the lower tail inequality of Chernoff bounds in (\ref{eqB1}), we get an upper bound for error probability $p_{\text{error}}^{\text{E}}$ that corresponds to the case of all equal $K$ input sequences of coherent states:
\begin{equation}
p_{\text{error}}^{\text{E}}=\Pr \left[ \sum\limits_{m=1}^{M}{X_{K}^{\text{E}}}\le r \right]\le \exp \left( -\frac{{{\left( r-\sum\limits_{m=1}^{M}{\text{E(}X_{K}^{\text{E}})} \right)}^{2}}}{2\ \sum\limits_{m=1}^{M}{\text{E(}X_{K}^{\text{E}})}} \right)\text{  valid if   }0<r<\sum\limits_{m=1}^{M}{\text{E(}X_{K}^{\text{E}})}.
\label{eqB19}
\end{equation}

Using the same definition for the referee threshold $r$ in (\ref{eqB5}), we may rewrite the two preceding inequalities (\ref{eqB18}) and (\ref{eqB19}) in a more convenient way for our purposes:
\begin{equation}
p_{\text{error}}^{\text{D}}\le \exp \left( -\frac{{{\left( \sum\limits_{m=1}^{M}{\left[ \text{E(}X_{K}^{\text{E}})-\text{E(}\tilde{X}_{K,m}^{\text{D}}) \right]} \right)}^{2}}}{\ \sum\limits_{m=1}^{M}{\left[ \text{2}\cdot \text{E(}X_{K}^{\text{E}})+6\cdot \text{E(}\tilde{X}_{K,m}^{\text{D}}) \right]}} \right),
\label{eqB20}
\end{equation}
\begin{equation}
p_{\text{error}}^{\text{E}}\le \exp \left( -\frac{{{\left( \sum\limits_{m=1}^{M}{\left[ \text{E(}X_{K}^{\text{E}})-\text{E(}\tilde{X}_{K,m}^{\text{D}}) \right]} \right)}^{2}}}{8\ \sum\limits_{m=1}^{M}{\text{E(}X_{K}^{\text{E}})}} \right).
\label{eqB21}
\end{equation}

Combining both correctness conditions of Chernoff bounds in (\ref{eqB18}) and (\ref{eqB19}), we know that $\sum\nolimits_{m=1}^{M}{\text{E(}\tilde{X}_{K,m}^{\text{D}})}<\sum\nolimits_{m=1}^{M}{\text{E(}X_{K}^{\text{E}})}$. This inequality has a clear meaning in the physical world: the last circuit output must lose photons when switching from equal input states to different input states. This is the desired circuit behaviour under normal realistic circumstances. Additionally, the aforesaid inequality allows an easy comparison between the denominators inside the exponential functions in (\ref{eqB20}) and (\ref{eqB21}). Clearly, inequality (\ref{eqB21}) imposes on the error probability an upper bound that is always greater than (\ref{eqB20}). As a consequence, for the rest of the present mathematical development, we shall focus on (\ref{eqB21}) only.

Following identical arguments as for deducing equations (\ref{eqB9}) and (\ref{eqB10}) for the other referee strategy, we can calculate now the expected amounts of clicks that appear in (\ref{eqB21}):
\begin{equation}
\begin{gathered}
  \sum\limits_{m=1}^{M}{\text{E(}\tilde{X}_{K,m}^{\text{D}})}\simeq M\cdot \left[ \delta \cdot \mu _{K}^{\text{E}}+(1-\delta )\cdot \mu _{K,\bar{P}}^{\text{D}} \right]+M\mu _{\text{dark}}^{{}}= \\
  =\delta \cdot g_{K}^{\text{E}}{{\left| \alpha  \right|}^{2}}+(1-\delta )\cdot g_{K,\bar{P}}^{\text{D}}{{\left| \alpha  \right|}^{2}}+M\mu _{\text{dark}}^{{}}, \\
\end{gathered}
\label{eqB23}
\end{equation}
\begin{equation}
\sum\limits_{m=1}^{M}{\text{E(}X_{K}^{\text{E}})}\simeq M\cdot \mu _{K}^{\text{E}}+M\mu _{\text{dark}}^{{}}=\,g_{K}^{\text{E}}{{\left| \alpha  \right|}^{2}}+M\mu _{\text{dark}}^{{}},
\label{eqB24}
\end{equation}
where now, for the particular referee strategy considered in this elaboration, gains are defined as
\begin{equation}
g_{K}^{\text{E}}=\frac{\mu _{K}^{\text{E}}}{{{\mu }_{\text{in}}}},\text{~~~~~~~}g_{K,\bar{P}}^{\text{D}}=\frac{\mu _{K,\bar{P}}^{\text{D}}}{{{\mu }_{\text{in}}}}.
\label{eqB22}
\end{equation}

By using again the conditions in (\ref{eqB18}) and (\ref{eqB19}) on (\ref{eqB23}) and (\ref{eqB24}), we obtain the following condition pertaining to the circuit gains:
\begin{equation}
\sum\limits_{m=1}^{M}{\text{E(}\tilde{X}_{K,m}^{\text{D}})}<\sum\limits_{m=1}^{M}{\text{E(}X_{K}^{\text{E}})}\text{  }\Rightarrow \text{  }g_{K,\bar{P}}^{\text{D}}<g_{K}^{\text{E}}.
\label{eqB25}
\end{equation}

We may  finally express the probability bound in (\ref{eqB21}) as a function of the gains and other relevant protocol parameters as
\begin{equation}
p_{\text{error}}^{{}}\le \exp \left( -\frac{{{\left( \sum\limits_{m=1}^{M}{\left[ \text{E(}X_{K}^{\text{E}})-\text{E(}\tilde{X}_{K,m}^{\text{D}}) \right]} \right)}^{2}}}{8\ \sum\limits_{m=1}^{M}{\text{E(}X_{K}^{\text{E}})}} \right)=\exp \left( -\frac{1}{8}\cdot \frac{{{(1-\delta )}^{2}}{{(g_{K}^{\text{E}}-g_{K,\bar{P}}^{\text{D}})}^{2}}{{\left| \alpha  \right|}^{4}}}{g_{K}^{\text{E}}{{\left| \alpha  \right|}^{2}}+M\mu _{\text{dark}}^{{}}} \right).
\label{eqB26}
\end{equation}

Including combined efficiency $\eta$ and realizing that the worst-case error upper bound occurs when vector $\bar{P}$ produces the maximum value of $g_{K,\bar{P}}^{\text{D}}$ (keep in mind that gain inequality in (\ref{eqB25}) must hold), we may solve (\ref{eqB26}) in order to upper bounding ${{\left| \alpha\right|}^{2}}$ as satisfying

\begin{equation}
{{\left| \alpha _{K}^{\text{bound}} \right|}^{2}}=\frac{4q+2{{\left[ 4{{q}^{2}}+2{{(1-\delta )}^{2}}{{\left( g_{K}^{\text{E}}-\max (g_{K,\bar{P}}^{\text{D}}) \right)}^{2}}M\mu _{\text{dark}}^{{}}\cdot \ln (1/p_{\text{error}}^{{}}) \right]}^{1/2}}}{\eta \ {{(1-\delta )}^{2}}{{\left( g_{K}^{\text{E}}-\max (g_{K,\bar{P}}^{\text{D}}) \right)}^{2}}},
\label{eqB27}
\end{equation}
with
\begin{equation}
q=g_{K}^{\text{E}}\cdot \ln (1/p_{\text{error}}^{{}}).
\label{eqB28}
\end{equation}
A closed-form equation for referee threshold $r$ may also be obtained from (\ref{eqB23}) and (\ref{eqB24}) by incorporating the maximum value of $g_{K,\bar{P}}^{\text{D}}$:
\begin{equation}
r=\frac{1}{2}{{\left| \alpha  \right|}^{2}}\left[ (1+\delta )\cdot g_{K}^{\text{E}}+(1-\delta )\cdot \max(g_{K,\bar{P}}^{\text{D}}) \right]+M\mu _{\text{dark}}^{{}}.
\label{eqB29}
\end{equation}

Let us finally emphasize again that quantum fingerprinting with this strategy of observing just one detector is only realizable if $g_{K}^{\text{E}}>\max (g_{K,\bar{P}}^{\text{D}})$.

\section{Transmitted information in the multi-party classical limit}
\label{apC}

An analytical lower bound on the amount of transmitted bits required in a classical $K$-user fingerprinting protocol is deduced in this appendix. We employ a simplified version of Claim 2.4 presented both in \cite{Babai} and in the supplementary material of \cite{Guan}. Though simplified, the 2-user claim upon which our $K$-user elaboration leverages is totally equivalent to those presented in \cite{Babai, Guan} for the 2-user scenario. Supplementary material of \cite{Guan} blends the proof of the claim with the final result itself, including additional notation that is not required to understand the final result. Meanwhile, the equivalent result in \cite{Babai} focuses on finding the scaling (information complexity cost) of fingerprinting, rather than finding a tight lower bound on the transmitted information.

~\\
\noindent \textbf{Claim C.1.} (From Claim 2.4 in \cite{Babai, Guan}) {Let $x$ ($y$) be a bit string owned by Alice (Bob) containing $N^{\rm A}$ ($N^{\rm B}$) bits. During the classical protocol, Alice (Bob) sends to the referee another bit string $F(x)$ ($F(y)$) containing $M^{\rm A}$ ($M^{\rm B}$) bits. The referee's task consists of computing \emph{any} boolean function $f(x,y)=\{0,\,1\}$ using not the original bit strings $x$ and $y$ but instead the two strings $F(x)$ and $F(y)$ comprising $M^{\rm A}$ and $M^{\rm B}$ bits, respectively, that she receives from Alice and Bob. This complete protocol involving Alice, Bob and the referee is assumed to be a so-called private-coin\footnote{``Private-coin'' means that both Alice and Bob are restricted to local unshared randomness only. They are allowed to share randomness neither with each other nor with the referee.} two-sided\footnote{``Two-sided error'' means that, when announcing the function outcome, the referee makes mistakes with probability $p_{\rm error}$ at most, and this probability is independent from the actual value of $f(x,y)$, which can be either 0 or 1. In contrast, a classical protocol in which the referee makes no mistakes for one of the two possible boolean values of $f(x,y)$  is called ``one-sided'' protocol.} error randomized protocol. We remark the fact that the herein presented classical protocol is \emph{not} only specific for classical fingerprinting, but it may also be used for \emph{any} boolean function $f(x,y)=\{0,\,1\}$ whatever it is.}
The claim states that, if the probability of error when the referee computes $f(x,y)$ is upper bounded by $p_{\rm error}$, then the following inequalities must simultaneously hold on the amounts of bits:
\begin{subequations}
\begin{gather}
  {{N}^{\rm A}}\le {{M}^{\rm A}}\left\lceil \frac{8\ln (2)\left( 1+{{M}^{\rm B}} \right)}{{{\left( 1-2\sqrt{{{p}_{\text{error}}}} \right)}^{2}}} \right\rceil, \label{eqC1a} \\[12pt]
  {{N}^{\rm B}}\le {{M}^{\rm B}}\left\lceil \frac{8\ln (2)\left( 1+{{M}^{\rm A}} \right)}{{{\left( 1-2\sqrt{{{p}_{\text{error}}}} \right)}^{2}}} \right\rceil.  \label{eqC1b}
\end{gather}
\label{eqC1}
\end{subequations}\\[0pt]

The approach we suggest for the multi-party scenario merely consists in thinking of a $K$-party classical protocol as a 2-party protocol, in which Alice plays the role of a certain amount of original users and Bob plays the role of the remaining users. As it will become clear following the complete development in this appendix, the specific approach that minimizes the amount of transmitted bits requires that Alice (or Bob) represents $\frac{K}{2}$ parties if $K$ is even or $\frac{K-1}{2}$ parties if $K$ is odd. Let us assume that the number of parties $K$ is odd. Under this initial assumption, each party represented by Alice sends $M_{k}^{\text{A}}$ bits, with $k=1,\ldots ,\frac{K-1}{2}$, so that Alice sends $\sum\nolimits_{k=1}^{\frac{K-1}{2}}{M_{k}^{\text{A}}}$ bits in total. Similarly, each of Bob's parties sends $M_{k}^\text{B}$ bits, with $k=1,\ldots ,\frac{K+1}{2}$, for a total of $\sum\nolimits_{k=1}^{\frac{K+1}{2}}{M_{k}^{\text{B}}}$ bits sent by Bob.

Let us note that the assumptions by virtue of which groups of original parties are represented by Alice and Bob may, in principle, pose a certain violation of local randomness, the quantification of which is beyond our aims. This observation does not invalidate, by any means, the final results here. This is so because we are finding a \emph{lower} bound for the private-coin $K\text{-user}$ protocol, and the assumption that certain pairs of users could have access to shared randomness would signify that they are actually transmitting \emph{less} information than the required amount. As the only consequence, our lower bound would not be as tight as with a pure private-coin model. In fact, the same approach was used, in another context, for the classical protocol in the proof of Lemma 3.2 of \cite{Fischer}: \emph{The K-player protocol induces a two-player protocol}.

Without loss of generality, let us suppose now that the average number of the bits sent by Bob's parties is greater than or equal to the average number of bits sent by the players represented by Alice, i.e.
\begin{equation}
\bar{M}_{{}}^{B}\ge \bar{M}_{{}}^{\text{A}},\text{~~~}\bar{M}_{{}}^{\text{A}}=\frac{2}{K-1}\sum\limits_{k=1}^{\frac{K-1}{2}}{M_{k}^{\text{A}}},\text{~~~}\bar{M}_{{}}^{\text{B}}=\frac{2}{K+1}\sum\limits_{k=1}^{\frac{K+1}{2}}{M_{k}^{\text{B}}}.
\label{eqC2}
\end{equation}
Applying (\ref{eqC1a}) and (\ref{eqC1b}) to Alice's and Bob's parties, assuming that each of the $K$ players owns exactly $N$ bits, we get
\begin{equation}
\cancel{\frac{K-1}{2}}N\le \cancel{\frac{K-1}{2}}{{\bar{M}}^{\text{A}}}\left\lceil \frac{4\ln (2)\left[ 2+(K+1){{{\bar{M}}}^{\text{B}}} \right]}{{{\left( 1-2\sqrt{{{p}_{\text{error}}}} \right)}^{2}}} \right\rceil \le \cancel{\frac{K-1}{2}}{{\bar{M}}^\text{B}}\left\lceil \frac{4\ln (2)\left[ 2+(K+1){{{\bar{M}}}^{\text{B}}} \right]}{{{\left( 1-2\sqrt{{{p}_{\text{error}}}} \right)}^{2}}} \right\rceil,
\label{eqC3}
\end{equation}
\begin{equation}
\cancel{\frac{K+1}{2}}N\le \cancel{\frac{K+1}{2}}{{\bar{M}}^{\text{B}}}\left\lceil \frac{4\ln (2)\left[ 2+(K-1){{{\bar{M}}}^{\text{A}}} \right]}{{{\left( 1-2\sqrt{{{p}_{\text{error}}}} \right)}^{2}}} \right\rceil \le \cancel{\frac{K+1}{2}}{{\bar{M}}^\text{B}}\left\lceil \frac{4\ln (2)\left[ 2+(K-1){{{\bar{M}}}^{\text{B}}} \right]}{{{\left( 1-2\sqrt{{{p}_{\text{error}}}} \right)}^{2}}} \right\rceil.
\label{eqC4}
\end{equation}
We take inequality (\ref{eqC4}) because it delivers a tighter lower bound on $\bar{M}_{{}}^{B}$, and we dismiss (\ref{eqC3}). Moreover, for simplicity, we will consider that all the $K$ parties each send the same amount of $M$ bits. We remark, though, that this latter simplification is not required to complete our development.

By replacing the term $K-1$ in (\ref{eqC4}) with $K$ to get rid of the ceiling function in an easy way, and, additionally, by completing the square we finally obtain
\begin{equation}
N\le \frac{4\ln (2)\left( 2+KM \right)M}{{{\left( 1-2\sqrt{{{p}_{\text{error}}}} \right)}^{2}}}<\frac{4\ln (2)}{{{\left( 1-2\sqrt{{{p}_{\text{error}}}} \right)}^{2}}}{{\left( M\sqrt{K}+\frac{1}{\sqrt{K}} \right)}^{2}}.
\label{eqC5}
\end{equation}
Solving for $M$ in the inequality above, we get a lower bound on the number of transmitted bits per user required in a classical $K$-user fingerprinting protocol as
\begin{equation}
M>\frac{\left( 1-2\sqrt{{{p}_{\text{error}}}} \right)\sqrt{N}}{2\sqrt{K\ln 2}}-\frac{1}{K} \text{~~[bits/user]}.
\label{eqC6}
\end{equation}
Following a similar procedure as above if $K$ is even, the same result in (\ref{eqC6}) is also valid as a lower bound.

\end{document}